\documentclass[3p]{elsarticle}
\usepackage{graphicx}
\usepackage{subfigure}
\usepackage{amsmath}
\usepackage{mathrsfs}
\usepackage{amssymb}
\usepackage{bm}
\usepackage{enumerate}
\usepackage{algorithm}
\usepackage{algorithmic}

\usepackage{lineno,hyperref}

\journal{XXXXX}

\begin{document}

\begin{frontmatter}

\title{Low-Latency Compression of Mocap Data Using Learned Spatial Decorrelation Transform\tnoteref{mytitlenote}}
\tnotetext[mytitlenote]{This research, which is carried out at
BeingThere Centre, collaboration among IMI of Nanyang Technological
University (NTU) Singapore, ETH Zrich, and UNC Chapel Hill, is
supported by the Singapore National Research Foundation (NRF) under
its International Research Centre @ Singapore Funding Initiative and
administered by the Interactive Digital Media Programme Office
(IDMPO). Ying He is partially supported by MOE2013-T2-2-011 and MOE
RG23/15.}

\author{Junhui Hou and Lap-Pui Chau}
\address{School of Electrical and Electronics
Engineering, Nanyang Technological University, Singapore, 639798}
\ead{jhhou@ntu.edu.sg, elpchau@ntu.edu.sg}

\author{Nadia Magnenat-Thalmann}
\address{Institute for Media Innovation,
Nanyang Technological University, Singapore, 639798}
\ead{nadiathalmann@ ntu.edu.sg}

\author{Ying He\corref{mycorrespondingauthor}}
\address{School of Computer Engineering, Nanyang Technological
University, Singapore, 639798.}
\cortext[mycorrespondingauthor]{Corresponding author}
\ead{yhe@ntu.edu.sg}

\begin{abstract}
  Due to the growing needs of motion capture (mocap) in movie, video games, sports, etc.,
  it is highly desired to compress mocap data for efficient storage and transmission.
  Unfortunately, the existing compression methods have either high latency or poor compression performance,
  making them less appealing for time-critical applications and/or network with limited bandwidth.
  This paper presents two efficient methods to compress mocap data with low latency.
  The first method processes the data in a frame-by-frame manner so that it is ideal for mocap data streaming.
  The second one is clip-oriented and provides a flexible trade-off between latency and compression
  performance. It can achieve higher compression performance while keeping the latency fairly low and controllable.
  Observing that mocap data exhibits some unique spatial characteristics, we learn an orthogonal transform to reduce the spatial redundancy.
  We formulate the learning problem as the least square of reconstruction error regularized by orthogonality and sparsity,
  and solve it via alternating iteration.
  We also adopt a predictive coding and temporal DCT for temporal decorrelation in the frame- and clip-oriented methods, respectively.
  Experimental results show that the proposed methods can produce higher compression performance at lower computational cost and latency than the state-of-the-art methods.
  Moreover, our methods are general and applicable to various types of mocap data.
\end{abstract}

\begin{keyword}
Motion capture, data compression, transform coding, low latency,
optimization
\end{keyword}

\end{frontmatter}


\section{Introduction}
  As a highly successful technique, motion capture (mocap) has been widely used to animate virtual characters in distributed virtual reality applications and networked games
  \cite{capin1999avatars,gutierrez2003controlling}.
  Due to the large amount of data and the limited bandwidth of communication network, congestion, packet loss, and delay often occur in mocap data transmission.
  Therefore, mocap data compression, specially lossy compression, is necessary to facilitate storage and transmission.

  Thanks to its smooth and coherent nature, mocap data exhibits high degree of temporal and spatial redundancy, making compression possible.
  To date, many mocap compression algorithms have been proposed (see Section \ref{sec:related work}).
  Among these approaches, most are \emph{sequence-based} (e.g., \cite{chattopadhyay2007human,gu2009,tournier2009,Lin2011,vavsa2014,Hou2014tvcg,hou2014})
  in that they process all the frames of a mocap sequence at a time.
  These methods are able to achieve high compression performance.
  However, such a good compression performance comes at a price of high latency,
  i.e., a large number of frames have to be captured and stored before compression, making them more suitable for efficient storage.
  On the other hand, the \emph{frame-based} (e.g., \cite{kwak2011hybrid}) approaches aim at time-critical applications (e.g., interactive applications)
  due to their no-latency nature.
  Unfortunately, the existing frame-based methods have poor compressing performance compared with the sequence-based methods,
  since they cannot explore spatial and temporal correlation well.
  As none of the sequence- and frame-based methods is perfect,
  it is natural to consider the \emph{clip-based} (e.g., \cite{arikan2006,liu2006segment,Chew2011}) methods
  which segment mocap data into short clips, providing a trade-off between latency and compression performance.

  In this paper, we present two efficient methods for compressing mocap data with low latency.
  The first method processes the data in a frame-by-frame manner, hereby compressing the data without any inherent latency at all.
  The second one is clip-based and can achieve higher compression performance while keeping the latency fairly low and controllable.
  Since mocap data exhibits some unique spatial characteristics, we propose a learned spatial decorrelation transform (LSDT) to explore the spatial redundancy.
  Taking the data content into account, the LSDT learns an orthogonal matrix via an $\ell_0$-norm regularized optimization.
  Due to its data adapted nature, the proposed LSDT outperforms the commonly used data-independent transforms,
  such as discrete cosine transform (DCT) and discrete wavelet transform (DWT), in terms of compression performance.
  We also adopt a predictive coding and temporal DCT for temporal decorrelation in the frame- and clip-based methods, respectively.
  We observe promising experimental results and demonstrate that our methods can produce higher compression performance at lower computational cost and latency than state-of-the-art.

  The rest of this paper is organized as follows: Section \ref{sec:related work} comprehensively reviews previous work on mocap data compression.
  Section \ref{sec:overview} gives the proposed frame- and clip-based methods.
  Section \ref{sec:prososed_T} shows the key component of the proposed methods, i.e., the learned spatial decorrelation
  transform, followed by the experimental results and discussion in Section \ref{sec:experiments}.
  Finally, Section \ref{sec:con} concludes this paper.

\section{Related Work}\label{sec:related work}
  All compression schemes aim at exploiting correlations among the data, so does mocap data compression.
  In terms of decorrelation techniques, the existing mocap data compression algorithms can be roughly classified into four groups,
  which are reviewed and analyzed as follows.

\subsection{Principal Component Analysis (PCA)}

  As a very popular technique, principal component analysis projects the data onto few principal orthogonal bases to convert data
  into a smaller set of values of linearly uncorrelated data.

  Breaking the mocap database into short clips that are approximated by B\'{e}zier curves,
  Arikan \cite{arikan2006} performed clustered PCA to reduce their dimensionality.
  Liu and McMillan \cite{liu2006segment} projected only the keyframes on the PCA bases and interpolated the other frames via spline functions.
  Motivated by the repeated characteristics of human motions,
  Lin \emph{et al}.~\cite{Lin2011} projected similar motion clips into PCA space and approximated them by interpolating functions with range-aware adaptive quantization.
  Observing that distortion to each of the joints causes a different overall distortion,
  V\'{a}\v{s}a and Brunnett~\cite{vavsa2014} proposed perception-driven error metric so that important joints have a higher precision than that of joints with small impact.
  They presented a Lagrange multiplier-based preprocessing for adjusting the joint precision.
  After Lagrangian equalization, the entire mocap sequence is projected into PCA pose space.
  Then, PCA is applied to short clips for further reducing the temporal coherence.

  Principal geodesic analysis (PGA) is a generalization of PCA for handling the case where the data is sampled from curved manifolds.
  Tournier \emph{et al}. \cite{tournier2009} presented a PGA-based method for the poses manifold in the configuration space of a skeleton,
  leading to a reduced, data-driven pose parameterization.
  Compression is then obtained by storing only the approximate parameterization along with the end-joints and root-joints trajectories.

  Although PCA can decorrelate mocap data very well, its bases are data-dependent and usually difficult to compress.
  Therefore, one has to explicitly store the orthogonal bases, which reduces the overall compression performance.
  Furthermore, PCA is usually applied to the whole mocap sequence (e.g., \cite{karni2004compression,vavsa2014}), resulting in a high latency.

\subsection{Discrete Wavelet and Cosine Transforms}

  DCT and DWT are commonly used techniques for converting correlated
  data into frequency domain,
  in which energy mainly concentrates on sparse frequencies (or most transform coefficients tend to zero).
  DCT and DWT have been widely adopted in some video/image coding standards~\cite{H264overview,hevc2012}.
  Moreover, they also have been exploited in the compression of 3D geometric data,
  e.g., static/dynamic meshes \cite{gu2002geometry,HoufacialGV,KGV2014} and mocap data \cite{preda2007optimized}, \cite{Chew2011}, \cite{kwak2011hybrid}.

  Kwak and Bajic \cite{kwak2011hybrid} applied 1D DCT to the predictive residuals between consecutive frames for exploiting the spatial coherence.
  In contrast, Preda \emph{et al.} \cite{preda2007optimized} applied 1D DCT/DWT to the residuals of motion compensation along the temporal dimension.
  Beaudoin \emph{et al}.~\cite{beaudoin2007} and Firouzmanesh \emph{et al}.~\cite{Firouzmanesh2011} adopted 1D DWT to trajectories of degrees of freedom and selected the sparse wavelet coefficients by a perceptual-based metric.
  Observing that neither 1D DCT nor 1D DWT considers the spatial and temporal correlation simultaneously,
  Chew \emph{et al}.~\cite{Chew2011} used Fuzzy C-means clustering to represent the mocap clips as 2D arrays, on which 2D DWT was applied.

  As pointed out in~\cite{Hou2014tvcg}, mocap data have some unique features that distinguish them from natural videos/images.
  For example, applying 1D DCT/DWT to each trajectory produces sparsity in the transform domain, since each trajectory is a smooth spatial curve.
  However, it does not make sense to apply 1D DCT/DWT to each mocap frame due to the lack of smoothness in the
  frame (see the analysis in Section \ref{sec:prososed_T}).

\subsection{Mocap Data Favored Transforms}

  As general-purpose transforms, DWT and DCT are data-\textit{independent} so that one does not need to store the bases.
  In contrast, data-driven transforms are adaptive to the input data, thus, they can take advantage of their intrinsic structure.
  However, the adaptiveness comes at a price of storing the basis functions explicitly.

  Zhu \emph{et al}.~\cite{zhu2012quaternion} proposed an elegant sparse decomposition model for the quaternion space that decomposes human rotational motion into a dictionary part and a weight part.
  As a result, a linear combination of 3D motion is equivalent to quaternion multiplication and the weight of linear combination is a power operation on quaternion.
  They showed that the transformed weights are sparse, leading to good compression performance.
  However, the quaternion space sparse representation is computationally expensive, diminishing its application to long motion sequences.
  Hou \emph{et al}.~\cite{hou2014} represented a motion sequence as a third-order tensor, which exhibits strong correlation within and across its slices.
  They performed the canonical polyadic (CP) tensor decomposition to explore correlation within and among clips to realize dimensionality reduction.
  Recently, Hou \emph{et al.}~\cite{Hou2014tvcg} proposed the mocap data tailored transform (MDTT),
  which partitions the input motion into clips and then computes a set of data-dependent orthogonal bases by minimizing the least square of distortions.
  Computational results show that MDTT significantly outperforms the existing techniques (e.g., \cite{arikan2006,Lin2011,zhu2012quaternion,tournier2009}) in terms of both compression performance and runtime.
  However, due to the overhead of explicitly storing the orthogonal bases, MDTT is less appealing for the short motion sequence.
  Note that all of the above-mentioned methods~\cite{zhu2012quaternion,hou2014,Hou2014tvcg} have very high latency due to their sequence-based nature.

\subsection{Indexing-based Methods}
   Chattopadhyay \emph{et al}. \cite{chattopadhyay2007human} proposed a smart indexing algorithm for exploiting structural information derived from the human skeleton,
  where each floating point number is represented as an integer index, based on the statistical distribution of the floating point numbers in a motion matrix.
  Gu \emph{et al}. \cite{gu2009} organized the markers into a hierarchy where each node corresponds to a meaningful part of the human body and coded each body part separately.
  Then, the motion sequence is represented as a series of motion pattern indices with respect to a predefined dataset including various patterns.

\section{Overview}
 \label{sec:overview}
  Given a mocap sequence of $F$ frames, we denote its $i$-th frame by $\mathbf{m}_i^d =[d_1^i~d_2^i~\cdots~d_J^i]^\textsf{T}\in \mathbb{R}^{J}$,
  where $J$ is the number of key points (markers) and  $d:=\{x,y,z\}$ stands for the $d$-dimensional coordinate.
  Then the $d$-component of the motion sequence is represented by a $J$-by-$F$ matrix $\mathbf{M}^d=[\mathbf{m}_1^d~\mathbf{m}_2^d~\cdots~\mathbf{m}_F^d ]\in \mathbb{R}^{J\times F}$.
  Each row of $\mathbf{M}^d$ corresponds to the $d$-trajectory of a key point.
  We partition $\mathbf{M}^d$ into non-overlapping clips of equal length, denoted by $\widetilde{\mathbf{M}}^d\in \mathbb{R}^{J\times L}$, where $L$ is the clip length.

  The primary goal of data compression is to reduce redundancy or correlation in the data.
  As pointed out in~\cite{Hou2014tvcg}, a typical mocap sequence exhibits strong spatial correlation due to the highly coordinated and structured nature of key points,
  and strong local temporal correlation since the object moves smoothly at a relatively small time scale.
  Therefore, mocap compression aims at eliminating both types of correlation as much as possible.
  In following sections, we present two low-latency and high-efficiency methods for compressing mocap data.

  \begin{figure}[t]
  \centering \subfigure[Frame-based method]{\label{subfig:frame based}
  \includegraphics[width=4.20in]{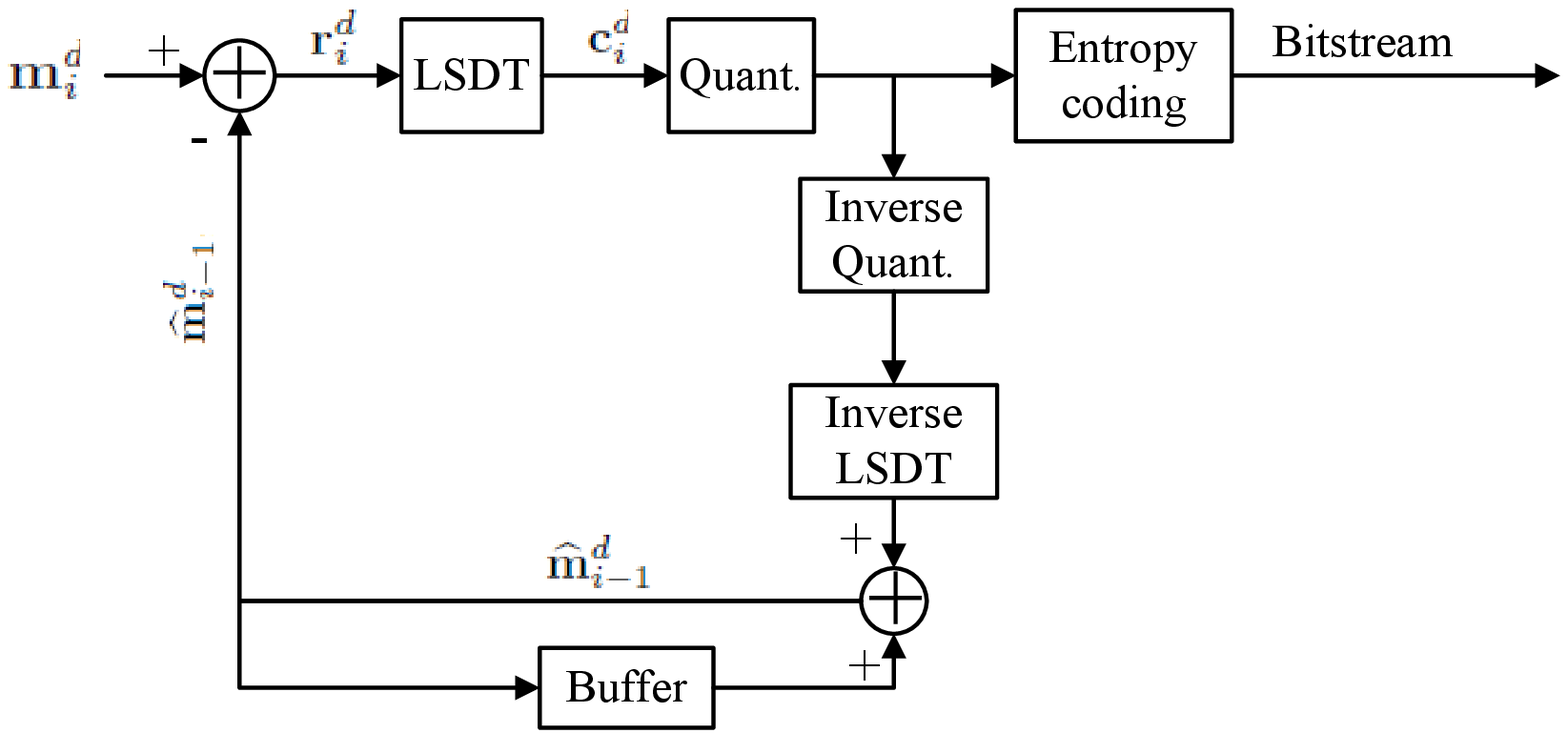}}\\
  \subfigure[Clip-based method]{\label{subfig:clip based}
  \includegraphics[width=4.20in]{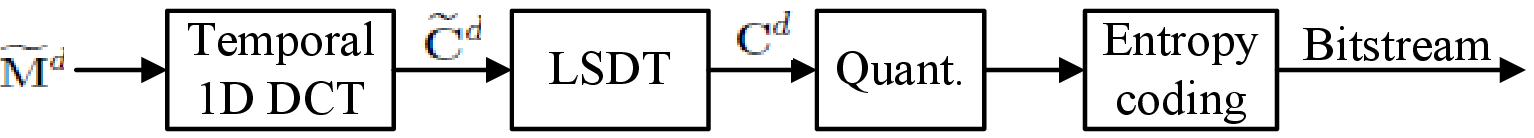}}
  \caption{The flowcharts of the proposed frame- and clip-based methods.
  } \label{fig:flowchart}
  \end{figure}

\subsection{Frame-based Method}
 \label{subsec:frame}

  As shown in Figure~\ref{subfig:frame based}, the frame-based method processes one frame at a time so that there is no inherent latency at all.
  Let us denote $\mathbf{B}^d$ the basis functions of the learned spatial decorrelation transform (LSDT) (to be presented in Section~\ref{sec:prososed_T}).
  For the first frame $\mathbf{m}_1$, we use $\mathbf{B}^d$ to remove its spatial correlation, i.e.,
  \begin{equation}
  \centering
  \mathbf{c}_1^d=\mathbf{B}^d\mathbf{m}_1^d.
  \end{equation}
  Then we adopt a simple predictive coding to the following frames to eliminate the temporal redundancy:
  the $i$-th frame is predicted only from the previous reconstructed one
  \begin{align}
  \mathbf{r}_i^d=\mathbf{m}_{i}^d-\widehat{\mathbf{m}}_{i-1}^d,~(i\geq 2)
  \end{align}
  where $\widehat{\mathbf{m}}_{i-1}^d$ is the reconstructed $(i-1)$-th frame, which is obtained by inverse quantization and inverse LSDT.
  Then, applying the spatial decorrelation transform $\mathbf{B}^d$ on the residual vector $\mathbf{r}^{d}_{i}$,
  we obtain
  \begin{equation}
  \centering
  \mathbf{c}_i^d=\mathbf{B}^d\mathbf{r}_i^d, \label{eqn:frameprojecting}
  \end{equation}
  where $\mathbf{c}_i^d\in \mathbb{R}^{J}$ are the transformed coefficients.

  Finally, we perform the hard thresholding operation and uniform quantization on $\mathbf{c}_i^d$.
  We store the following information for reconstruction:
  (1) the locations and values of nonzero elements, which are further entropy-coded using lossless coding, i.e., Huffman codes;
  (2) the number of nonzero elements in each coefficient vector, which is encoded using fixed-length encoding.

\subsection{Clip-based Method}

\label{subsec:clip}
  The frame-based scheme has no inherent latency at the price of relatively low compression performance,
  since it cannot fully exploit the temporal coherence.
  The clip-based scheme, in contrast, processes $L$ consecutive frames at a time, leading to better temporal decorrelation.
  With a proper $L$, the clip-based algorithm is a trade-off between latency and compression performance.

  Figure~\ref{subfig:clip based} shows the flowchart of the clip-based scheme.
  Let $\widetilde{\mathbf{M}}^d\in\mathbb{R}^{J\times L}$ be a clip of length $L$.
  Each row of $\widetilde{\mathbf{M}}^d$ corresponds to the $d$
  dimensional trajectory of a key point, i.e., a spatial curve.
  Thus, applying the 1D DCT to the rows of $\widetilde{\mathbf{M}}^d$ to explore the temporal correlation (see the analysis in Section \ref{sec:prososed_T}), we obtain
  \begin{equation}
  \widetilde{\mathbf{C}}^d=\widetilde{\mathbf{M}}^d\mathbf{U}_t ,
  \end{equation}
  where $\mathbf{U}_t\in \mathbb{R}^{L\times L}$ is the 1D DCT matrix.
  We then apply the LSDT to $\widetilde{\mathbf{C}}^d$ to further remove its spatial redundancy,
  \begin{equation}
  \mathbf{C}^d=\mathbf{B}^d\widetilde{\mathbf{C}}^d.
  \end{equation}
  Finally, we adopt the same quantization and entropy coding used in the frame-based method to encode $\mathbf{C}^d$ into bit stream.
  The sequence can be reconstructed by inverse quantization and inverse transform.

\section{Learned Spatial Decorrelation Transform}
  \label{sec:prososed_T}
  DCT and DWT decorrelate the data by converting it from spatial domain to frequency domain in a sparse form.
  They have been widely used for image and video compression \cite{H264overview,skodras2001jpeg}.
  DCT is suitable for signals which can be approximately modeled as a first-order Markov process (Markov-I) with the correlation coefficient 1,
  while DWT is particularly desired to piecewise signals \cite{wang2012introduction}.
  Note that each row of $\mathbf{M}^d$ corresponds to the $d$-dimensional trajectory of a key point,
  which can be viewed as Markov-I.
  Thus, it is reasonable to employ DCT to exploit the coherence within them.
  However, since the key points are organized in an irregular, tree-like structure (i.e., skeleton graph),
  the elements of $\mathbf{m}_i^d$ may not be correlated with their neighbors,
  meaning that columns of $\mathbf{M}^d$ do not follow Markov-I.
  Also note that the columns of $\mathbf{M}^d$ do not exhibit the piecewise smooth characteristic either.
  As a result, it does not make sense to apply DCT or DWT for de-correlation among the rows of $\mathbf{M}^d$.
  We refer readers to \cite{Hou2014tvcg} for quantitative analysis.
  As pointed out in \cite{lowdimension,tan2013human,tan2014motion}, mocap data lies in a relatively lower dimensional space,
  which are spanned by a set of specific bases.
  Based on the above analysis,
  we propose to learn an orthogonal transform to span the subspace of mocap data as much as possible.

  Given $N$ training frames $\{\mathbf{m}_i\}_{i=1}^{N}$, $\mathbf{m}_i\in\mathbb{R}^{J\times 1}$,
  the learned spatial decorrelation transform (LSDT) aims at finding an orthogonal matrix $\mathbf{B}\in\mathbb{R}^{J\times J}$
  so that it can transform each training frame into a sparse vector.
  We formulate the learning problem as follows:
  \begin{align}
  &\min_{\mathbf{B}^d\in \mathbb{R}^{J\times J}\atop\{\mathbf{e}_i^d\}\in\mathbb{R}^{J}}\sum_{i=1}^N \left\|\mathbf{B}^d\mathbf{m}_i^d-\mathbf{e}_i^d\right\|_2^2 \nonumber\\
  &~~~~~\textrm{subject~to}~~~\mathbf{B}^d{\mathbf{B}^d}^{\textsf{T}}={\mathbf{B}^{d}}^{\textsf{T}}\mathbf{B}^d=\mathbf{I},~~
  \left\|\mathbf{e}_i^d\right\|_0\leq P, \label{equ:p1}
  \end{align}
  where the $\ell_0$-norm $\|\mathbf{e}_i\|_0$ counts the number of non-zero entries in the transform coefficient of the $i$-th training sample,
  $P$ is the user-specified parameter controlling the sparsity in $\mathbf{e}_i$,
  and $\mathbf{I}\in \mathbb{R}^{J\times J}$ is the identity matrix.
  The orthogonality constraint on $\mathbf{B}^d$ allows us to obtain the inverse LSDT easily.
  Observe that the optimization problem in Equation (\ref{equ:p1}) is non-convex due to the non-convex constraints.
  We develop an alternating iterative method, which alternately solves the following two subproblems until convergence.

\subsection{The Sparse Vector Subproblem}

  With fixed $\mathbf{B}^d$, let
  $\mathbf{g}_i^d\triangleq\mathbf{B}^d\mathbf{m}_i^d$. The sparse vector subproblem is equivalent to the summation of multiple independent univariate minimization problems, in which the $i$-th one is written as
  \begin{equation}
  \centering \min_{\{\mathbf{e}_i^d\}}
  \left\|\mathbf{g}_i^d-\mathbf{e}_i^d\right\|_2^2~~{\rm subject~to}~~ \left\|\mathbf{e}_i^d\right\|_0\leq P.
  \label{equ:subp1}
  \end{equation}
  Obviously, the minimization is achieved only when $\mathbf{e}_i^d$ contains the largest $P$ entries (in magnitude) of $\mathbf{g}_i^d$
  which are at the corresponding locations.
  Therefore, we can compute $\mathbf{e}_i^d$ by setting the $(J-P)$ smallest (in magnitude) entries of $\mathbf{B}^d\mathbf{m}_i^d$ to zero:
  \begin{equation}
  \centering
  \mathbf{e}_i^d=\mathcal{T}\left(\mathbf{g}_i^d,J-P\right),
  \label{eqn:subp1-solution}
  \end{equation}
  where $\mathcal{T}$ is the truncating operation.

\subsection{The Orthogonal Matrix Subproblem}

  Given fixed sparse vectors $\mathbf{e}_i^d$, $i=1,\ldots,N$, let us denote $\mathbf{E}^d=[\mathbf{e}_1^d,\ldots,\mathbf{e}_N^d]$ the matrix representation.
  The orthogonal matrix subproblem is
  \begin{align}
  &\min_{\mathbf{B}^d}
  \left\|\mathbf{B}^d\mathbf{M}^d-\mathbf{E}^d\right\|_F^2~~
   {\rm
  subject~to}~\mathbf{B}^d{\mathbf{B}^d}^{\textsf{T}}={\mathbf{B}^{d}}^{\textsf{T}}\mathbf{B}^d=\mathbf{I},
  \label{eqn:subp2}
  \end{align}
  where $\|\cdot\|_F$ is the Frobenius norm of matrix and $\mathbf{M}^d$ is the matrix representation of all training frames.
  Observe that
  \begin{align}
  &\left\|\mathbf{B}^d\mathbf{M}^d-\mathbf{E}^d\right\|_F^2={\rm Tr}\left(\left(\mathbf{B}^d\mathbf{M}^d-\mathbf{E}^d\right)\left(\mathbf{B}^d\mathbf{M}^d-\mathbf{E}^d\right)^{\textsf{T}}\right)\nonumber\\
  &={\rm Tr}\left(\mathbf{M}^d{\mathbf{M}^d}^{\textsf{T}}\right)-2{\rm Tr}\left(\mathbf{B}^d\mathbf{M}^d{\mathbf{E}^d}^{\textsf{T}}\right)+{\rm Tr}\left(\mathbf{E}^d{\mathbf{E}^d}^{\textsf{T}}\right),\nonumber
  \label{eqn:expend}
  \end{align}
  where $\mathrm{Tr}$ is the matrix trace.

  Ignoring the first and third terms which are constant, the minimization problem in (\ref{eqn:subp2}) is equivalent to
  \begin{equation}
  \max_{\mathbf{B}^d} {\rm Tr}\left(\mathbf{B}^d\mathbf{M}^d{\mathbf{E}^d}^{\textsf{T}}\right)~~{\rm~subject~to}~\mathbf{B}^d{\mathbf{B}^d}^{\textsf{T}}={\mathbf{B}^{d}}^{\textsf{T}}\mathbf{B}^d=\mathbf{I}.
  \label{eqn:subp2-2}
  \end{equation}
  Factoring $\mathbf{M}^d{\mathbf{E}^d}^{\textsf{T}}$ using the singular value decomposition (SVD), we obtain
  $\mathbf{M}^d{\mathbf{E}^d}^{\textsf{T}}=\widetilde{\mathbf{U}}^d\mathbf{S}^d\widetilde{\mathbf{V}}^{d^\textsf{T}}$,
  where $\widetilde{\mathbf{U}}^d,~\widetilde{\mathbf{V}}^d\in \mathbb{R}^{J\times J}$ are two orthogonal matrices, and $\mathbf{S}^d$ is a diagonal matrix.

  Then we can rewrite the objective function as
  \begin{displaymath}
  {\rm Tr}\left(\mathbf{B}^d\mathbf{M}^d{\mathbf{E}^d}^{\textsf{T}}\right)={\rm
  Tr}\left(\mathbf{B}^d\widetilde{\mathbf{U}}^d\mathbf{S}^d\widetilde{\mathbf{V}}^{d^\textsf{T}}\right)
  ={\rm Tr}\left(\widetilde{\mathbf{B}}^d\widetilde{\mathbf{U}}^d\mathbf{S}^d\right),
  \end{displaymath}
  where $\widetilde{\mathbf{B}}^d=\widetilde{\mathbf{V}}^{d^\textsf{T}}\mathbf{B}^d$ is still an orthogonal matrix.

  Since $\mathbf{S}^d$ is a diagonal matrix, maximizing (\ref{eqn:subp2-2}) is equivalent to maximize the diagonal entries of $\widetilde{\mathbf{B}}^d\widetilde{\mathbf{U}}^d$.
  With Cauchy-Schwartz inequality, the $i$-th diagonal entry of $\widetilde{\mathbf{B}}^d\widetilde{\mathbf{U}}^d$ is
  \begin{displaymath}
  \sum_{j=1}^{J}\widetilde{\mathbf{B}}_{ij}^d\widetilde{\mathbf{U}}_{ji}^d\leq
  \sqrt{\sum_{j=1}^J\widetilde{\mathbf{B}}_{ij}^{d^2}\sum_{j=1}^J\widetilde{\mathbf{U}}_{ji}^{d^2}}=1.
  \end{displaymath}
  The last equation comes from the fact that both $\widetilde{\mathbf{B}}$ and $\widetilde{\mathbf{U}}$ are orthogonal matrices.
  Therefore, the objective function in (\ref{eqn:subp2-2}) is maximized when $\widetilde{\mathbf{B}}^d\widetilde{\mathbf{U}}^d=\mathbf{I}$,
  leading to
  \begin{equation}
  \centering
  \mathbf{B}^d=\widetilde{\mathbf{V}}^d\widetilde{\mathbf{U}}^{d^\textsf{T}}.
  \label{eqn:subp2-solution}
  \end{equation}

  Algorithm \ref{Alg:LOT} shows the pseudocode of the LSDT algorithm.
  In each iteration, the truncating operation (line 4) and matrix multiplication (lines 6 and 7) take $O(J\log J)$ and $O(2NJ^2)$ time, respectively.
  Singular value decomposition has an $O(J^3)$ time complexity.
  Putting it all together, the time complexity of Algorithm 1 is $\mathcal{O}(KNJ^2+KNJ\log J+KJ^3)$.
  Although there is no theoretical guarantee of the convergence of our algorithm,
  each subproblem does have an exact solution and we observe that it converges in a few hundred iterations on training datasets (see Section \ref{subsec:imapct}).

  \begin{algorithm}
  \caption{Computing LSDT Bases for Mocap Data}
  \textbf{Input}: training samples $\{\mathbf{m}_i\}_{i=1}^N$, the sparsity parameter $P$ and the maximum number of iterations $K$\\
  \textbf{Output}: the orthogonal matrix $\mathbf{B}^d$
  \begin{algorithmic}[1]
  \STATE initialize $\mathbf{B}^d$ using an orthogonal matrix (e.g., DCT or DWT bases)
  \FOR {$iter \leftarrow 1:K$}
    \FOR {$i\leftarrow 1:N$}
   \STATE update $\mathbf{e}_i^d$ using (\ref{eqn:subp1-solution})
   \ENDFOR
   \STATE factor $\mathbf{M}^d{\mathbf{E}^d}^\textsf{T}$ using SVD
   \STATE update $\mathbf{B}^d$ using (\ref{eqn:subp2-solution})
   \ENDFOR
  \end{algorithmic}
  \label{Alg:LOT}
  \end{algorithm}

\section{Experimental Results and Discussion}
 \label{sec:experiments}
  We implement our methods in MATLAB using only 200 lines of codes and evaluate them on the CMU Mocap Database\footnote{http://mocap.cs.cmu.edu/},
  in which each frame consists of $J=31$ key points (i.e., joints of the human skeleton) sampled at 120 frames per second (fps).
  We store each coordinate of the original data as a 32-bit float and hereby represent one key point using 96 bits.
  Table \ref{tab:sequences} describes the training and test motion sequences and their lengths.

  The compression distortion $D$ is measured by the average Euclidean distance between the original joint location
  $\mathbf{p}_{i,j}:=\{x_{i,j},y_{i,j},z_{i,j}\}^\textsf{T}$ and the reconstructed location
  $\widehat{\mathbf{p}}_{i,j}:=\{\widehat{x}_{i,j},\widehat{y}_{i,j},\widehat{z}_{i,j}\}^\textsf{T}$ (in cm),
  \begin{equation}
  \centering D=\frac{1}{JF}\sum_{i=1}^J \sum_{j=1}^F \left\|\mathbf{p}_{i,j}-\widehat{\mathbf{p}}_{i,j}\right\|_2.
  \end{equation}
  The compression ratio (CR) is the ratio between the original data size and the compressed data size.
  The compression is determined by the quantization bit, that is, a larger quantization bit induces smaller distortion at a smaller CR.

  \begin{table}
  \centering \caption{Description of training sequences and test sequences.}
  \setlength\tabcolsep{1pt}
  \label{tab:sequences}
  \begin{small}
  \begin{tabular}{l|l|l|l}
  \hline
  Sequence& $F$& Size (kB) &Description  \\
  \hline
  \hline
  86\_02 & 10,617&3,856.9&walk, squats, run, stretch, jumps,punches, and drinking\\
  \hline
  56\_04 & 6767  & 2,458.3 & fists up, wipe window, grab, walk, throw punches, yawn, stretch, jump\\
  \hline
  15\_05 & 22948 &  8,336.5 & wash windows, paint, hand signals, dance, dive, twist, boxing\\
  \hline
  \hline 14\_08 & 2,625  & 953.6   & jump up to grab\\
  \hline 15\_04 & 22,549 & 8,191.6 & dance, the twist, boxing \\
  \hline 17\_08 & 6,179  & 2,244.7 & muscular person's walk\\
  \hline 17\_10 & 2,783  & 1,011   & boxing\\
  \hline 41\_07 & 7,536  & 2,737.7 &climb, step over, jump over\\
  \hline 49\_02 & 2,085  & 757.4   &jump, hop on one foot\\
  \hline 56\_07 & 9,420  & 3,422.1 &yawn, stretch, walk, run, halt\\
  \hline 85\_12 & 4,499  & 1,634.4 & jumps, flips, breakdance\\
  \hline 86\_05 & 8,340  & 3,029.7 & walking, jumping, punching\\
  \hline
  \end{tabular}
  \end{small}
  \end{table}

\subsection{Training the LSDT Bases}

 \label{subsec:imapct}

  \begin{figure}[t]
  \centering
  \subfigure[1D DCT bases $\mathbf{U}_t$ ]{\includegraphics[width=1.55in]{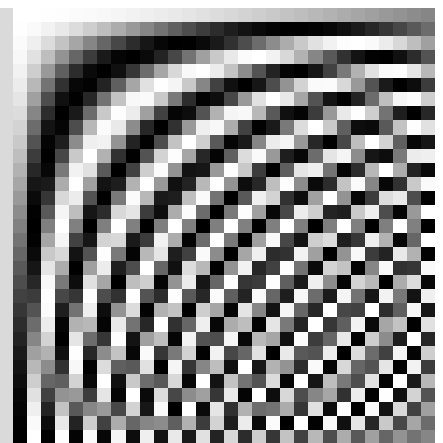}}
  \subfigure[LSDT bases $\mathbf{B}^x$]{\includegraphics[width=1.55in]{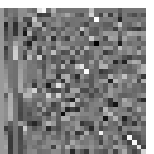}}
  \subfigure[LSDT bases $\mathbf{B}^y$]{\includegraphics[width=1.55in]{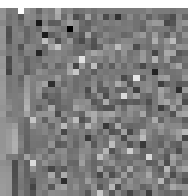}}
  \subfigure[LSDT bases $\mathbf{B}^z$]{\includegraphics[width=1.55in]{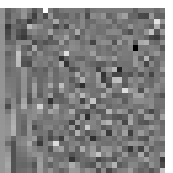}}
  \caption{Visualization of the 1D DCT and LSDT bases, where the greyscale color indicates the normalized function value.
  In each square matrix, a column corresponds to one basis function and frequencies increase from left to right.
  } \label{fig:basis}
  \end{figure}
  \begin{figure}[h]
  \centering \subfigure[$\mathbf{B}^x$]{
  \includegraphics[width=2.0in]{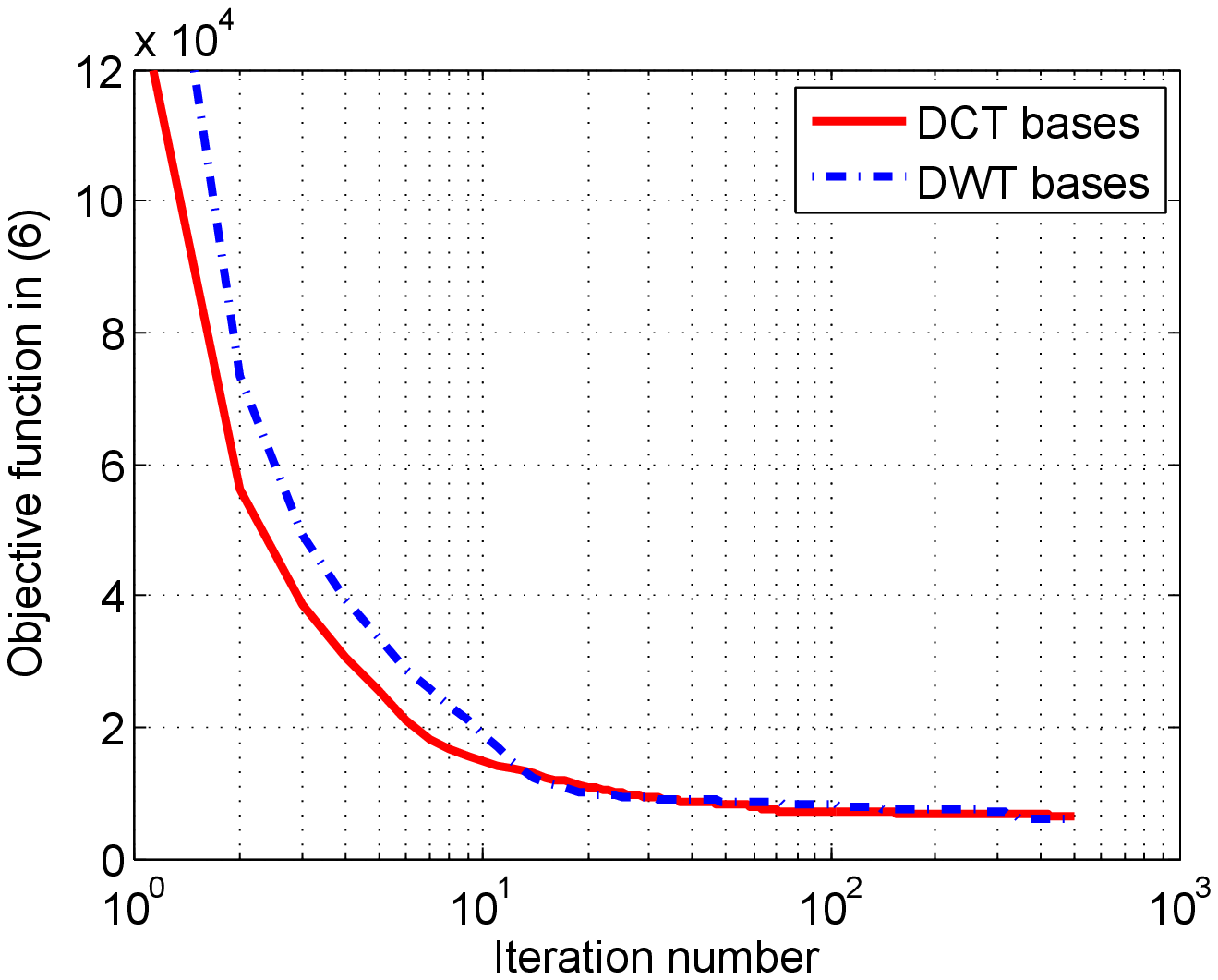}}
  \subfigure[$\mathbf{B}^y$]{
  \includegraphics[width=2.0in]{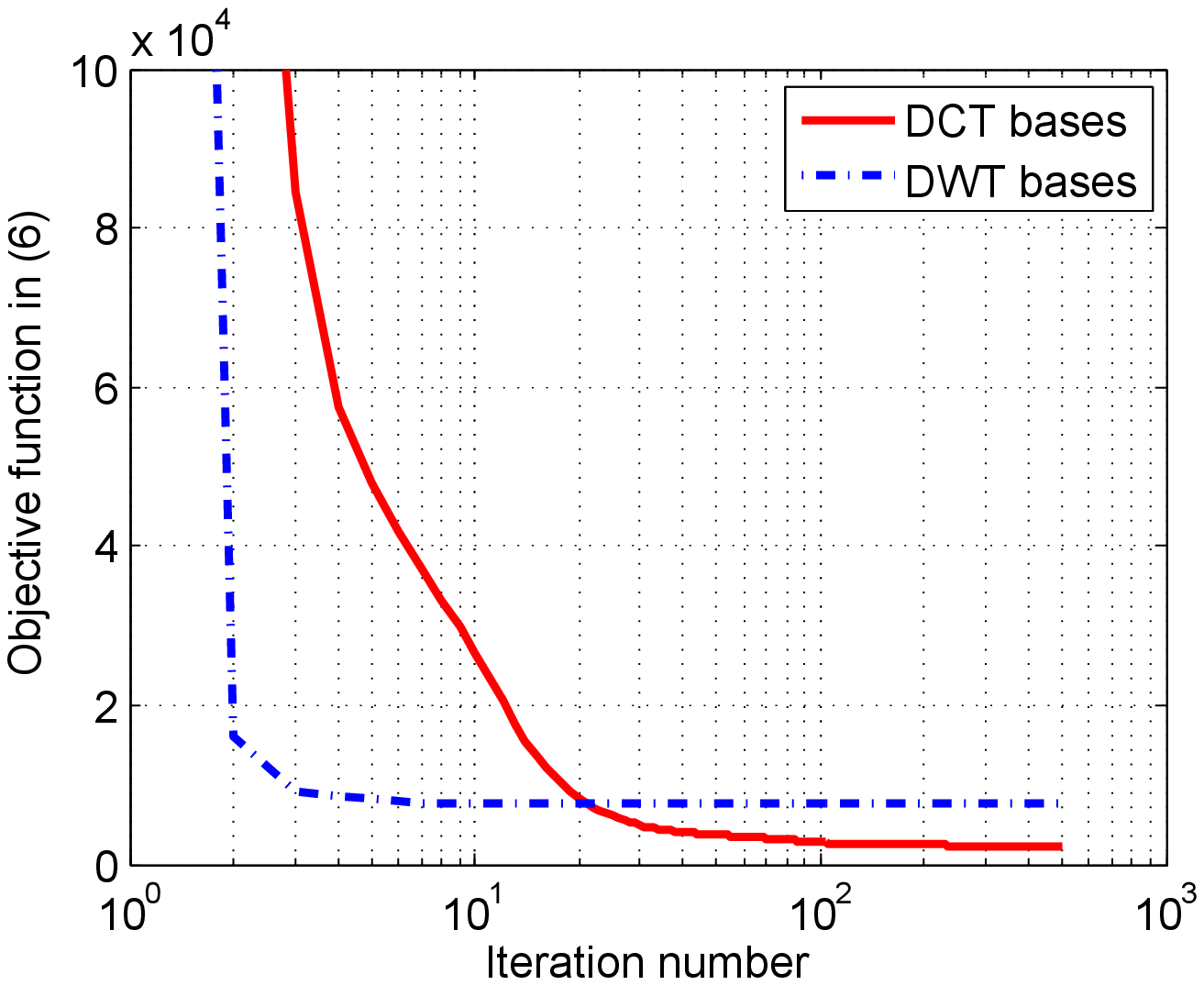}}
  \subfigure[$\mathbf{B}^z$]{
  \includegraphics[width=2.0in]{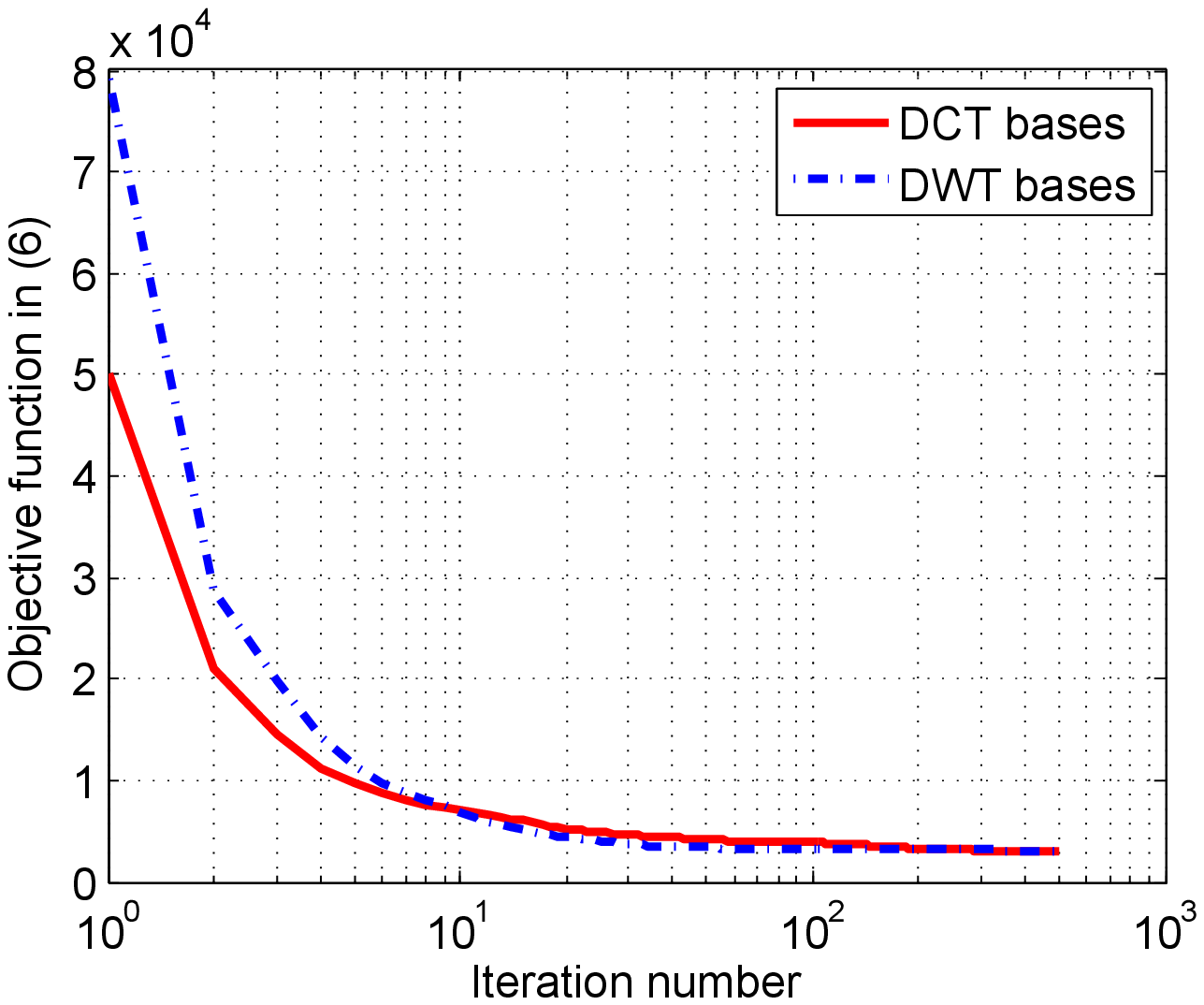}}
  \caption{Convergence plots of Algorithm \ref{Alg:LOT} with two different initializations.
  \# training frames $N$=10,617; sparsity parameter $P$=8.
  (a), (b), and (c) correspond to $x$, $y$, and $z$-coordinates, respectively.
  } \label{fig:behavior}
  \end{figure}
\begin{figure}[h]
\centering
\includegraphics[width=1.5in]{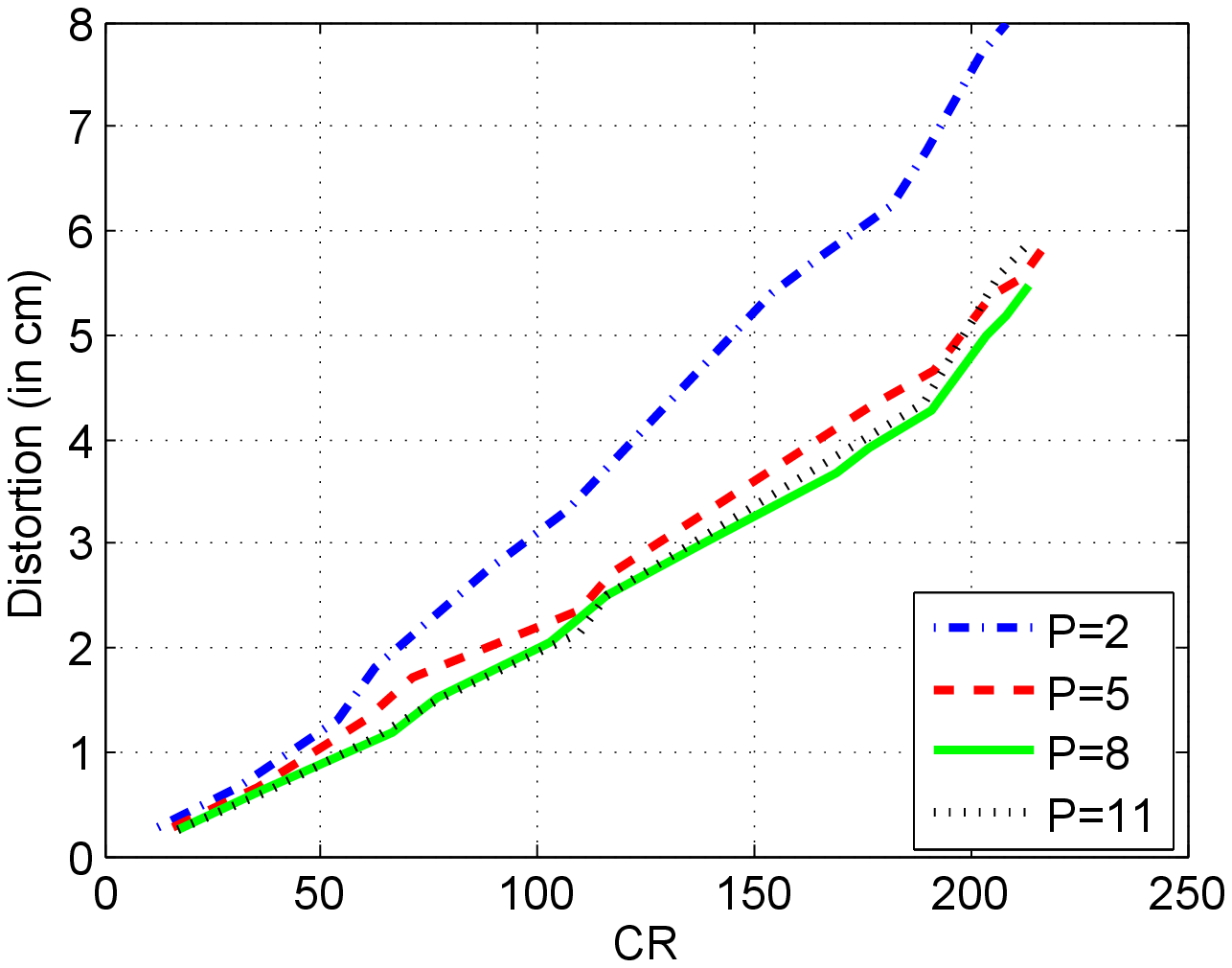}
\includegraphics[width=1.5in]{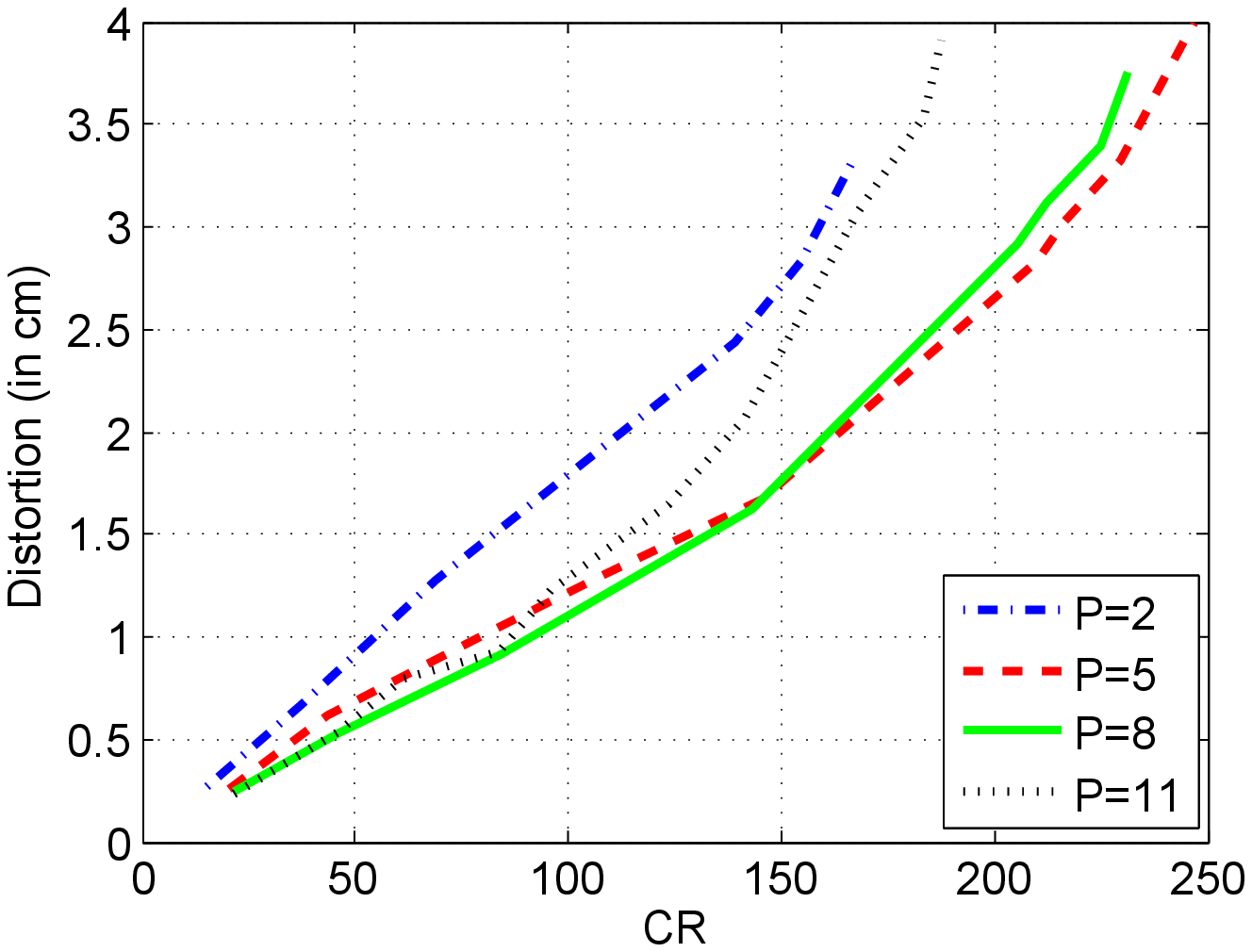}
\includegraphics[width=1.5in]{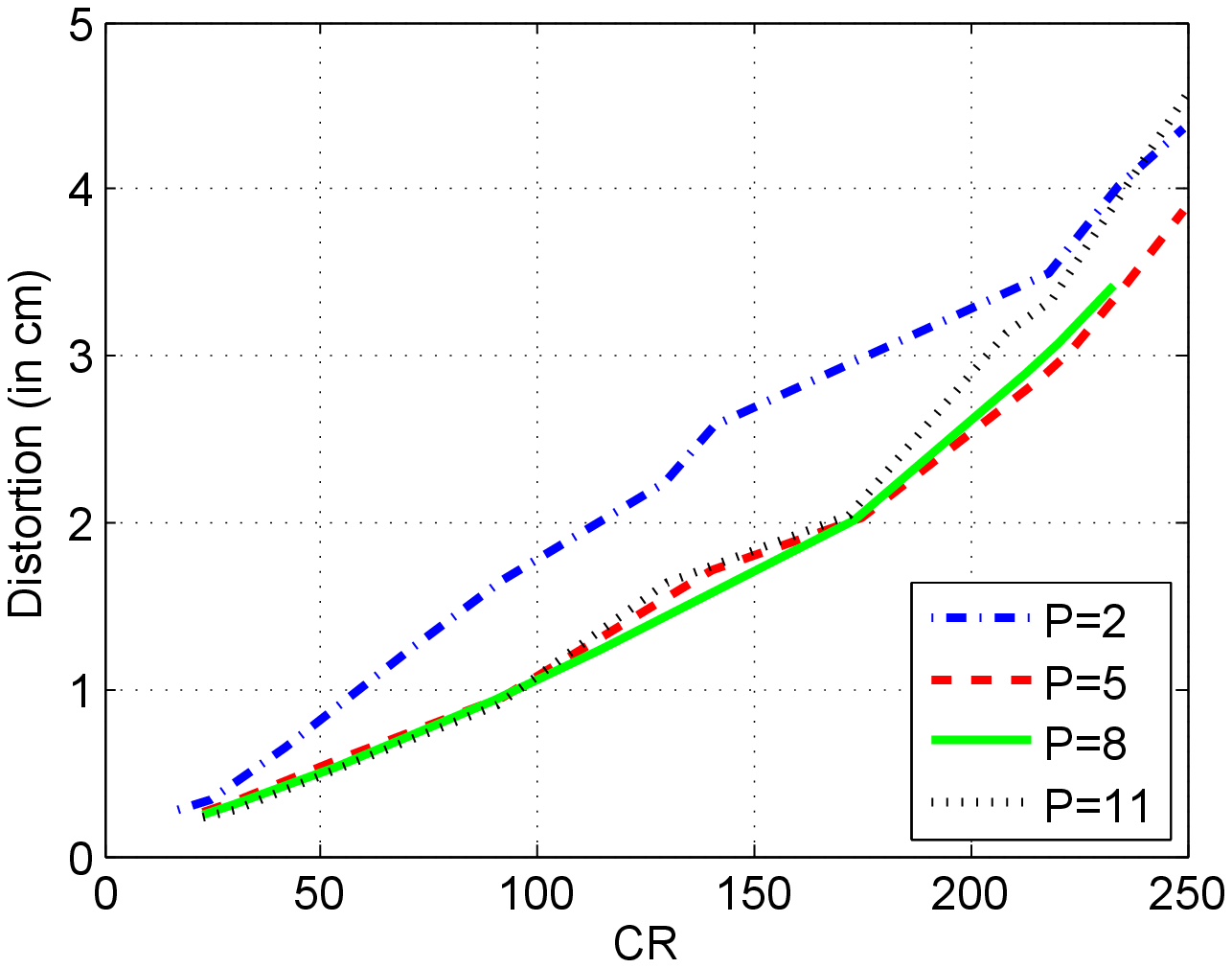}
\includegraphics[width=1.5in]{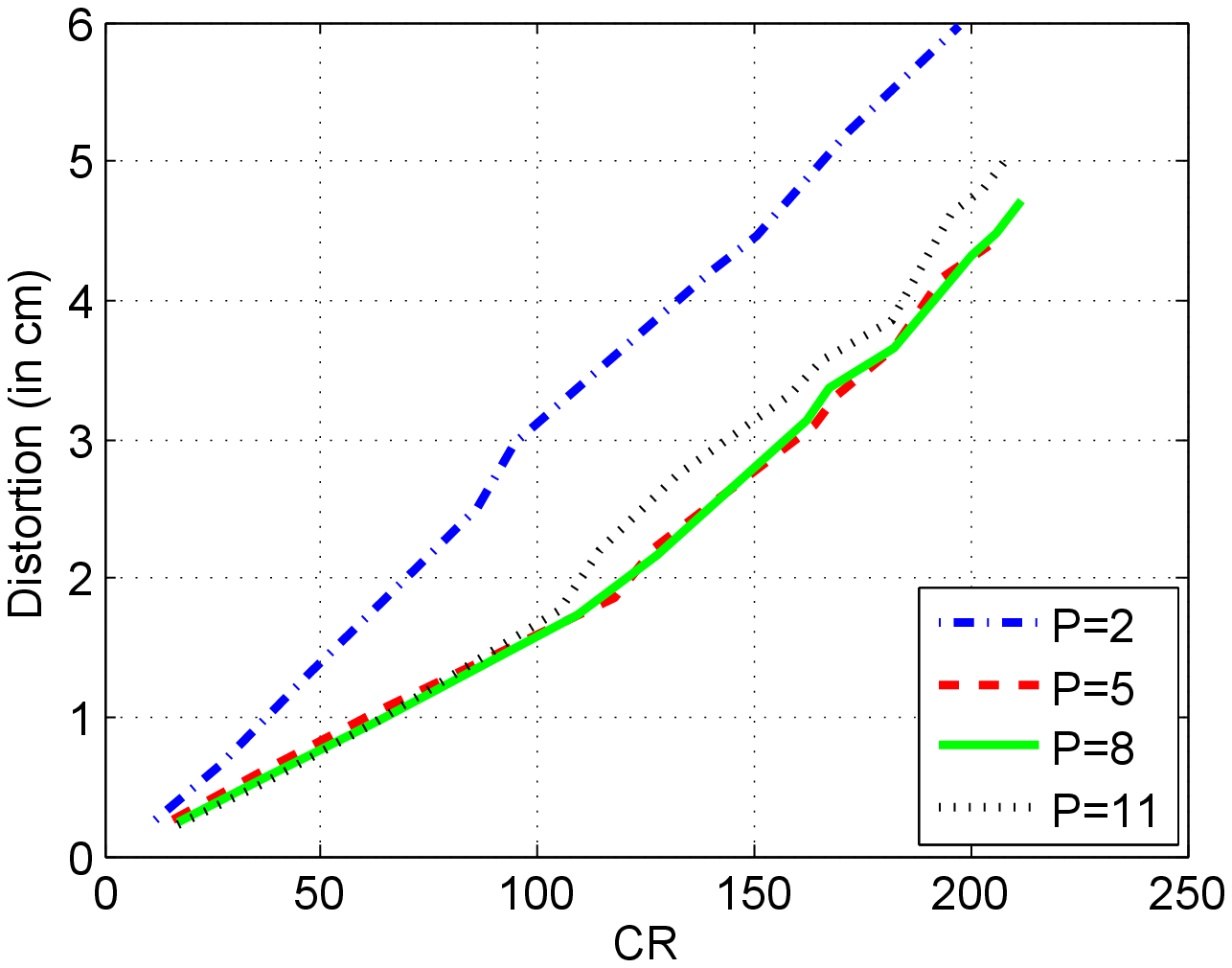}\\
\subfigure[14\_08]{
\includegraphics[width=1.5in]{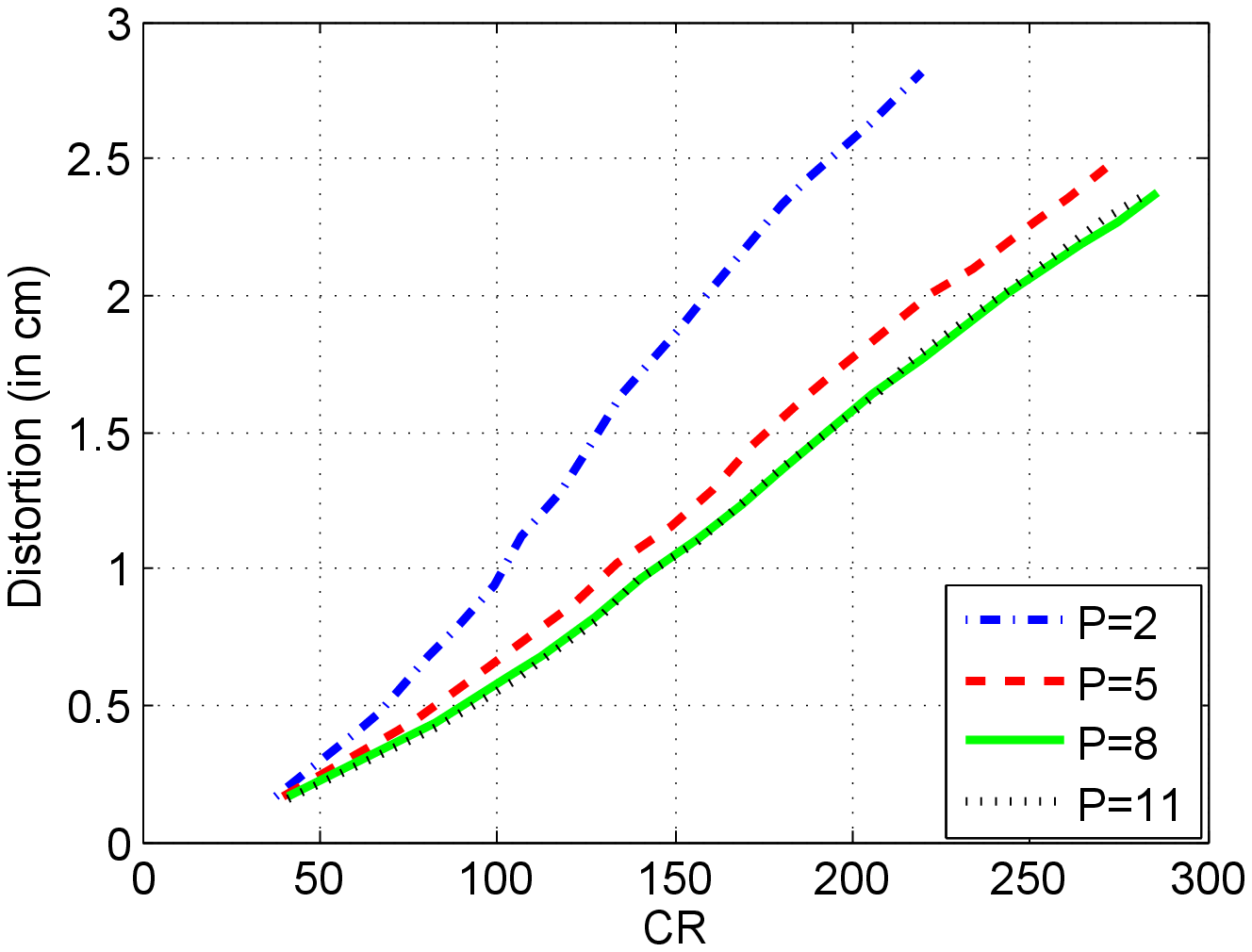}}
\subfigure[17\_08]{
\includegraphics[width=1.5in]{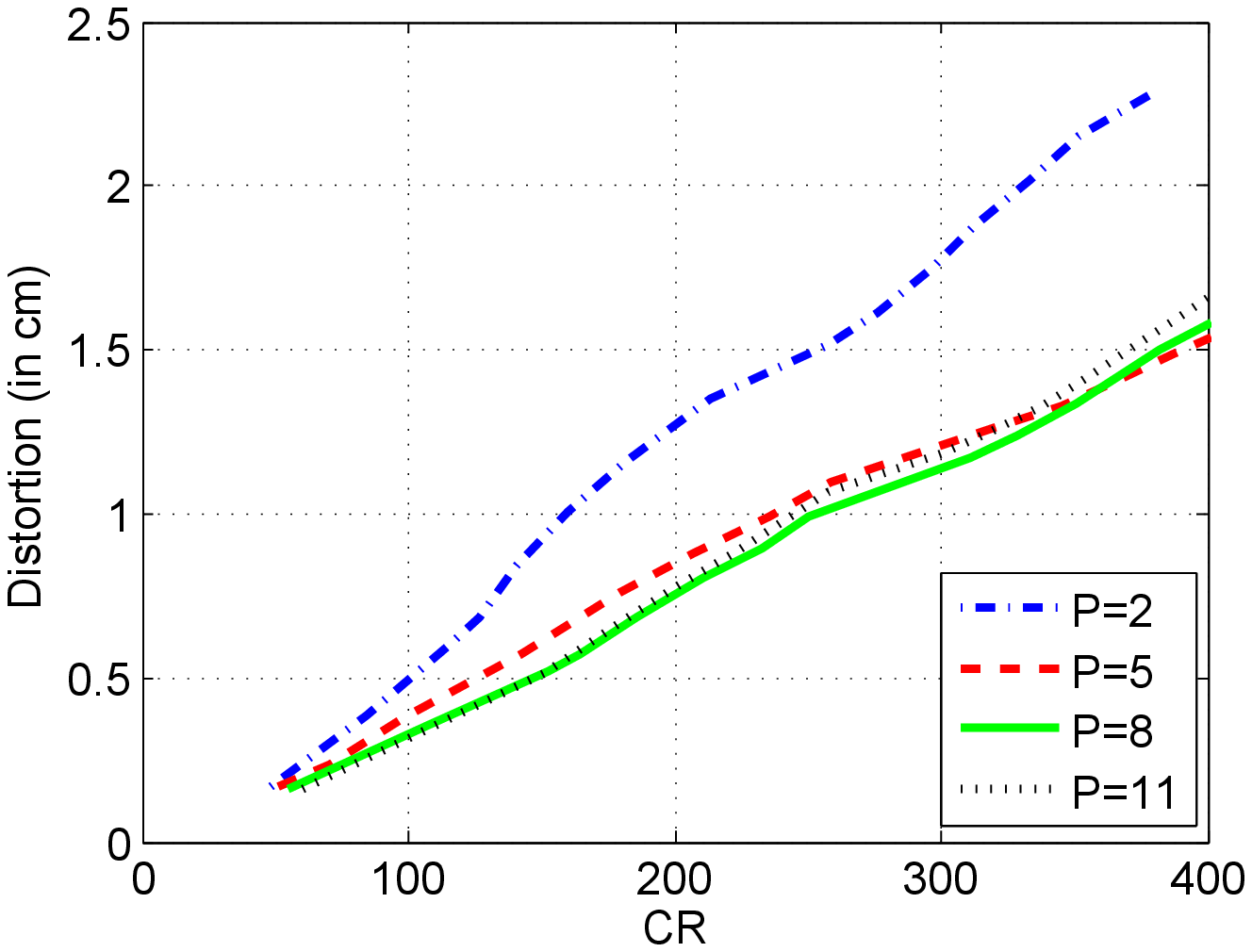}}
\subfigure[41\_07]{
\includegraphics[width=1.5in]{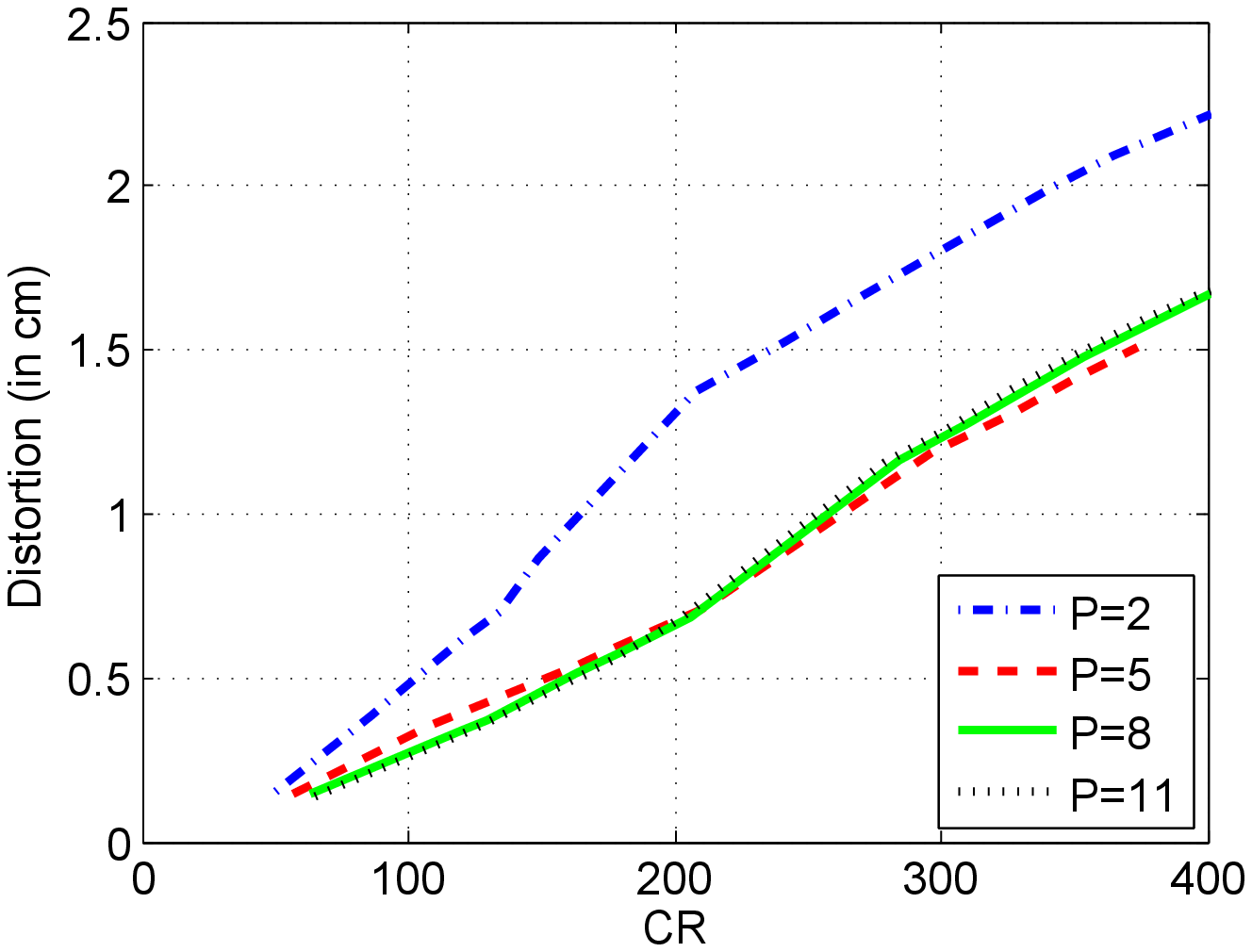}}
\subfigure[49\_02]{
\includegraphics[width=1.5in]{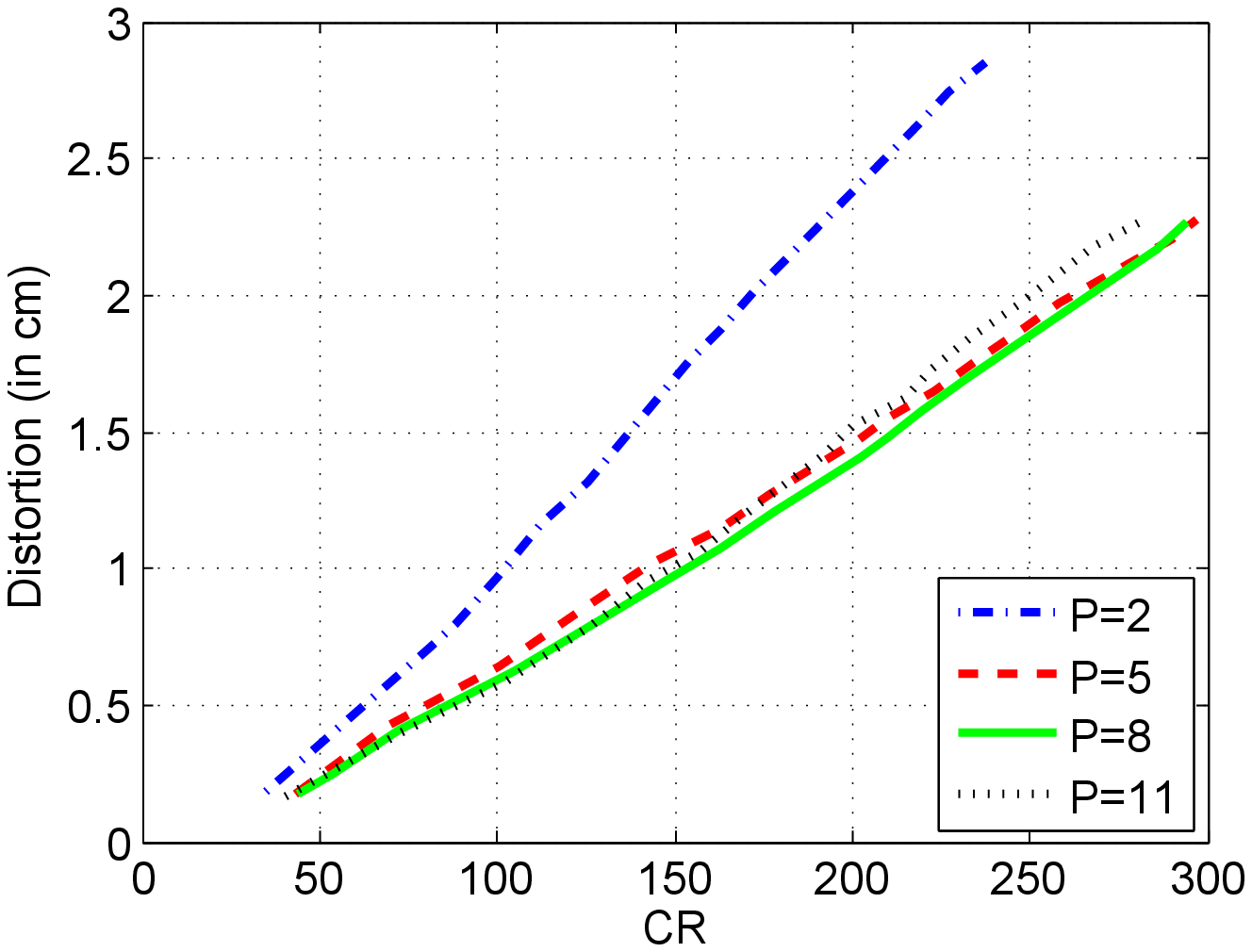}}
\caption{The impact of the sparsity parameter $P$ on the overall
compression performance. The top and bottom rows correspond to the
frame- and clip-based ($L=240$) schemes, respectively.
$\mathbf{B}^d$ is initialized using the DCT bases.}
\label{fig:P-CR-D}
\end{figure}
  We take sequences ``86\_02" ``56\_04",  and ``15\_05" as the training datasets, which consist of various types of human motion.
  It is worth noting that more training frames can generate better performance, but the computational cost also increases.
  Thus, it is a tradeoff between quality and efficiency.

  The LSDT bases training algorithm (cf. Algorithm \ref{Alg:LOT}) is an iterative algorithm.
  We evaluate the convergence rate of the training algorithm on two types of initializations, 1D DCT bases and 1D DWT bases realized by the 3-level ``Haar" wavelet.
  As Figure \ref{fig:behavior} shows, the objective function converges to almost the same value after a few hundred iterations,
  meaning that the output of Algorithm \ref{Alg:LOT} is intrinsic, which does not depend on initialization.
  Figure~\ref{fig:basis} also visualizes the bases of 1D DCT and LSDT to show the difference between them.

  The parameter $P$, specifying the sparsity of transform coefficients during the learning procedure,
  directly affects the structure of the learned orthogonal matrix $\mathbf{B}^d$, which in turn controls the compression performance.
  In the training process, we set $P$ to four different values: 2, 5, 8, and 11.
  Then, the learned orthogonal matrices under different $P$ are tested in the frame- and clip-based methods, respectively.
  For both schemes, four randomly chosen sequences with various motion characteristics and lengths are compressed,
  and the results are shown in Figure \ref{fig:P-CR-D}, where we can see that the best compression performance is achieved  when the value of $P$ is equal to 8.

  \subsection{Evaluating the Spatial Decorrelation Transforms}
  \label{subsec:evaluation_sdt}
\begin{figure}[t]
\centering \subfigure[14\_08]{
\includegraphics[width=1.5in]{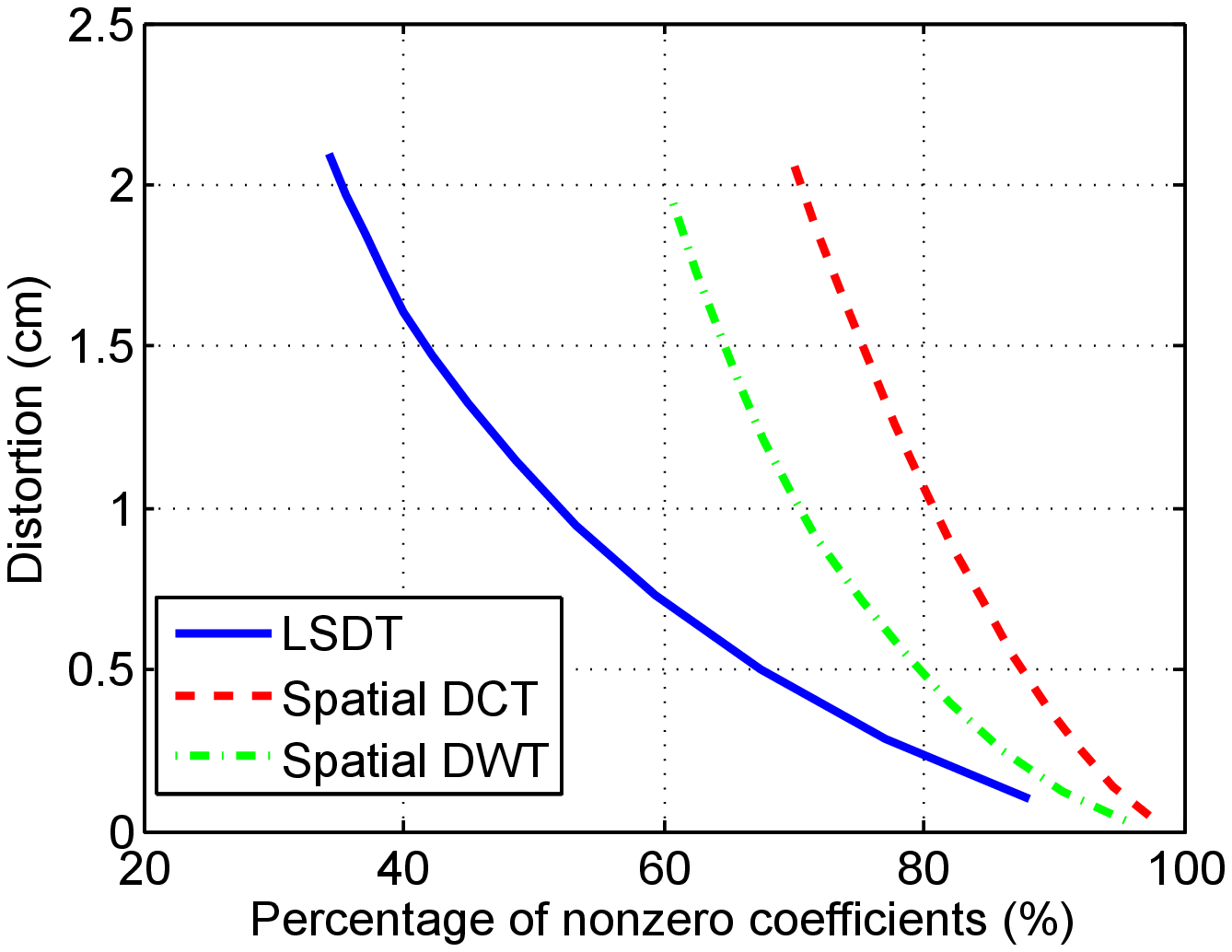}}
\subfigure[17\_08]{
\includegraphics[width=1.5in]{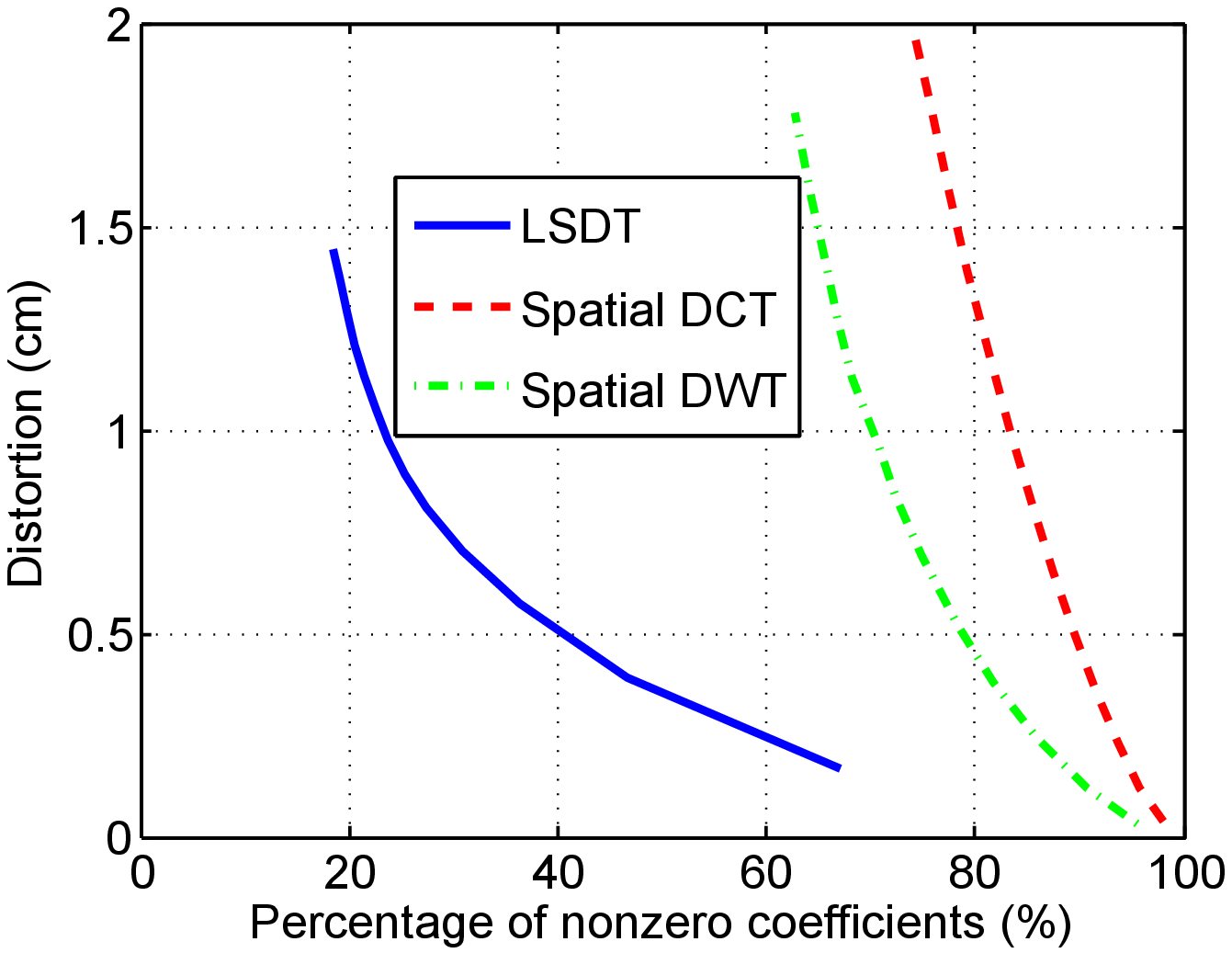}}
\subfigure[41\_07]{
\includegraphics[width=1.5in]{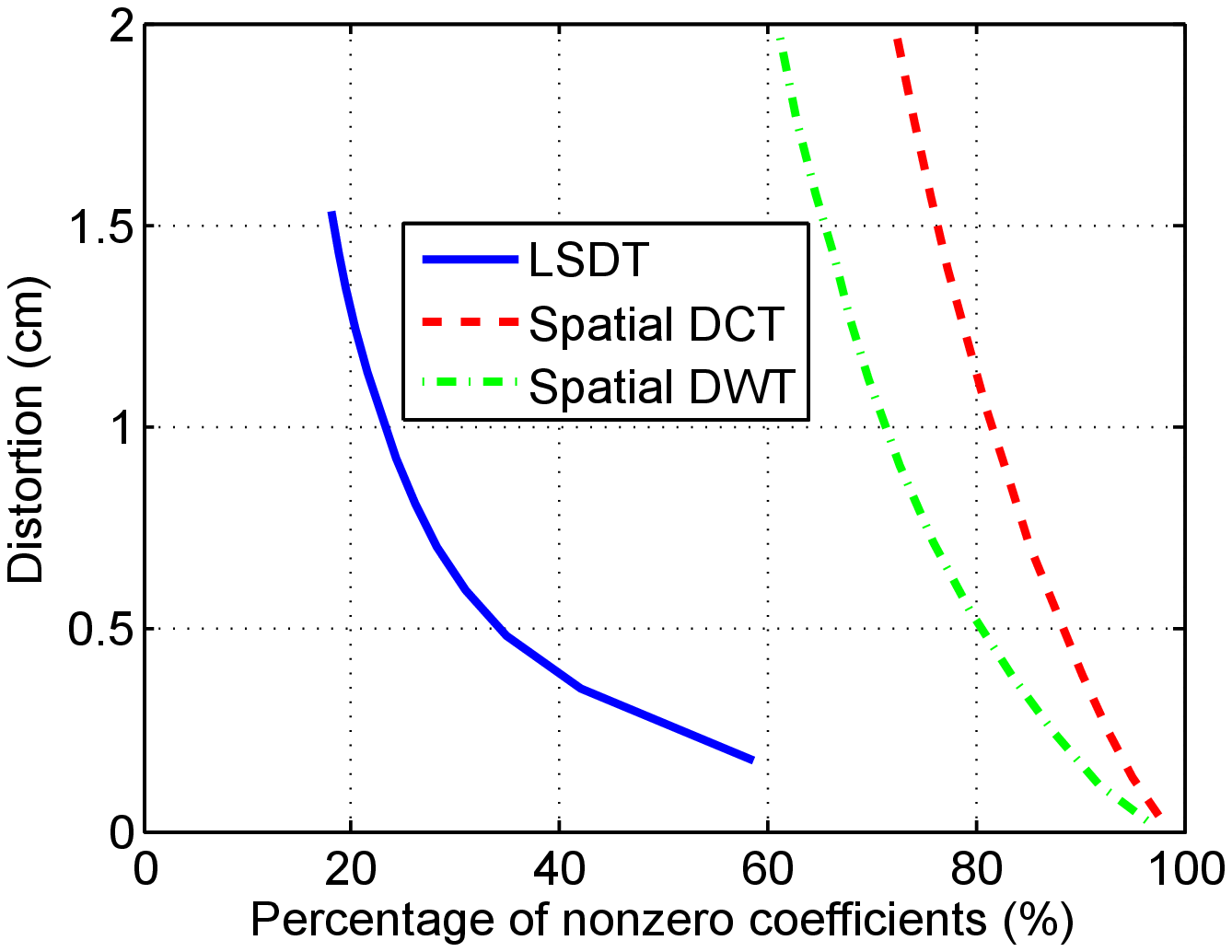}}
\subfigure[56\_07]{
\includegraphics[width=1.5in]{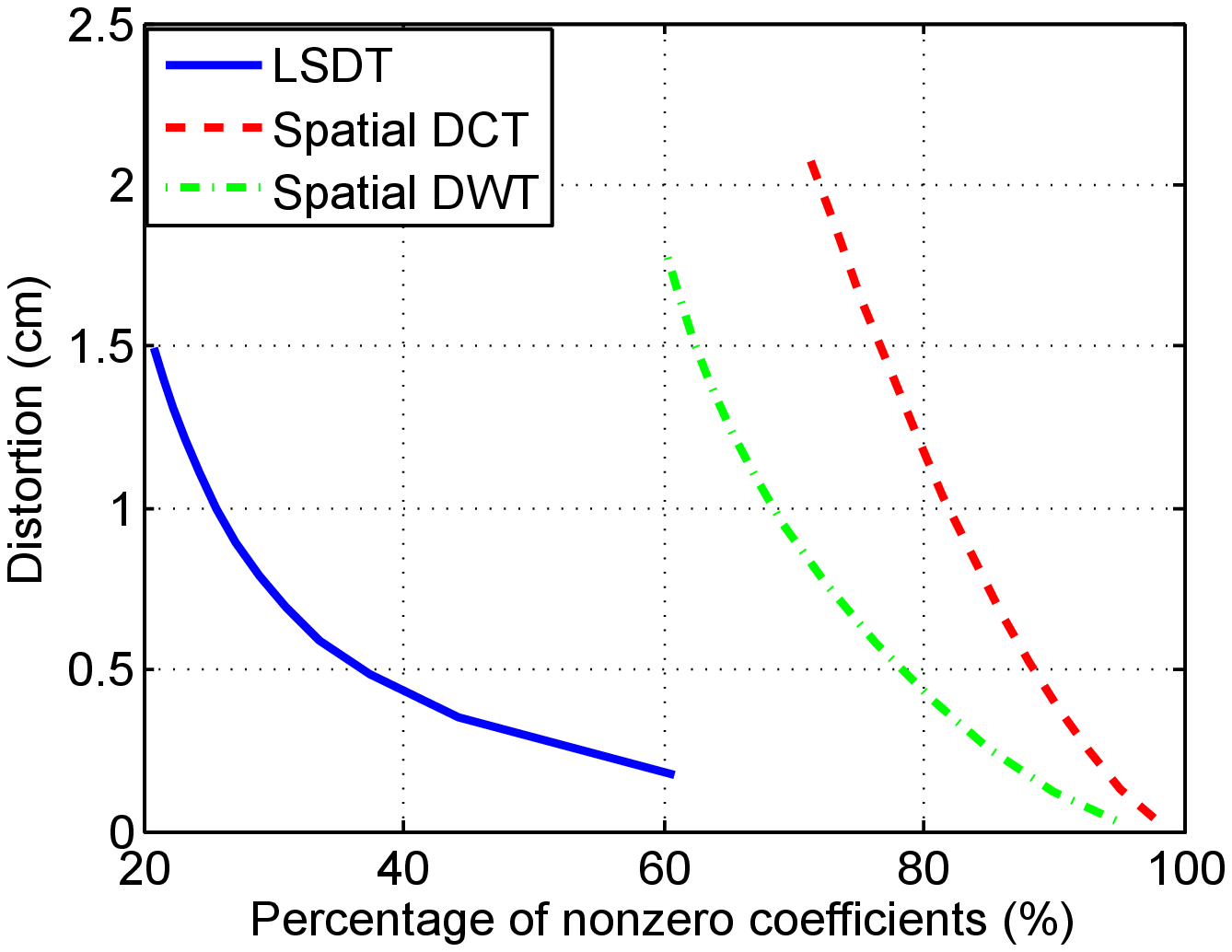}}
\caption{Evaluating the performance of spatial decorrelation of the
proposed LSDT. The horizontal axis shows the percentage of nonzero
transformed coefficients. LSDT performs the best among the three
SDTs.} \label{fig:sparsity-distortion}
\end{figure}

  We compare the performance of several spatial decorrelation transforms, including LSDT, spatial DCT, and spatial DWT.
  We apply each transform to the $x$, $y$ and $z$ components of each frame separately,
  and examine the relationship between the percentage of nonzero transform coefficients and the distortion.
  As Figure \ref{fig:sparsity-distortion} shows, given the same number of nonzero transformed coefficients,
  the distortions produced by LSDT are consistently much smaller than those of DCT and
  DWT, meaning that LSDT concentrates energy (or spatially decorrelated mocap data) better than DCT and DWT.

\begin{figure}[t]
\centering \subfigure[15\_04]{
\includegraphics[width=1.5in]{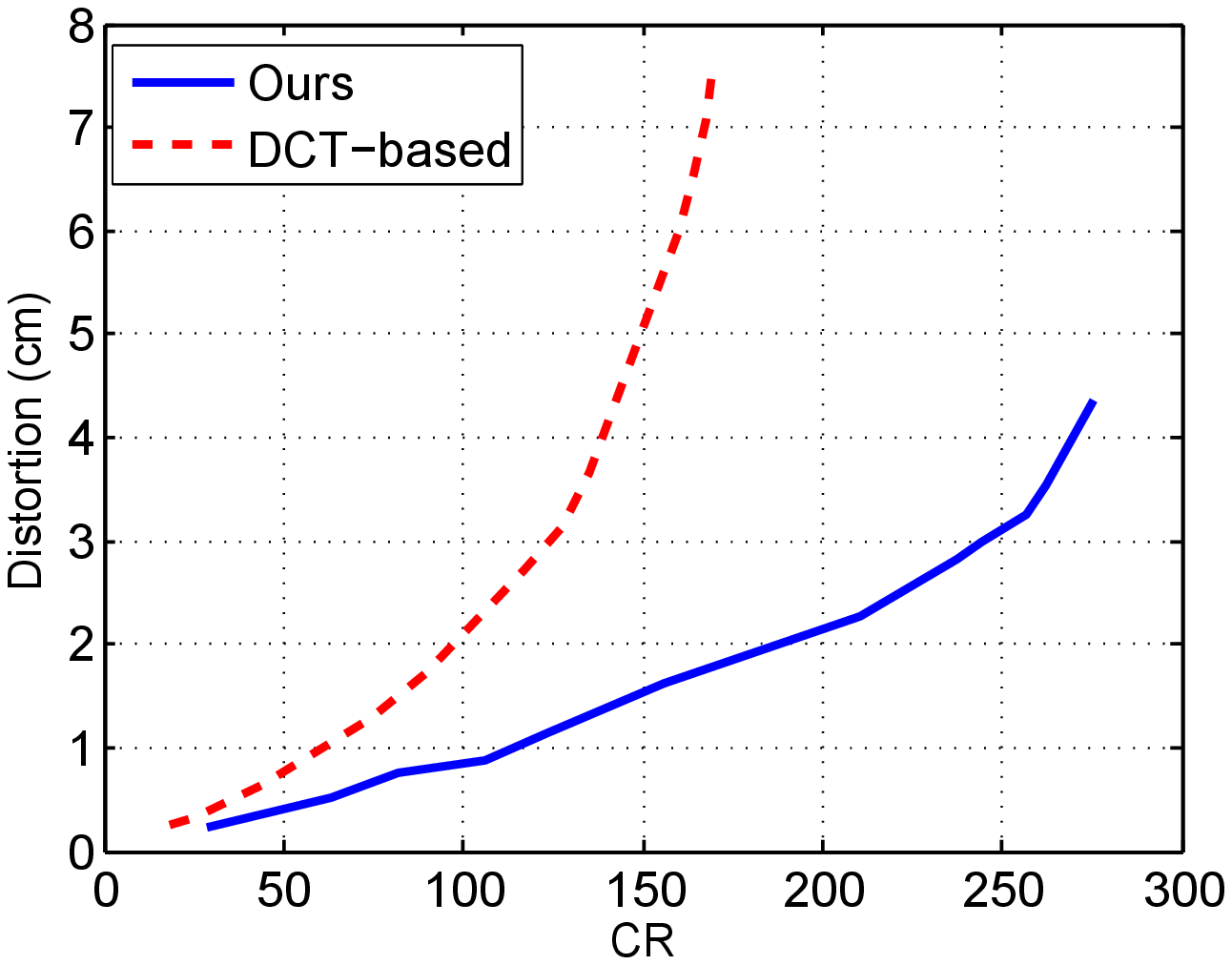}}
\subfigure[17\_08]{
\includegraphics[width=1.5in]{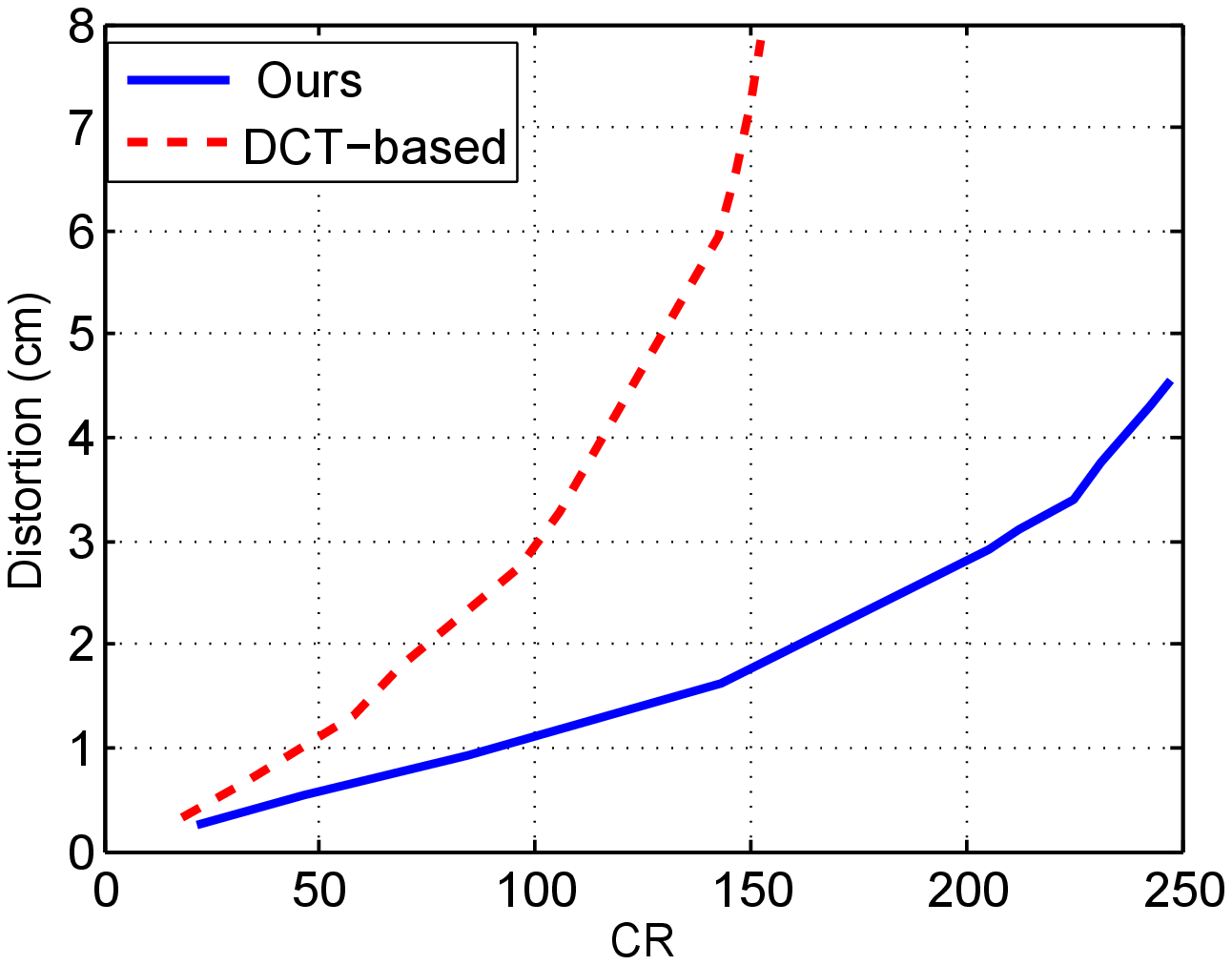}}
\subfigure[17\_10]{
\includegraphics[width=1.5in]{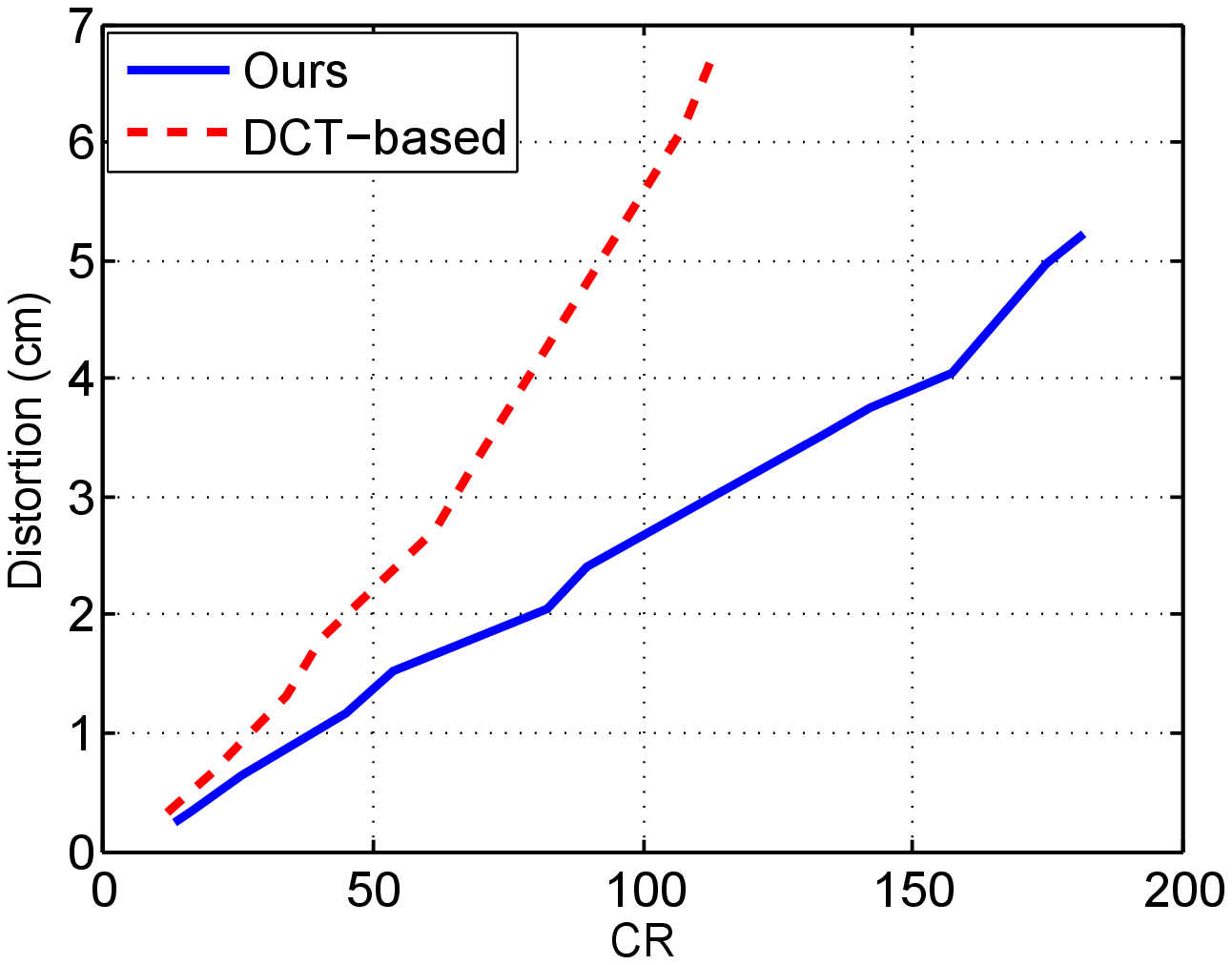}}
\subfigure[41\_07]{
\includegraphics[width=1.5in]{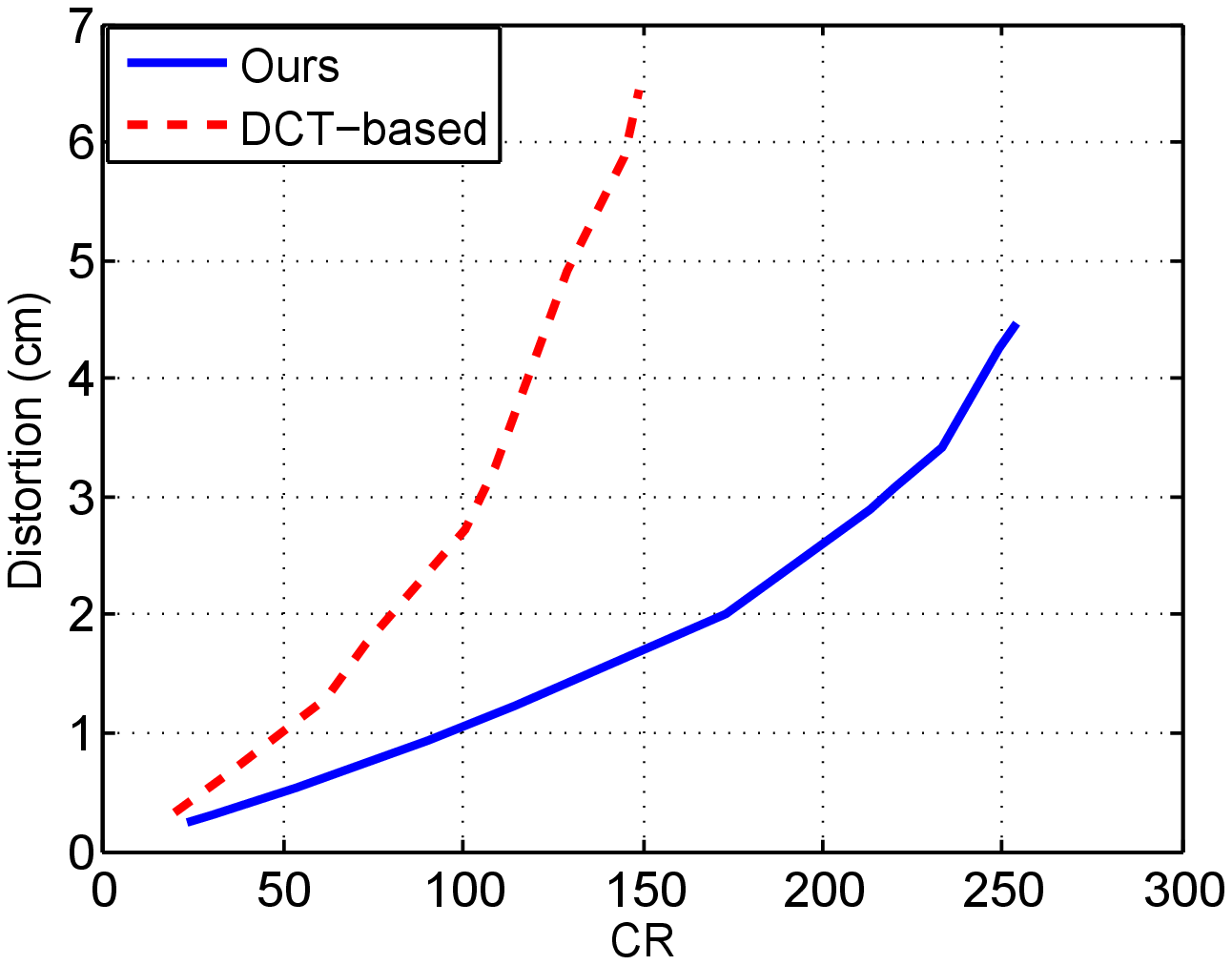}}
\subfigure[49\_02]{
\includegraphics[width=1.5in]{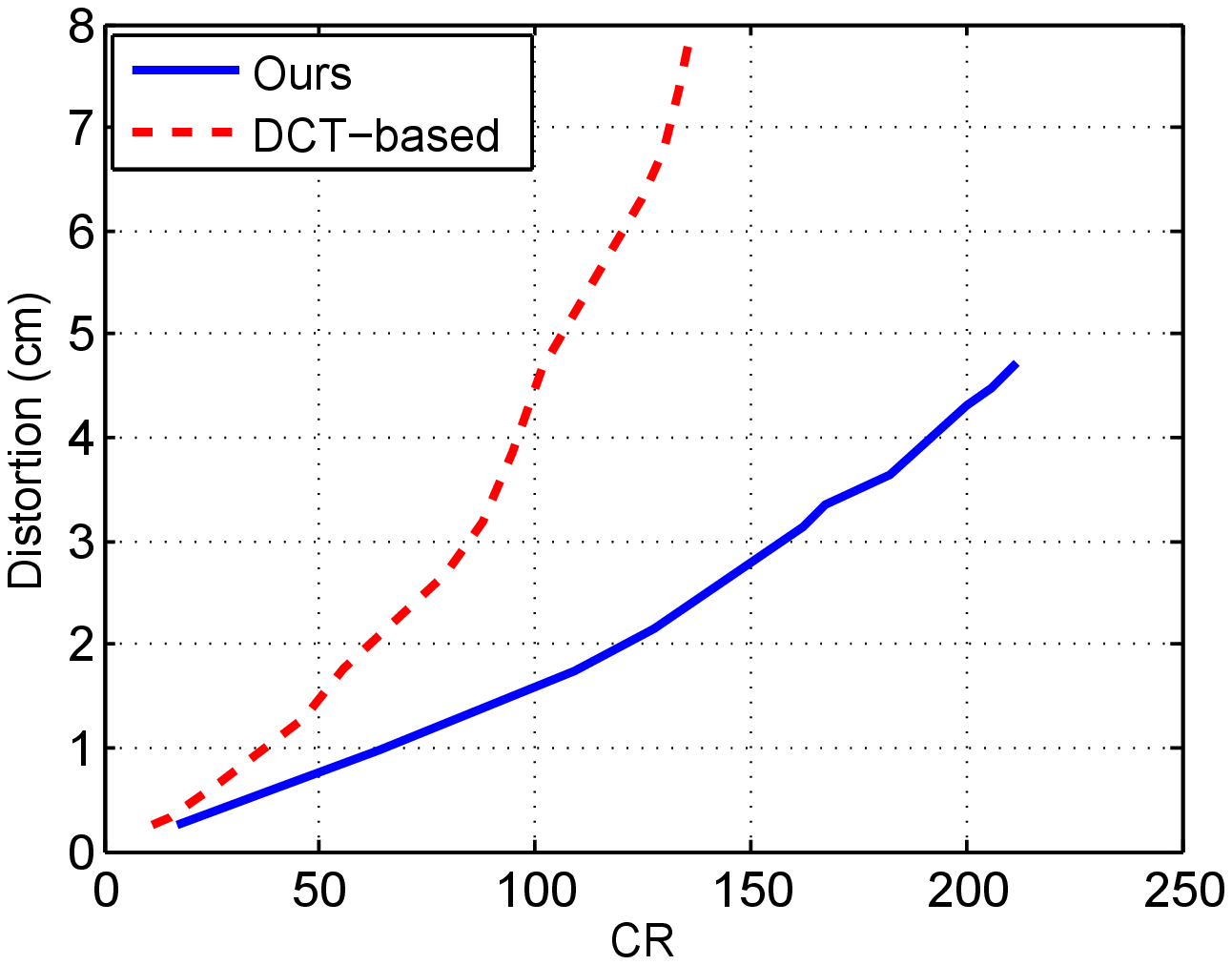}}
\subfigure[56\_07]{
\includegraphics[width=1.5in]{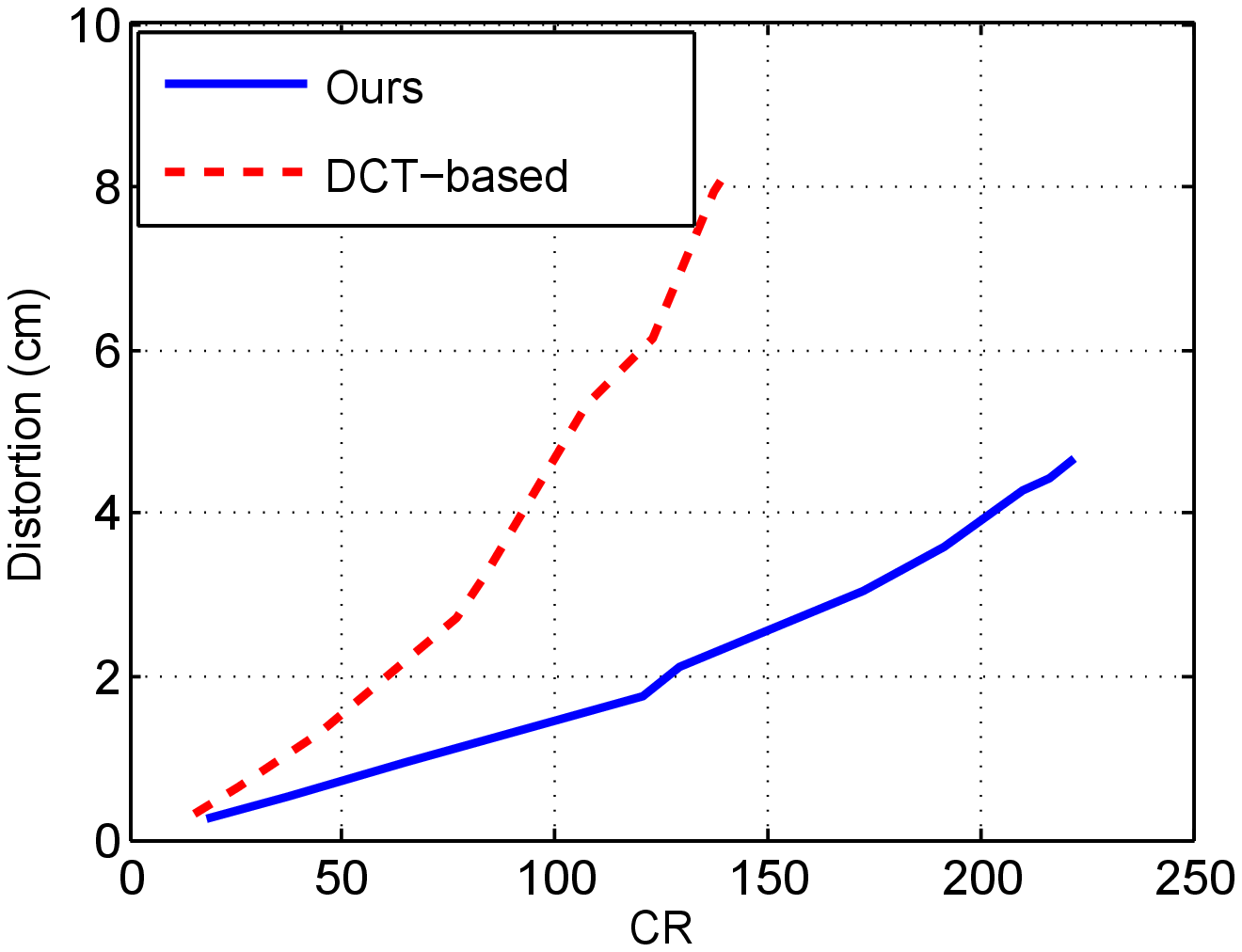}}
\subfigure[85\_12]{
\includegraphics[width=1.5in]{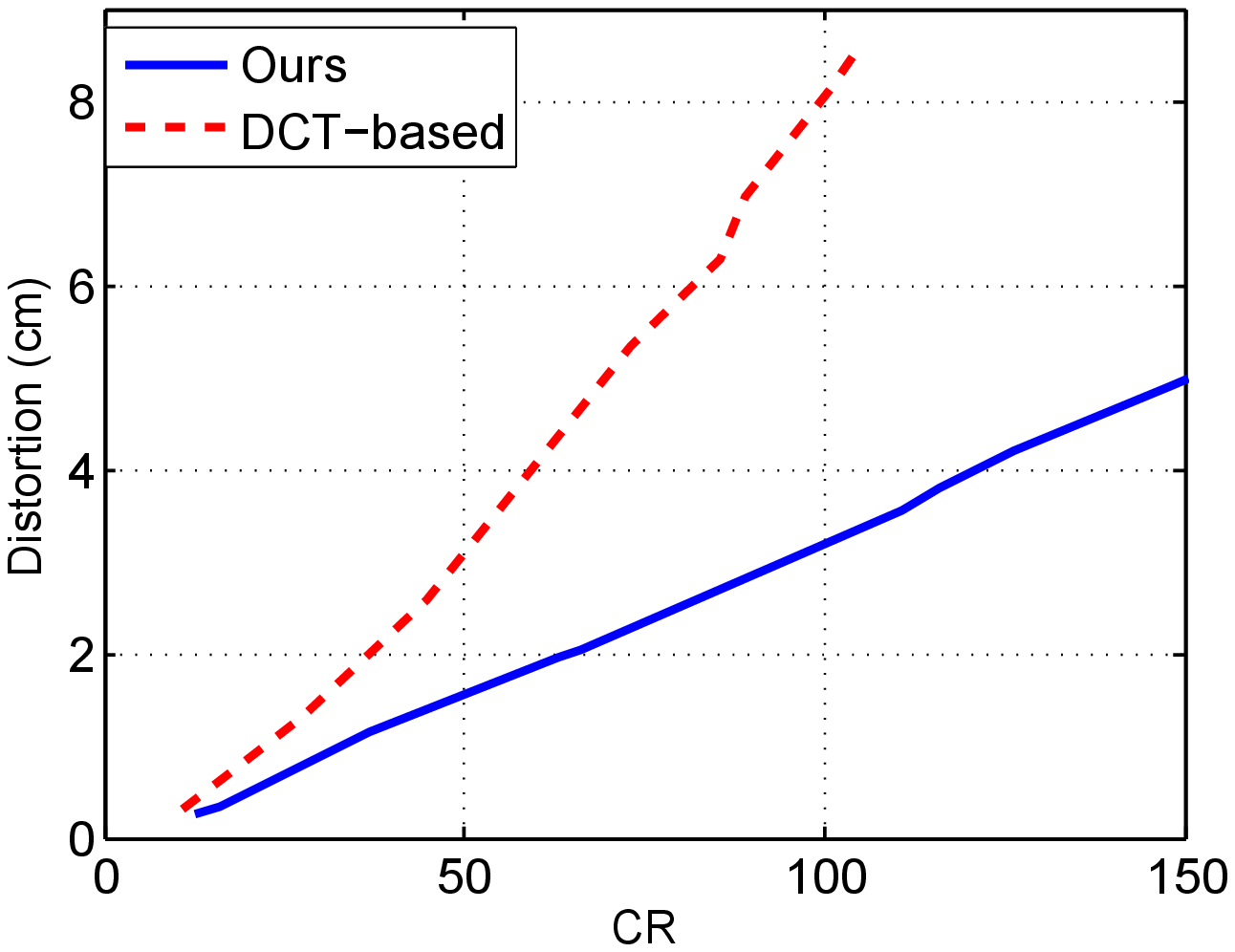}}
\subfigure[86\_05]{
\includegraphics[width=1.5in]{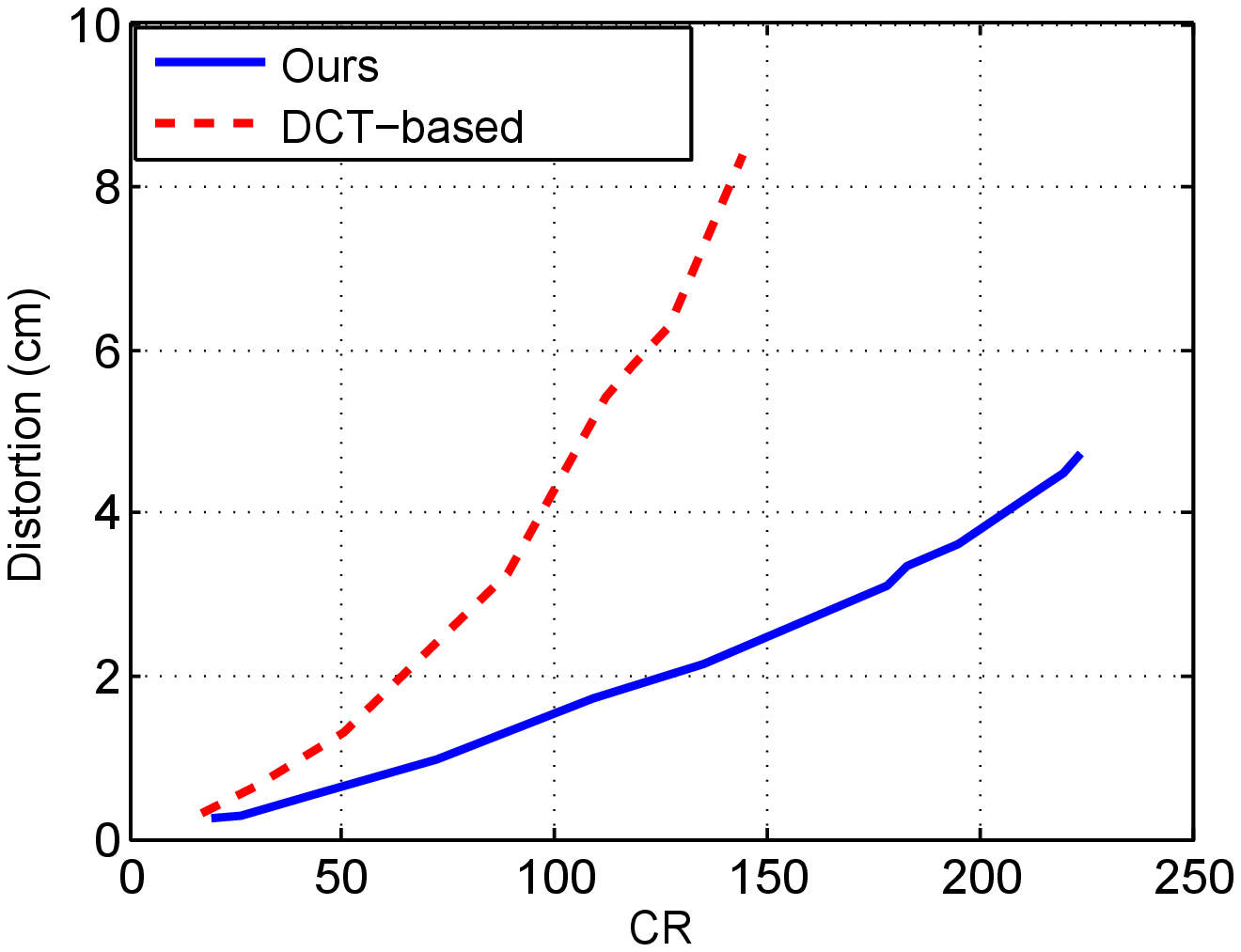}}
\caption{Comparison of compression performance of frame-based
methods.}\label{fig:CR-D frame}
\end{figure}

\begin{table}[t]\footnotesize
\centering \caption{Qualitative comparison of various mocap
compression methods. The latency is measured in number of frames.
$\#p$: the number of parameters used in the encoding process; $F_s$:
the number of frames in a mocap sequence; $F_c$: the numbers of
frames in a short clip, and $F_s\gg F_c$. Note that the quantization
parameter is not included in $\#p$ for all methods.}
\label{tab:latency}
\begin{tabular}{c|l||c|c|c|c|c}
\hline
 Category & Method & Latency & $\#p$ & Computational & Implementation & Compression \\
  & & & & cost& & performance\\
\hline \hline
\cline{2-7}     & V\'{a}\v{s}a and Brunnett \cite{vavsa2014}      & $F_s$   & 5 &   high & fair & high \\
\cline{2-7}              &     Lin \emph{et al}.   \cite{Lin2011}        & $F_s$   & 3  & fair &difficult &medium \\
\cline{2-7}    PCA-based &     Arikan   \cite{arikan2006}     & $F_c$   & 3  & fair  &fair & low\\
\cline{2-7}              &     Liu \emph{et al}.  \cite{liu2006segment} & $F_c$   & 3  & fair &  fair & low\\
\cline{2-7}              &    Tournier \emph{et al}.    \cite{tournier2009}   & $F_s$   & 2  & high& fair  &medium\\
\hline       DCT/ & Kwak and Bajic \cite{kwak2011hybrid} & 0 & 0    &  low&  easy & low\\
\cline{2-7} DWT-based        &   Chew \emph{et al}.     \cite{Chew2011}       & $F_c$   & 2  &fair& easy  &medium\\
\cline{2-7}         &   Firouzmanesh \emph{et al}.   \cite{Firouzmanesh2011}&$F_c$   & 3  &  low & easy  &low\\
\hline            &     Zhu \emph{et al}. \cite{zhu2012quaternion}&$F_s$   & 3  & high& difficult & medium\\
\cline{2-7}   Mocap Data Favored              & Hou \emph{et al}.  \cite{Hou2014tvcg}    & $F_s$   & 2   &  low & easy &high\\
\cline{2-7}        Transform          & Hou \emph{et al}. \cite{hou2014}        & $F_s$   & 2 &    fair& fair &medium\\
\cline{2-7}       & \textbf{Our frame-based method}& \textbf{0} & \textbf{0}& \textbf{low} & \textbf{easy} & \textbf{high}\\
\cline{2-7}       & \textbf{Our clip-based method} & $\textbf{\emph{F}}_c$& \textbf{1}& \textbf{low}& \textbf{easy} & \textbf{high}\\
\hline      Indexing-based         &Chattopadhyay \emph{et al}.\cite{chattopadhyay2007human}  & $F_s$   & 3   &  fair & fair & low\\
\cline{2-7}         &     Gu \emph{et al}.  \cite{gu2009}         & $F_s$   & 4  & fair& fair &low\\
\hline
\end{tabular}
\end{table}

\subsection{Compression Performance}

  Figure~\ref{fig:CR-D frame} shows the CR-distortion (CR-D) curves of
  the frame-based scheme.
  As Section~\ref{subsec:evaluation_sdt} shows, our data-adapted LSDT is  superior to the data-independent 1D DCT for spatial decorrelation.
  Therefore, it is not surprising that our frame-based scheme significantly outperforms the 1D DCT-based method \cite{kwak2011hybrid} in terms of compression performance.
  We observe that with a relatively high CR, our frame-based scheme can reduce up to 70\% distortion of \cite{kwak2011hybrid}.

\begin{figure}[t]
\centering \subfigure[15\_04]{
\includegraphics[width=1.5in]{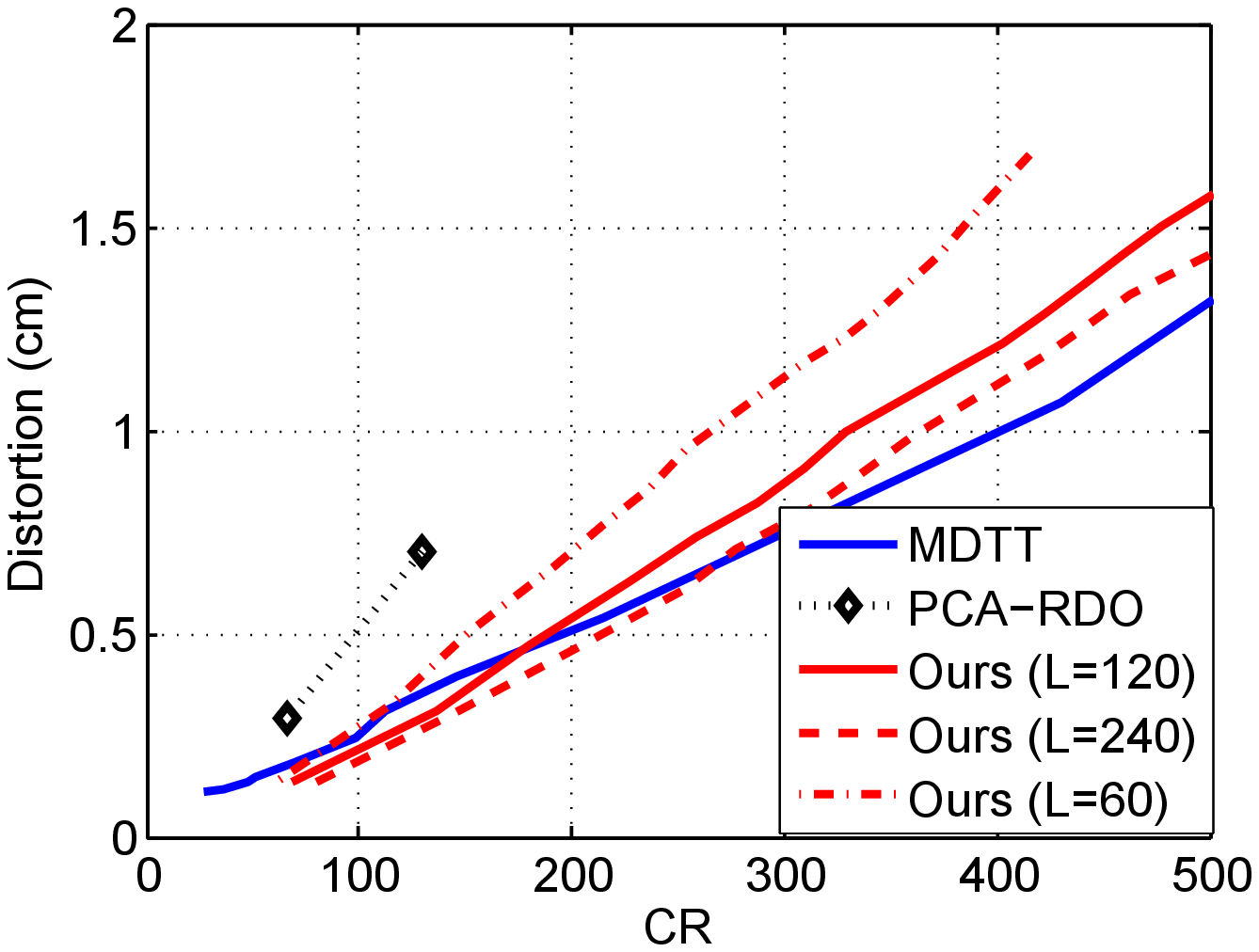}}
\subfigure[17\_08]{
\includegraphics[width=1.5in]{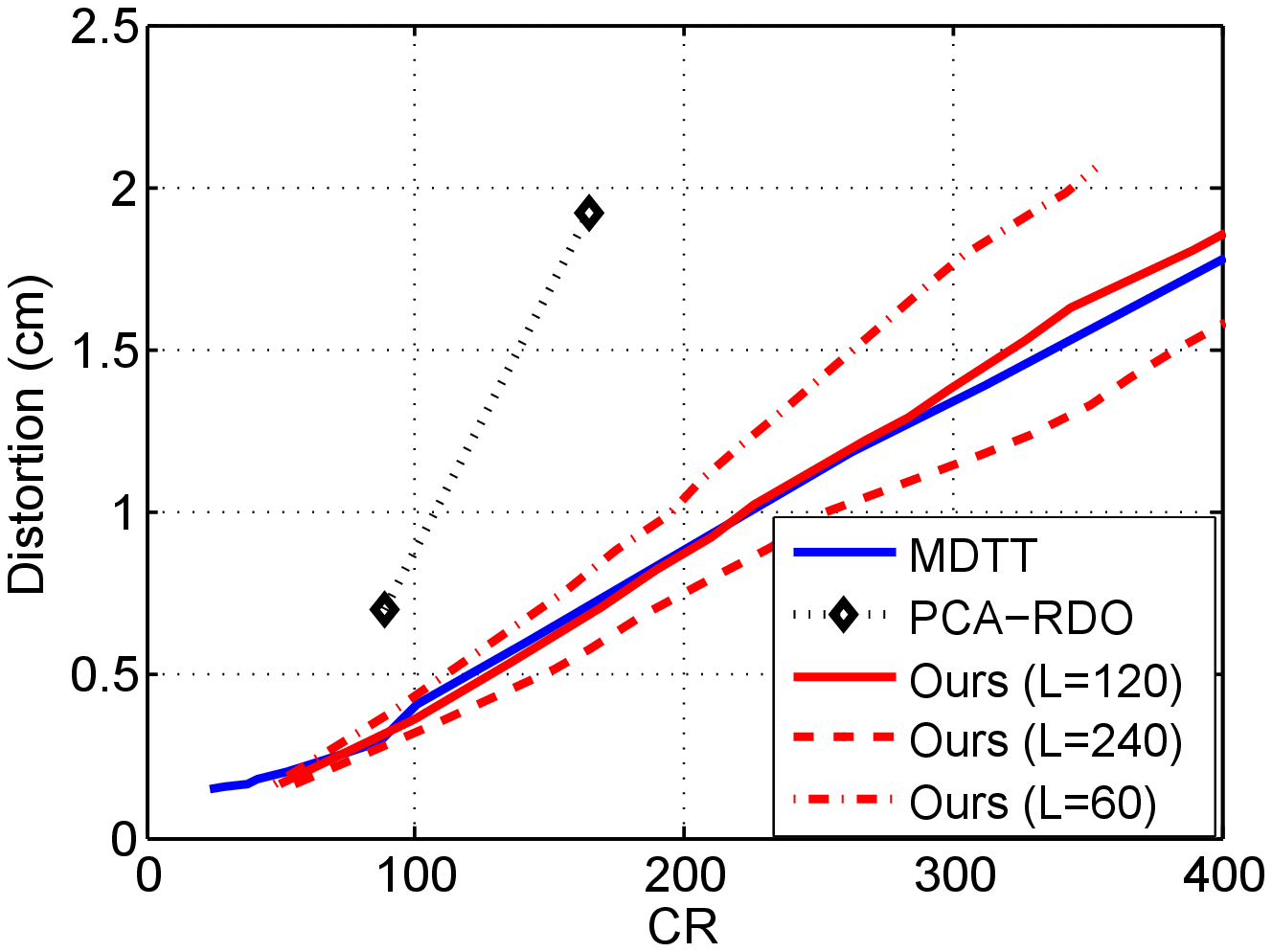}}
\subfigure[17\_10]{
\includegraphics[width=1.5in]{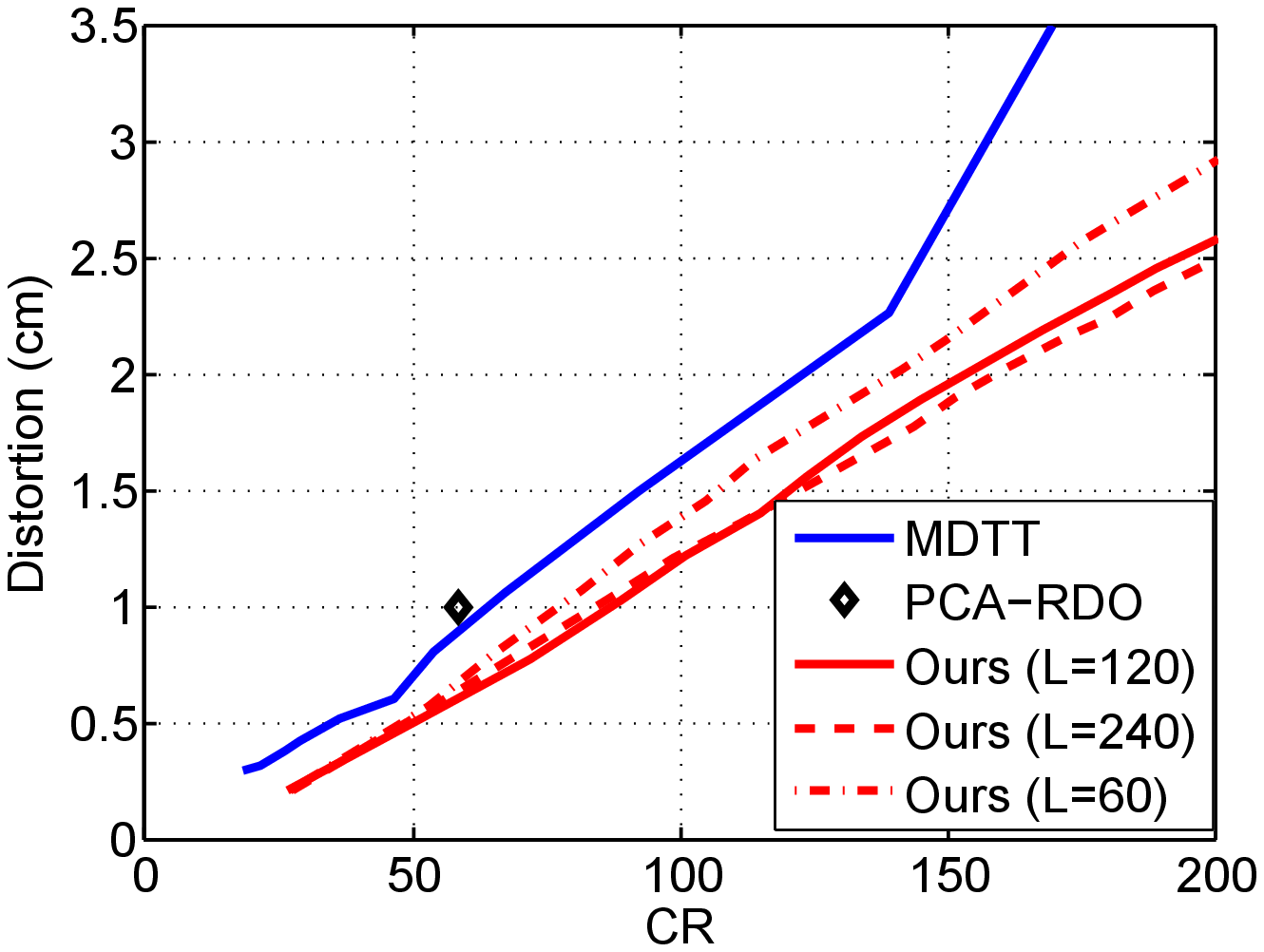}}
\subfigure[41\_07]{
\includegraphics[width=1.5in]{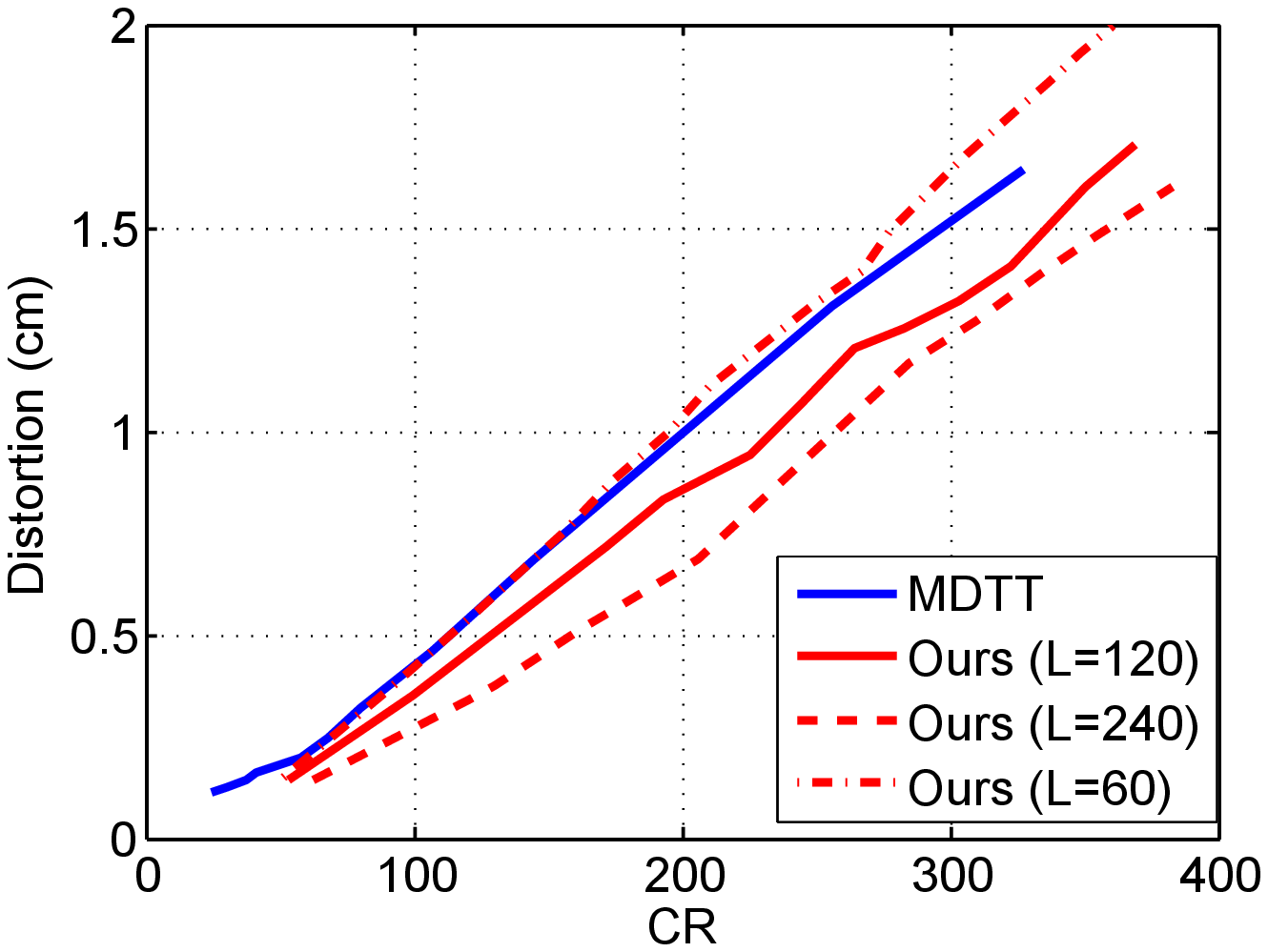}}
\subfigure[49\_02]{
\includegraphics[width=1.5in]{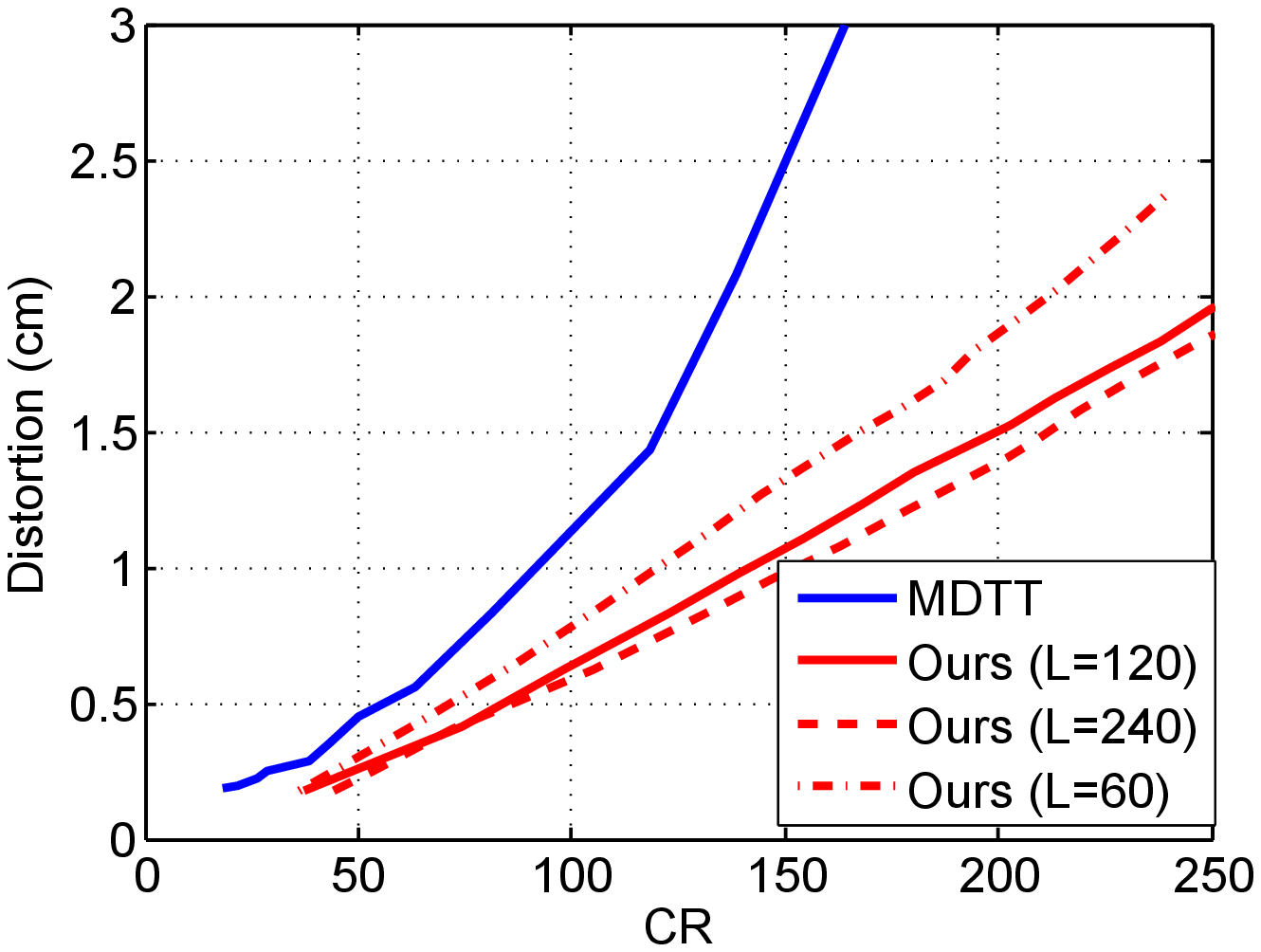}}
\subfigure[56\_07]{
\includegraphics[width=1.5in]{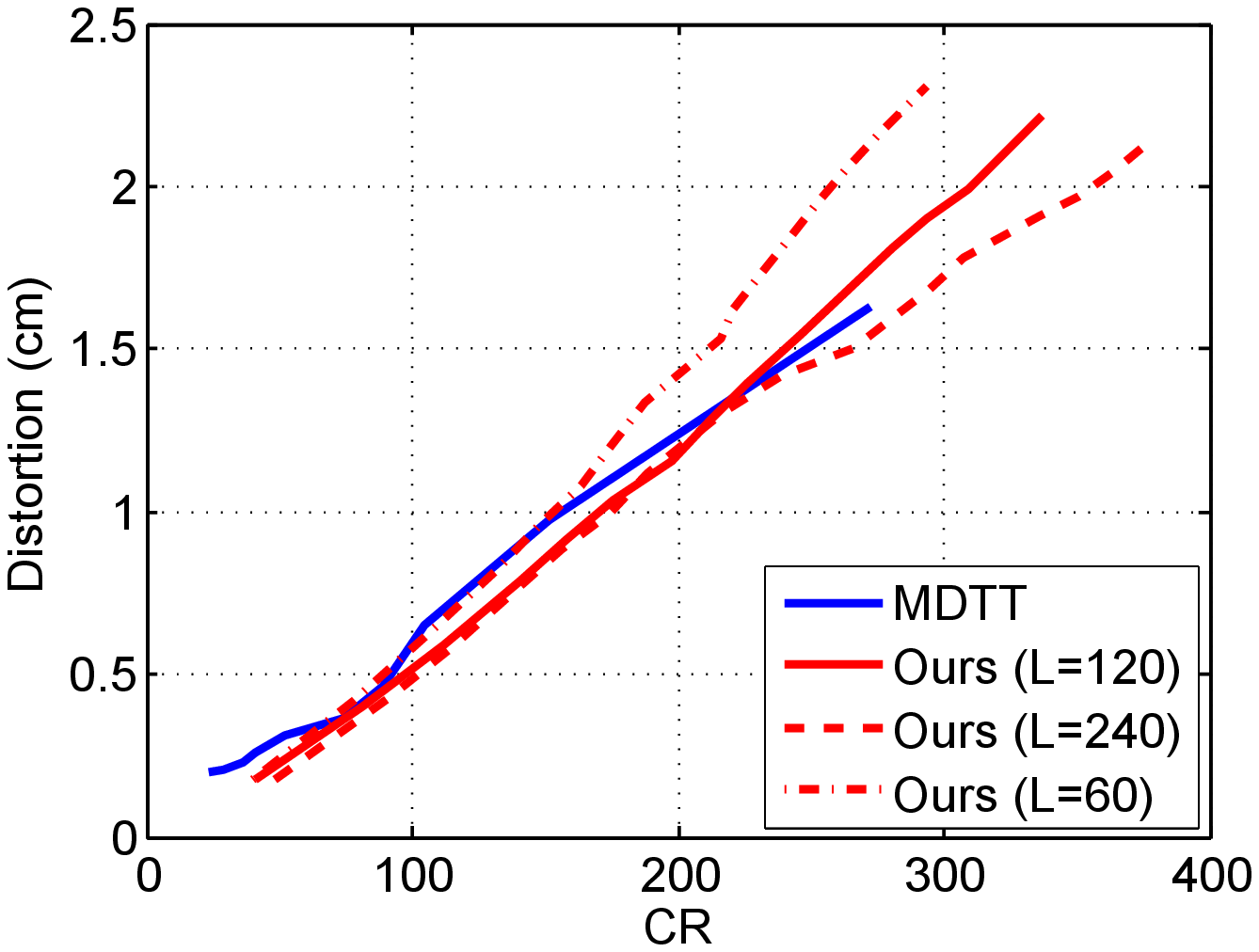}}
\subfigure[85\_12]{
\includegraphics[width=1.5in]{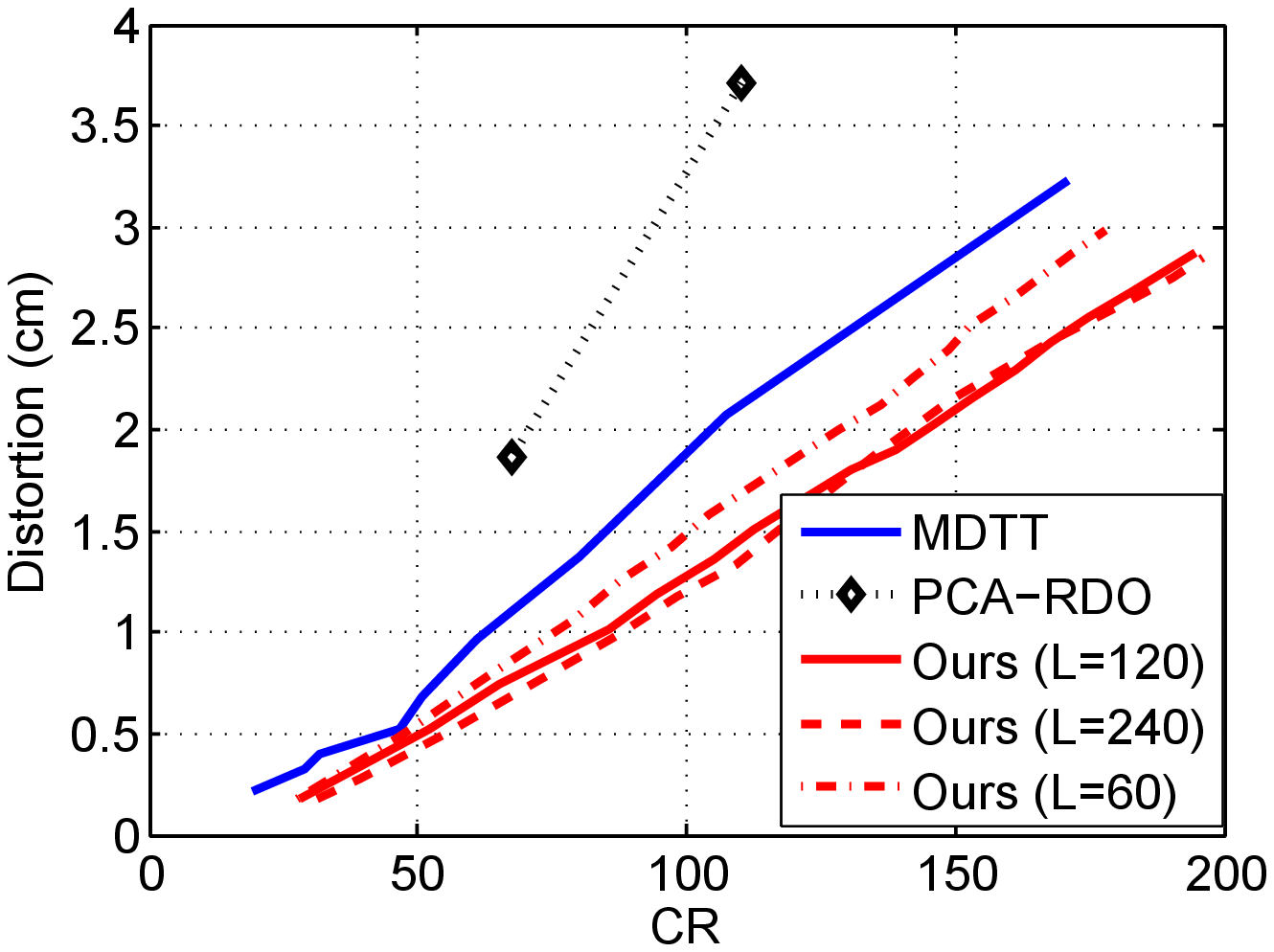}}
\subfigure[86\_05]{
\includegraphics[width=1.5in]{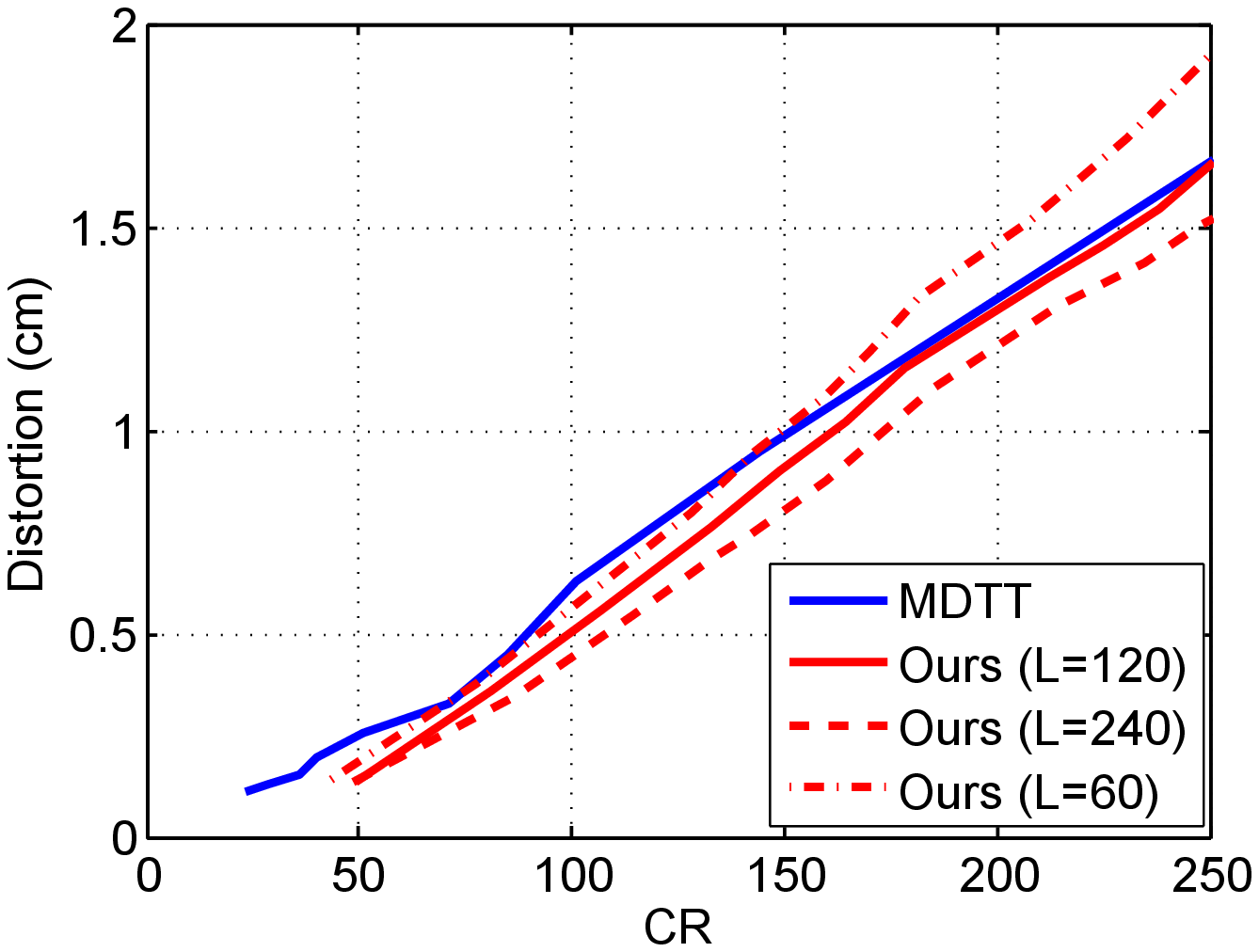}}
\caption{Compression performance of our clip-based schemes and the
state-of-the-art methods, such as the PCA-RDO method
\cite{vavsa2014} and the MDTT method~\cite{Hou2014tvcg}. For MDTT,
we adopt equal segmentation with $L=240$ and
follow~\cite{Hou2014tvcg} to set the other parameters. The results
of PCA-RDO were taken from \cite{vavsa2014} .} \label{fig:CRDclip}
\end{figure}

  Figure \ref{fig:CRDclip} shows the CR-D curves of the clip-based scheme, from which we observe the following:
 \begin{enumerate}
\item
  As expected, the clip-based scheme has much better compression performance than the frame-based scheme,
  since it can exploit the temporal coherence better.
  At the same time, users can easily control the latency for the clip-based scheme.
  Taking the CMU mocap data which are sampled at 120 fps as example, the clip length $L=120$ (resp. 240) means 1 second (resp. 2 seconds) latency.

\item
  The compression performance of the proposed clip-based scheme can be improved by increasing the clip length (or latency).
  More specifically, when $L$ ranges from 60 to 120, the trajectories in a clip still remain smooth and have small variation (due to the small duration),
  causing the DCT coefficients to be distributed at similar locations, which can then be encoded using a similar number of bits.
  Since the number of clips in the sequence decreases, the total number of bits to encode one sequence (i.e., the sum of
  bits for all clips in one sequence) is significantly reduced,
  leading to higher compression performance.
  However, the improvement is little when the value of $L$ increases from 120 to 240.
  The reason is that the joint trajectories change more
  significantly. As a result, the DCT coefficients are spread out, requiring more bits for encoding.
  Although the number of clips decreases, the total number of bits for one sequence only drops slightly.
  \end{enumerate}

\begin{figure}[thp]
\centering
\includegraphics[width=3.2in]{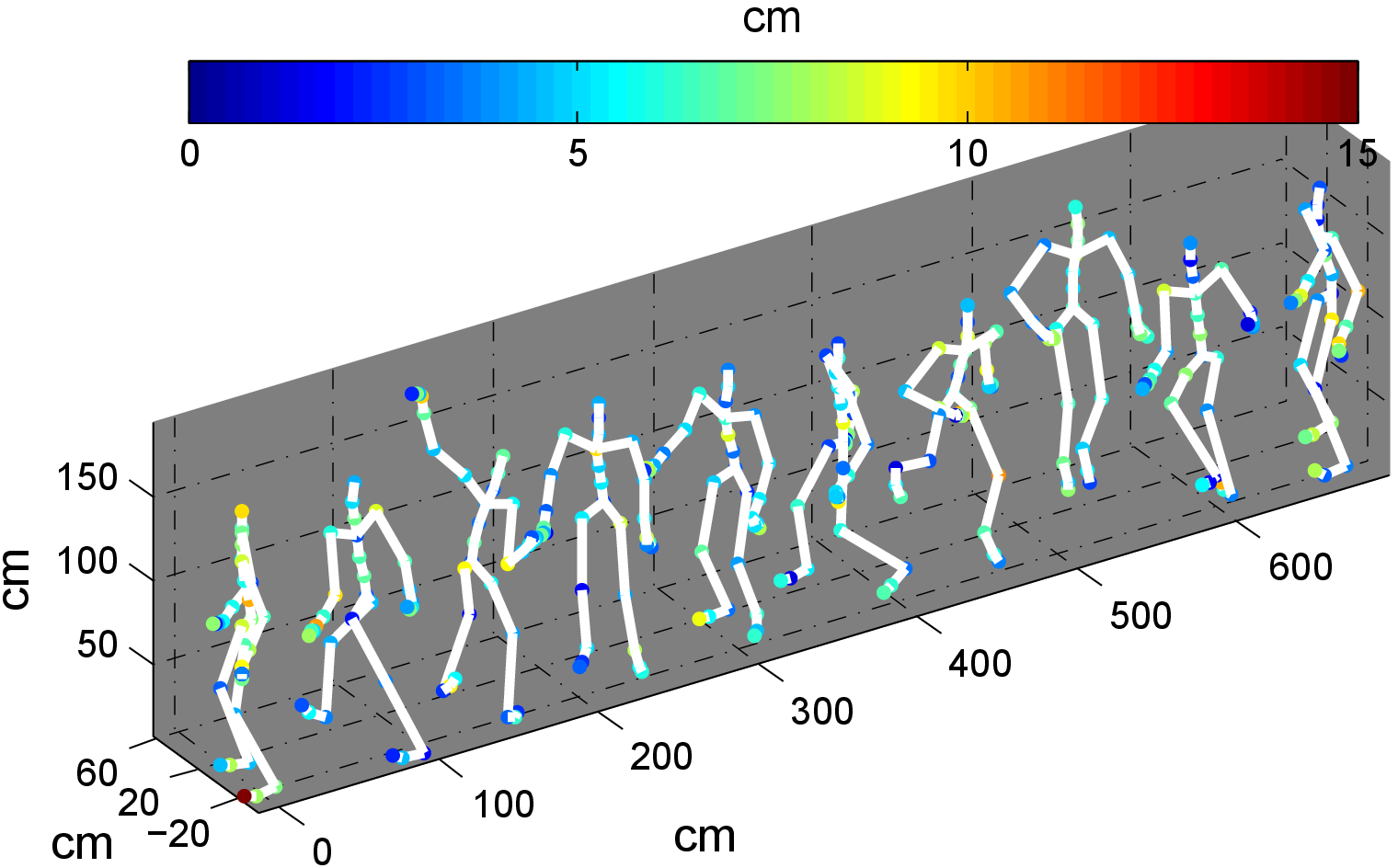}
\includegraphics[width=3.2in]{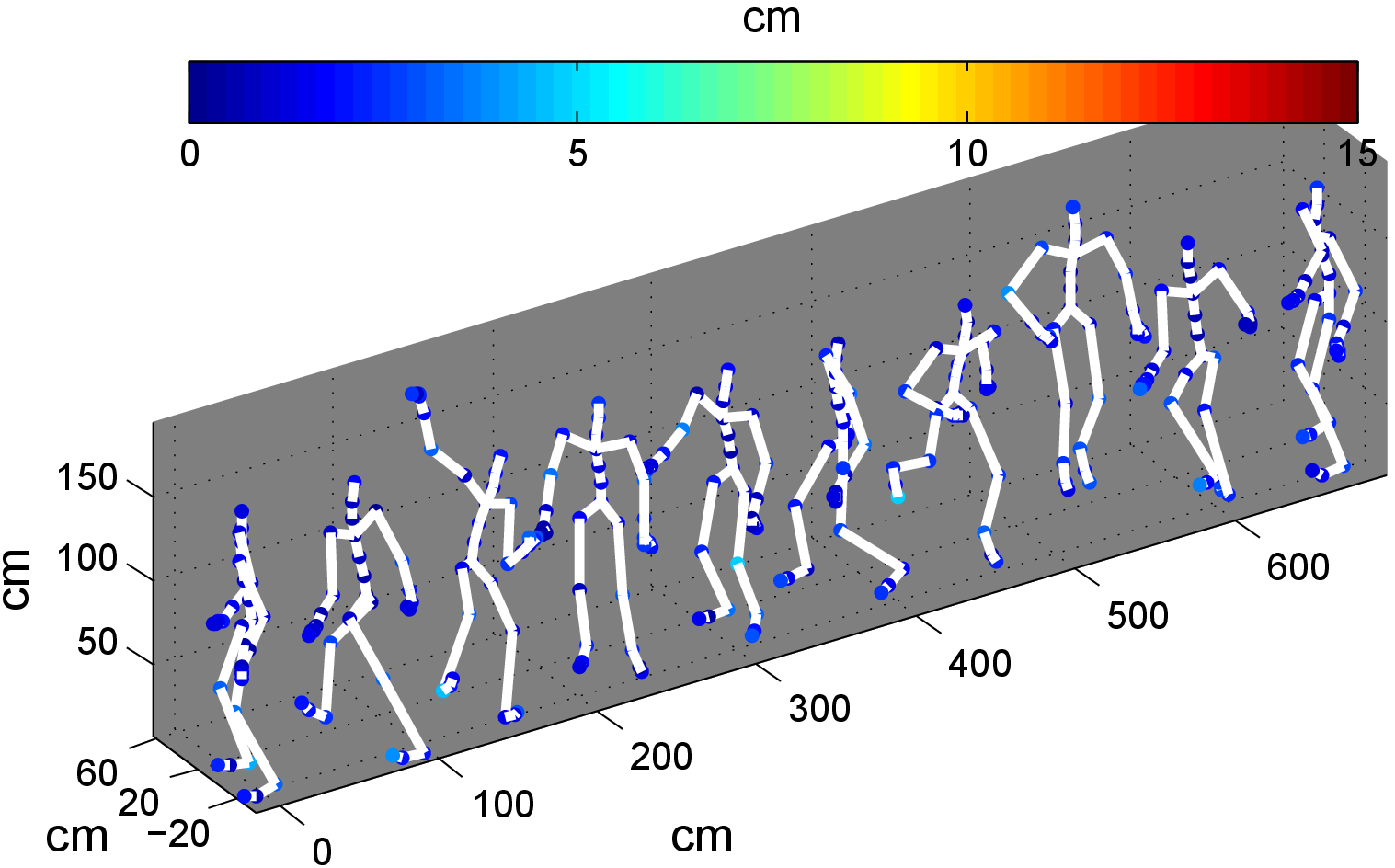}\\
\makebox[2.3in]{{\small (a) 56\_07, CR=105}}\\
\includegraphics[width=3.2in]{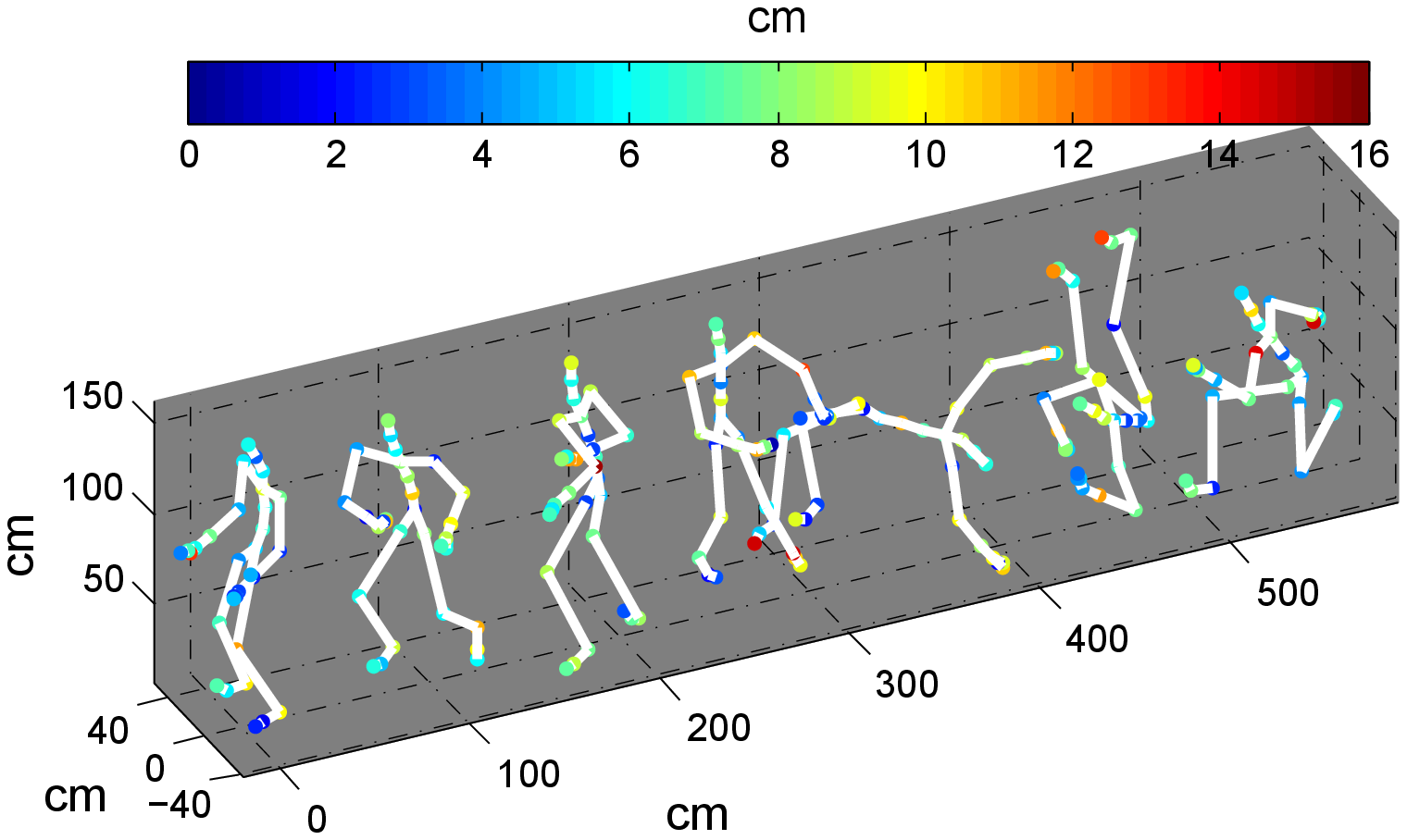}
\includegraphics[width=3.2in]{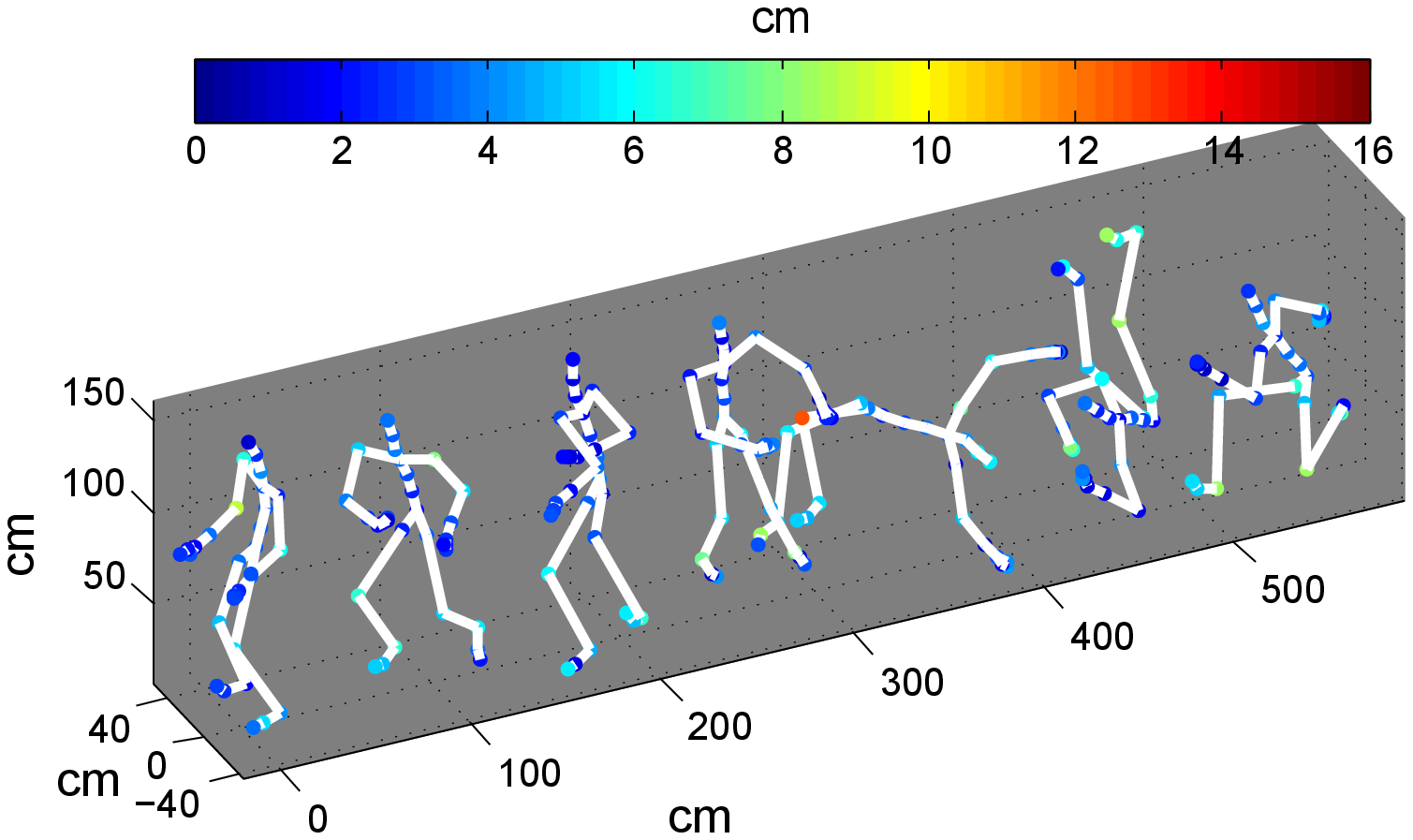}\\
\makebox[2.3in]{{\small (b) 85\_12, CR=90}}\\
\includegraphics[width=3.2in]{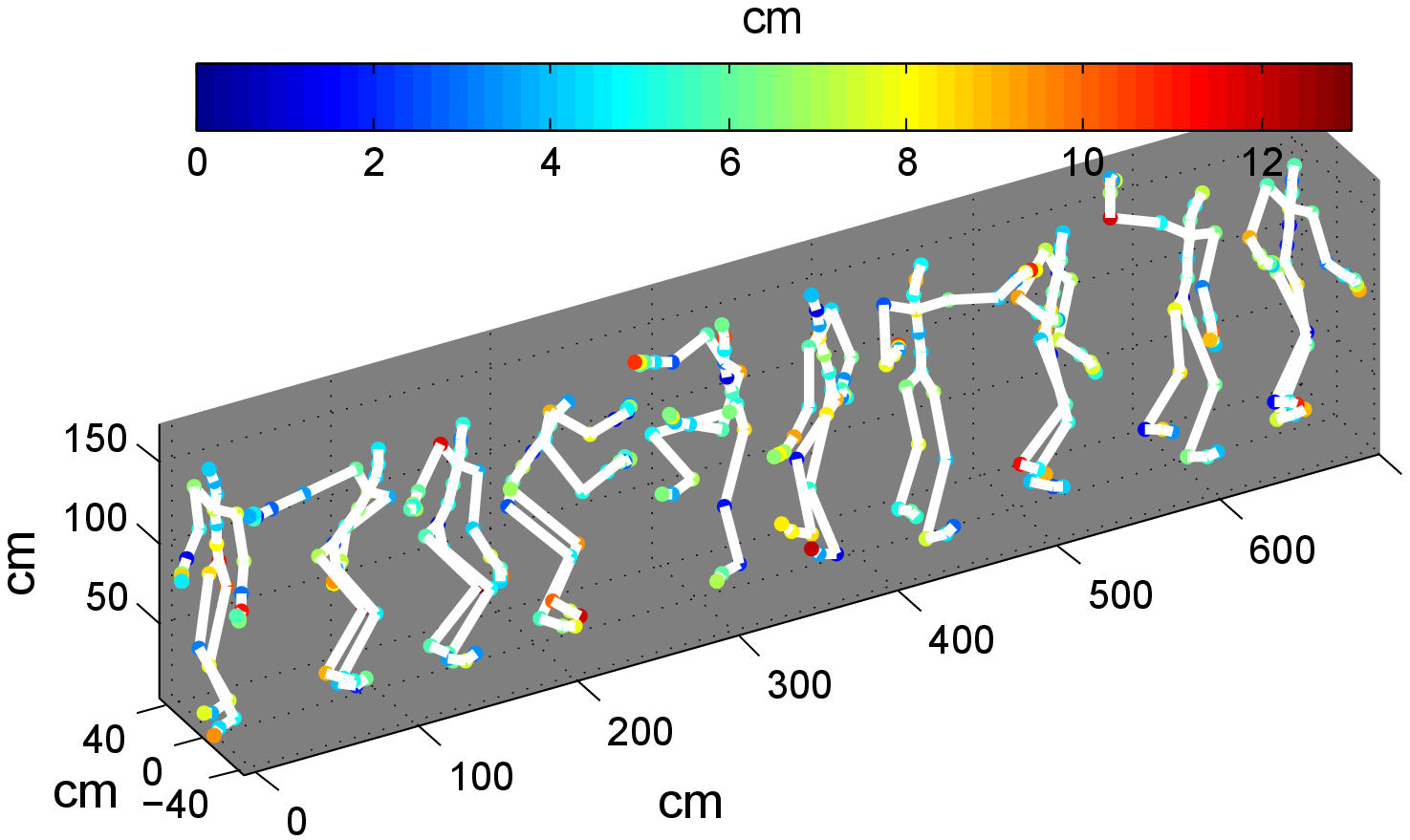}
\includegraphics[width=3.2in]{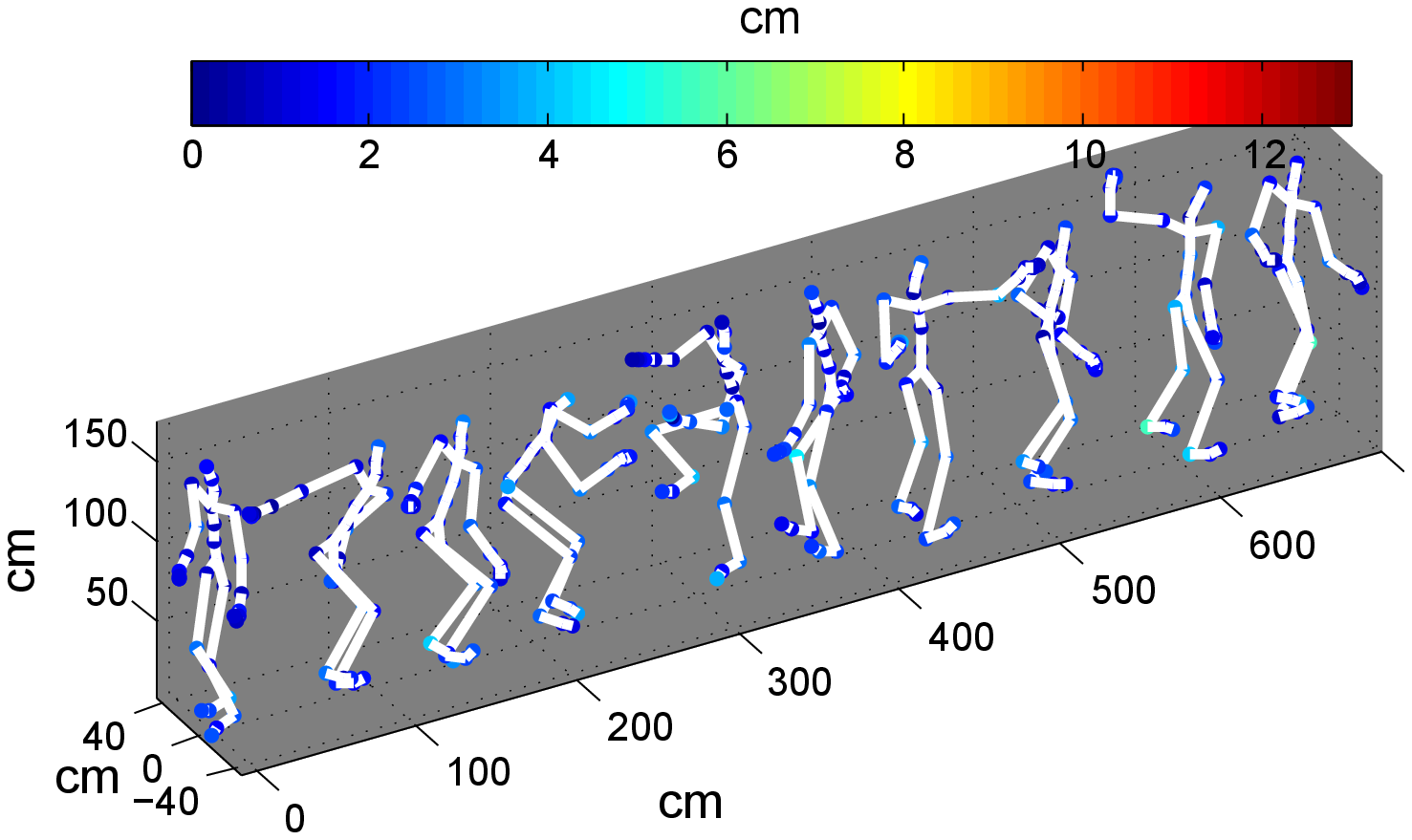}\\
\makebox[2.3in]{{\small (c) 86\_05, CR=120}}\\ \caption{Visual
results comparison of frame-based schemes. The distortions are
colored as heat map, and the frames are uniformly extracted from the
sequences. Left: the DCT-based method in \cite{kwak2011hybrid};
Right: our frame-based scheme.} \label{fig:visual frame}
\end{figure}
\begin{figure}[thp]
\centering
\includegraphics[width=3.2in]{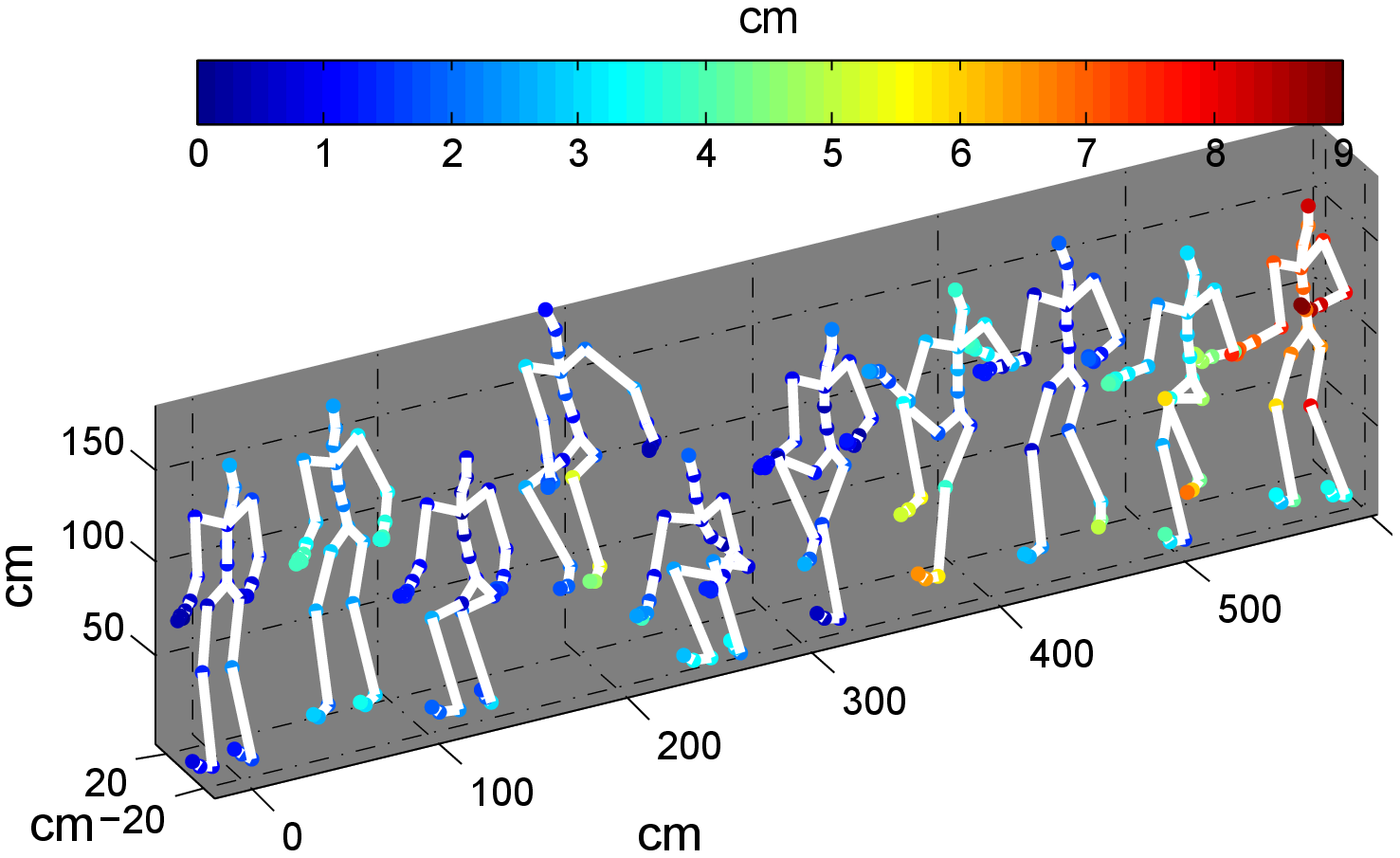}
\includegraphics[width=3.2in]{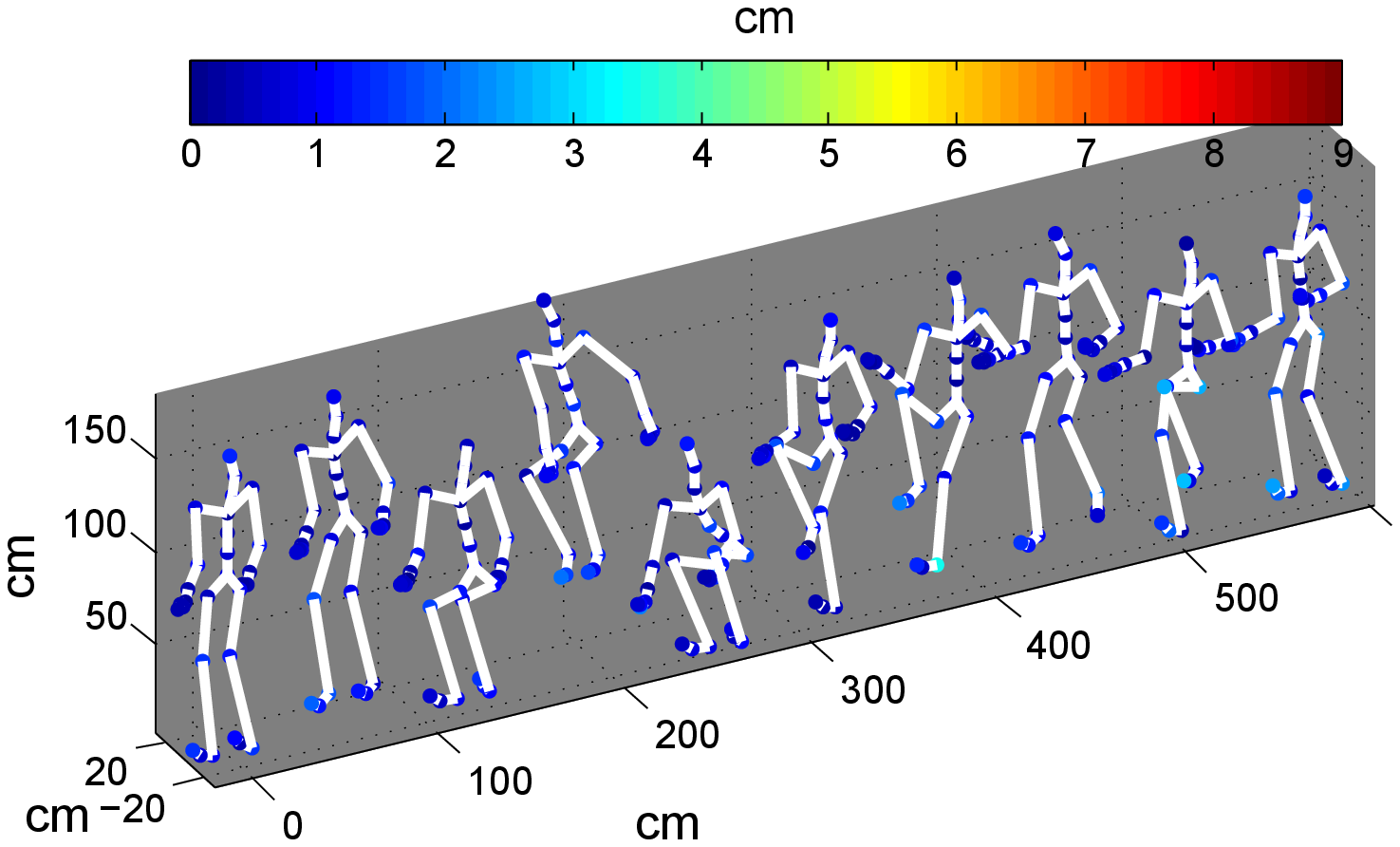}\\
\makebox[2.3in]{\small {(a) 49\_02, CR=138}}\\
\includegraphics[width=3.2in]{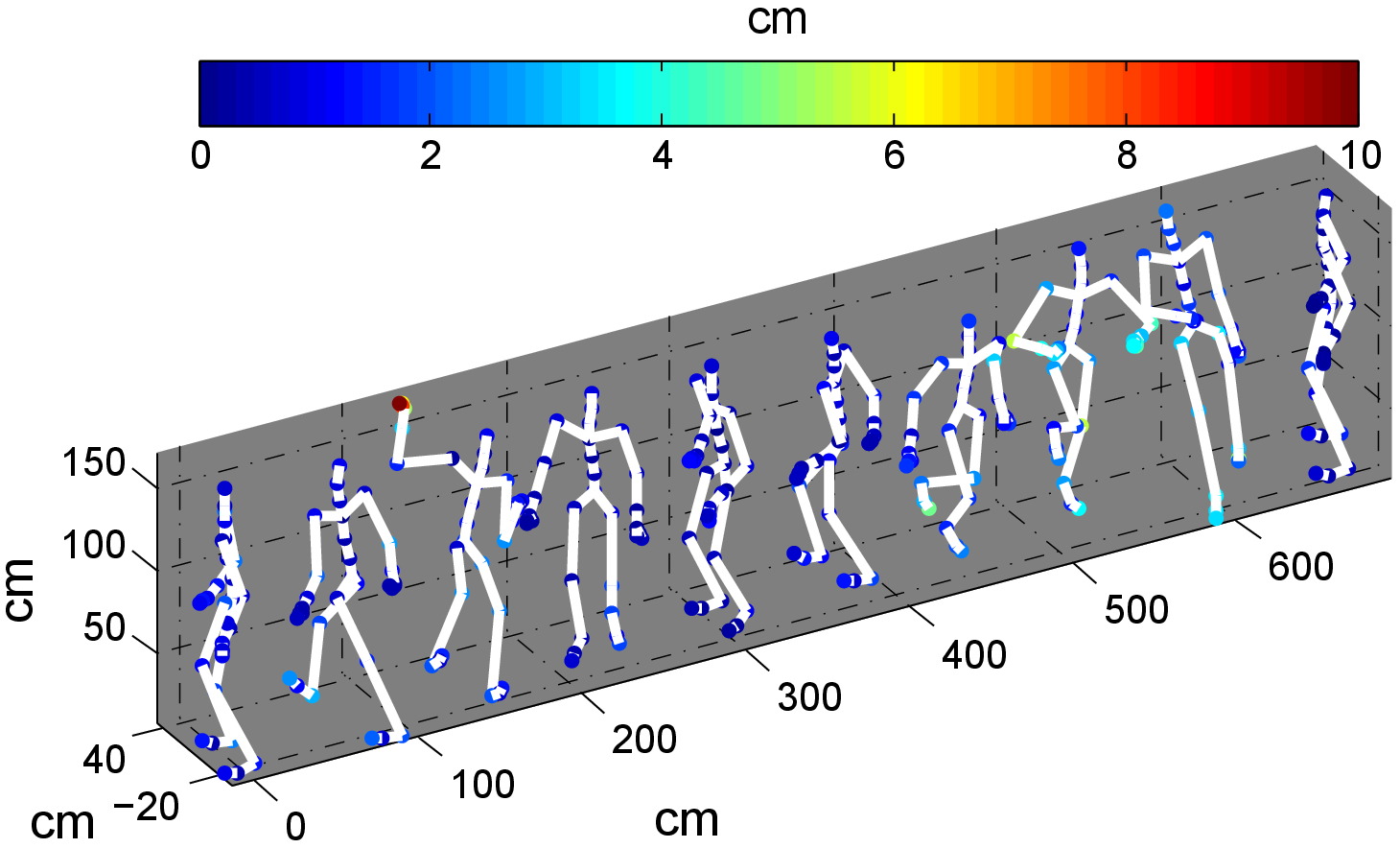}
\includegraphics[width=3.2in]{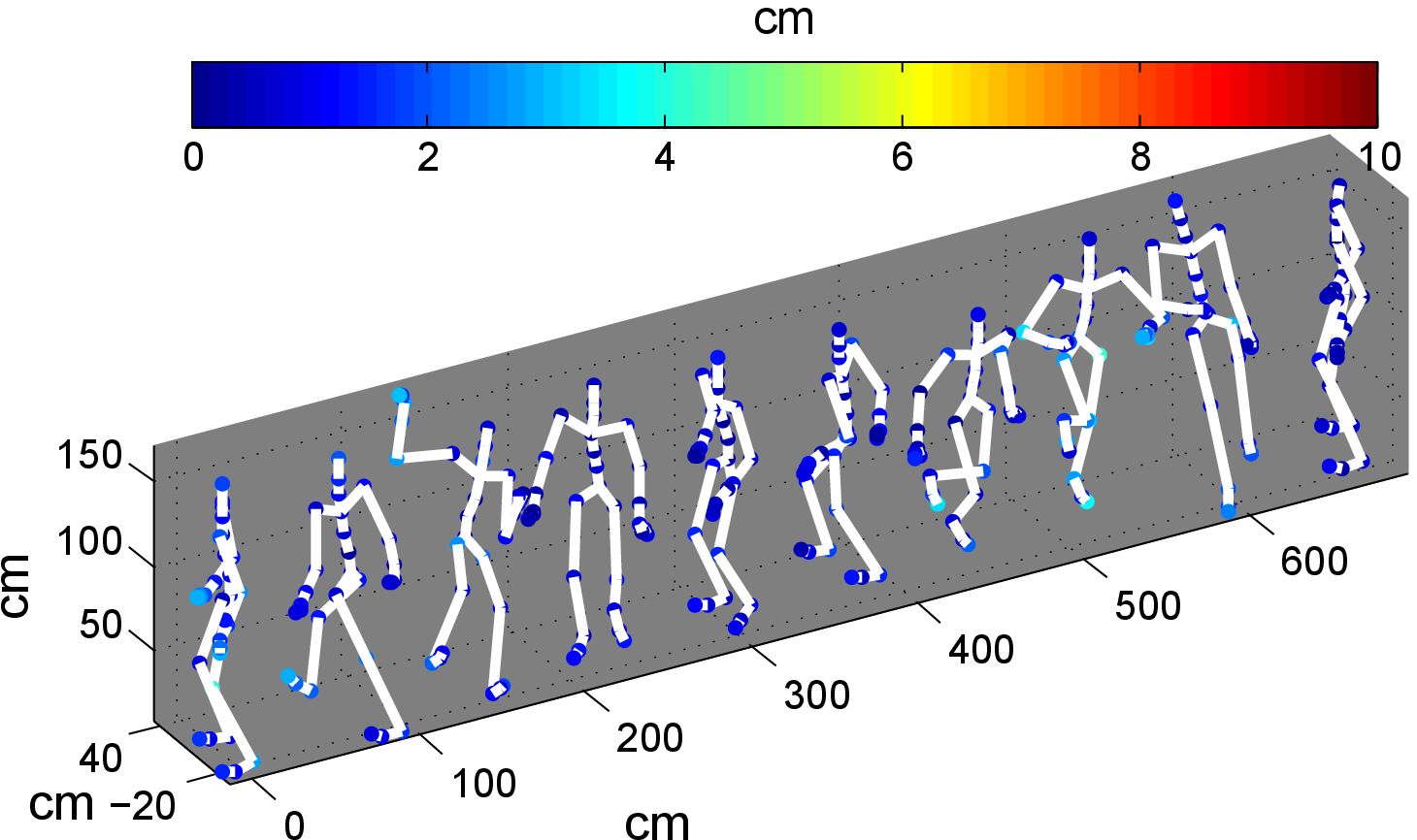}\\
\makebox[2.3in]{{\small (b ) 56\_07, CR=171}}\\
\includegraphics[width=3.2in]{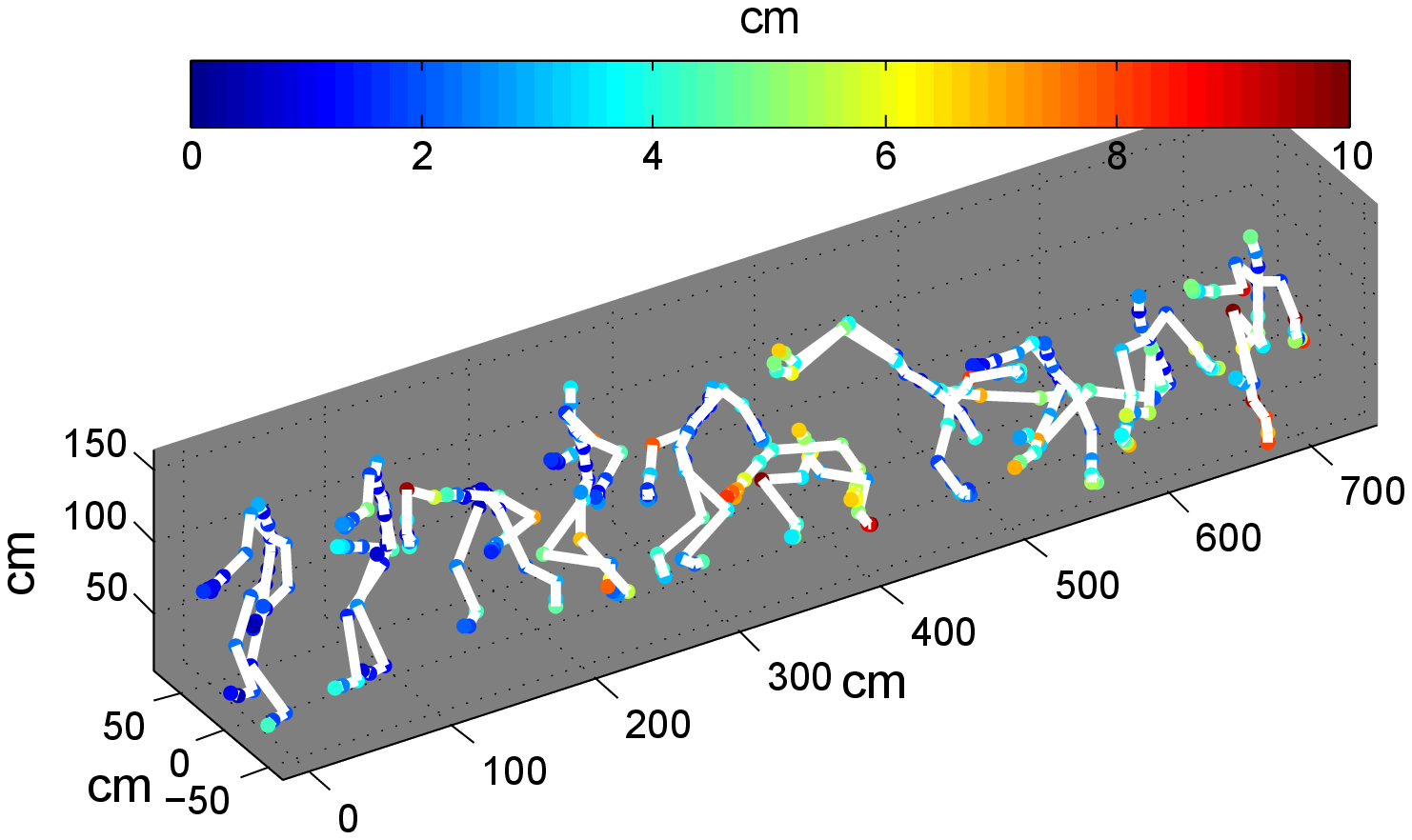}
\includegraphics[width=3.2in]{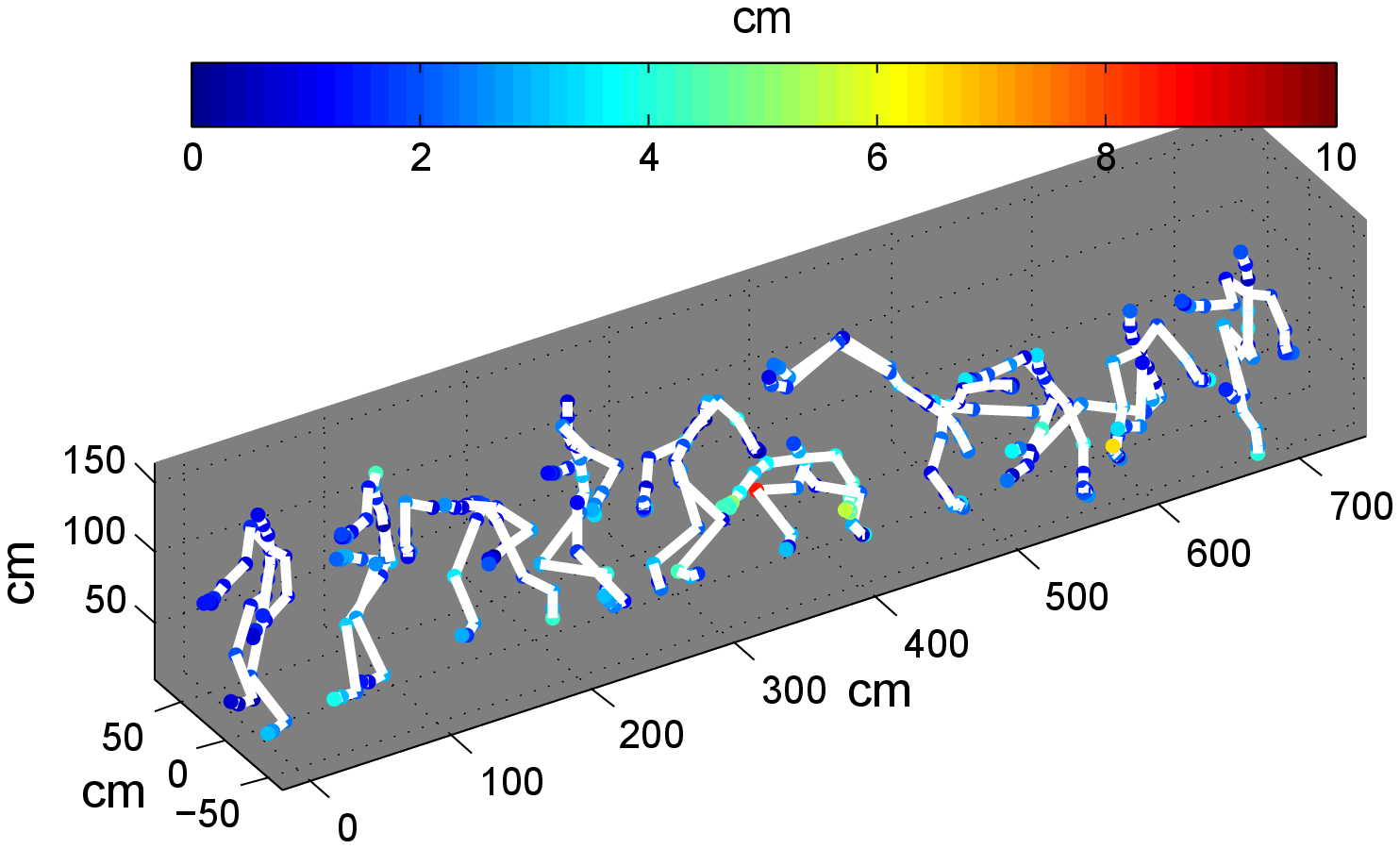}
\makebox[2.3in]{{\small (c)  85\_12, CR=133}}\\ \caption{Visual
results comparison of our clip-based scheme and the MDTT method
\cite{Hou2014tvcg}. The joint distortions are colored in heat map,
and the frames are uniformly extracted from the test sequences.
Left: MDTT;  Right: our clip-based scheme.} \label{fig:visual clip}
\end{figure}

\subsection{Comparison}
   Table~\ref{tab:latency} qualitatively compares our methods with the existing works in terms of latency, computational cost, implementation, compression performance,
   and the number of parameters used in the \textbf{encoding} process.
   Note that \textbf{all} methods have a quantization parameter to specify the number of bits used to quantize a coefficient.
   We do not include this quantization parameter in Table~\ref{tab:latency}, since it is a fixed parameter according to bandwidth.
   Also note that the sparsity parameter $P$ in our method appears only in the \textbf{training} stage.

   In this subsection, we compare our clip-based scheme with only two works, namely PCA-Rate Distortion Optimization (PCA-RDO) method~\cite{vavsa2014},
    and the equal segmentation case of Mocap Data Tailored Transform (MDTT) method~\cite{Hou2014tvcg}, which represent the state-of-the-art.
    See \cite{vavsa2014} and \cite{Hou2014tvcg} for detailed performance evaluation on earlier works \cite{arikan2006,Lin2011,tournier2009,gu2009,zhu2012quaternion}.

\subsubsection{Comparison with the MDTT Method}

  Both our clip-based algorithm and the MDTT method~\cite{Hou2014tvcg} apply temporal DCT to each trajectory for \textit{temporal}
  decorrelation.
  The two methods differ fundamentally in spatial decorrelation.
  For each mocap sequence, the MDTT method segments the motion sequence into short clips, and compute a set of orthogonal basis functions tailored for all clips together, resulting in better
  decorrelation at the price of a large latency and overhead for storing the data-dependent basis functions.
  Within our clip-based method, the LSDT bases are adapted to all mocap data, therefore, there is no need to store the bases for each sequence.

  The MDTT method adopts low-rank approximation, which is a linear approximation, to reduce the dimension of transformed coefficients.
  In contrast, the LSDT makes the transform coefficients sparse by quantization, which is a nonlinear approximation and more flexible.
  It has been pointed out in \cite{donoho1998data,cohen2002importance} that the nonlinear approximation outperforms the linear approximation in data compression.

  From the CR-D curves in Figure \ref{fig:CRDclip}, we observe the MDTT~\cite{Hou2014tvcg} has better performance than our scheme for long motion sequences (e.g., 15\_04 and 56\_07),
  where the overhead of storing MDTT bases (compared with the transformed coefficients) is very small so that it can be ignored.
  However, for short sequences (e.g., 17\_10 and 49\_02),
  the space usage for storing the basis functions in the MDTT is comparable to that of the transformed coefficients,
  leading to a large overhead.
  As a result, its compression performance is not as good as ours.
  For remaining sequences, the MDTT method is comparable to ours.

  Our clip-based method and the MDTT method have similar runtime performance,
  which can process more than 10,000 frames per second on an Intel Core i7-3770 CPU (3.40 GHz).

  In summary, both methods have merits.
  The mocap tailored transform is suitable for long motion sequences in the applications where large latency is tolerated,
  while our methods work for both short and long sequences and are desired for time-critical applications such as streaming.

\subsubsection{Comparison with the PCA-RDO Method}
  The PCA-RDO method~\cite{vavsa2014} is a PCA-based approach, which adopts PCA twice.
  In the first round, it applies PCA to the entire motion sequence to obtain reduced orthogonal basis of pose space.
  This PCA, called posed space PCA, is to explore the spatial correlation.
  Then, applying PCA to clips, it obtains orthogonal basis for joint trajectories.
  The second PCA, called temporal PCA, is for temporal decorrelation.
  With two rounds of PCA, the data dimension is reduced significantly.
  V{\'a}{\v{s}}a and Brunnett~\cite{vavsa2014} also proposed a general preprocessing step based on Lagrange multipliers,
  which allows the user to optimize with respect to various error metrics.

  Our clip-based method and the PCA-RDO method differ in several aspects:
  First, the PCA-RDO method is sequence-based, thus, it has large latency, whereas ours is clip-based and has low latency.
  Second, it is known that compression of the PCA's orthogonal basis is difficult,
  although their method adopts an advanced predictive coding \cite{vavsa2009cobra}.
  As Figures~\ref{fig:CRDclip}(a)(b)(c)(g) show, our clip-based scheme consistently outperforms the PCA-RDO method~\cite{vavsa2014} in terms of compression performance.
  Third, similar to the MDTT method, the PCA-RDO method is also low-rank approximation-based. So, it is not as flexible as ours.
  Fourth, the PCA-RDO algorithm has high computational cost and we observe that the speed of our clip-based method is 3 to 4 times faster than theirs.
  Last but not least, tuning the parameters of the PCA-RDO method is tedious and non-intuitive.
  In contrast, within our clip-based method, the user only needs to specify the clip length $L$, which directly controls the latency.

 Finally, Figures \ref{fig:visual frame} and \ref{fig:visual clip} show some visual results of our methods, the DCT-based, and the MDTT to further demonstrate the advantage of our methods.

\subsection{Discussion}
  We formulate the LSDT problem as a least square with \textit{orthogonal} constraint.
  In fact, a non-orthogonal matrix $\mathbf{B}^d$ may produce even better compression performance.
  However, one has to employ other constraints (e.g., using the determinant of $\mathbf{B}^d$ and Frobenius norm of
  $\mathbf{B}^d$) to ensure the learned matrix invertible (i.e.,
  ensure existence of the inverse transform) and a small condition
  number. Correspondingly, the optimization problem becomes
  complicated and it is difficult to solve.

\section{Conclusion}
  \label{sec:con}
  We presented frame- and clip-based methods for compressing mocap data with low latency.
  Taking advantage of the unique spatial characteristics,
  we proposed learned spatial decorrelation transform to effectively reduce the spatial redundancy in mocap data.
  Due to its data adaptive nature, LSDT outperforms the commonly used \textit{data-independent} transforms,
  such as discrete cosine transform and discrete wavelet transform, in terms of the decorrelation performance.
  Experimental results show that the proposed methods can produce higher compression ratios at a
  lower computational cost and latency than the state-of-the-art methods.

  In our current implementation, we compress 3D \emph{position-based} mocap data defined on a skeleton graph.
  However, it is straightforward to apply our methods to other types of mocap data, such as facial expressions, hand gestures and motion of human bodies.
  In the future, we will extend our methods to compress mocap data represented by Euler angles.
  Due to the nonlinear nature of angles, the hierarchical structure may produce significant accumulation errors in the compressed data~\cite{arikan2006,Chew2011}.
  We will seek effective data-driven techniques to tackle this challenge.

\bibliography{refs}

\begin{thebibliography}{10}
\expandafter\ifx\csname url\endcsname\relax
  \def\url#1{\texttt{#1}}\fi
\expandafter\ifx\csname urlprefix\endcsname\relax\def\urlprefix{URL }\fi
\expandafter\ifx\csname href\endcsname\relax
  \def\href#1#2{#2} \def\path#1{#1}\fi

\bibitem{capin1999avatars}
T.~Capin, I.~Pand{\v{z}}i{\'c}, N.~Magnenat-Thalmann, D.~Thalmann, Avatars in
  Networked Virtual Environments, John Wiley \& sons, 1999.

\bibitem{gutierrez2003controlling}
M.~Gutierrez, F.~Vexo, D.~Thalmann, Controlling virtual humans using pdas, in:
  Proceedings of the 9th International Conference on Multi-Media Modeling,
  2003, pp. 150--166.

\bibitem{chattopadhyay2007human}
S.~Chattopadhyay, S.~Bhandarkar, K.~Li, Human motion capture data compression
  by model-based indexing: A power aware approach, IEEE Transactions
  onVisualization and Computer Graphics 13~(1) (2007) 5--14.

\bibitem{gu2009}
Q.~Gu, J.~Peng, Z.~Deng, Compression of human motion capture data using motion
  pattern indexing, Computer Graphics Forum 28~(1) (2009) 1--12.

\bibitem{tournier2009}
M.~Tournier, X.~Wu, N.~Courty, E.~Arnaud, L.~Reveret, Motion compression using
  principal geodesics analysis, Computer Graphics Forum 28~(2) (2009) 355--364.

\bibitem{Lin2011}
I.-C. Lin, J.-Y. Peng, C.-C. Lin, M.-H. Tsai, Adaptive motion data
  representation with repeated motion analysis, IEEE Transactions on
  Visualization and Computer Graphics 17~(4) (2011) 527--538.
\newblock \href {http://dx.doi.org/10.1109/TVCG.2010.87}
  {\path{doi:10.1109/TVCG.2010.87}}.

\bibitem{vavsa2014}
L.~V{\'a}{\v{s}}a, G.~Brunnett, Rate-distortion optimized compression of motion
  capture data, Computer Graphics Forum 33~(2) (2014) 283--292.

\bibitem{Hou2014tvcg}
J.~Hou, L.-P. Chau, N.~Magnenat-Thalmann, Y.~He, Human motion capture data
  tailored transform coding, IEEE Transactions on Visualization and Computer
  Graphics 21~(7) (2015) 848--859.

\bibitem{hou2014}
J.~Hou, L.-P. Chau, N.~Magnenat-Thalmann, Y.~He, Scalable and compact
  representation for motion capture data using tensor decomposition, IEEE
  Signal Processing Letters 21~(3) (2014) 255--259.
\newblock \href {http://dx.doi.org/10.1109/LSP.2014.2299284}
  {\path{doi:10.1109/LSP.2014.2299284}}.

\bibitem{kwak2011hybrid}
C.-H. Kwak, I.~V. Bajic, Hybrid low-delay compression of motion capture data,
  in: Proceedings of IEEE International Conference on Multimedia and Expo
  (ICME), 2011, pp. 1--6.

\bibitem{arikan2006}
O.~Arikan, Compression of motion capture databases, ACM Transactions on
  Graphics 25~(3) (2006) 890--897.

\bibitem{liu2006segment}
G.~Liu, L.~McMillan, Segment-based human motion compression, in: Proceedings of
  the ACM SIGGRAPH/Eurographics SCA, 2006, pp. 127--135.

\bibitem{Chew2011}
B.-S. Chew, L.-P. Chau, K.-H. Yap, A fuzzy clustering algorithm for virtual
  character animation representation, IEEE Transactions onMultimedia 13~(1)
  (2011) 40--49.
\newblock \href {http://dx.doi.org/10.1109/TMM.2010.2082512}
  {\path{doi:10.1109/TMM.2010.2082512}}.

\bibitem{karni2004compression}
Z.~Karni, C.~Gotsman, Compression of soft-body animation sequences, Computers
  \& Graphics 28~(1) (2004) 25--34.

\bibitem{H264overview}
T.~Wiegand, G.~Sullivan, G.~Bjontegaard, A.~Luthra, Overview of the h.264/avc
  video coding standard, IEEE Transactions on Circuits and Systems for Video
  Technology 13~(7) (2003) 560--576.
\newblock \href {http://dx.doi.org/10.1109/TCSVT.2003.815165}
  {\path{doi:10.1109/TCSVT.2003.815165}}.

\bibitem{hevc2012}
G.~Sullivan, J.~Ohm, W.-J. Han, T.~Wiegand, Overview of the high efficiency
  video coding ({HEVC}) standard, IEEE Transactions on Circuits and Systems for
  Video Technology 22~(12) (2012) 1649--1668.
\newblock \href {http://dx.doi.org/10.1109/TCSVT.2012.2221191}
  {\path{doi:10.1109/TCSVT.2012.2221191}}.

\bibitem{gu2002geometry}
X.~Gu, S.~J. Gortler, H.~Hoppe, Geometry images, ACM Transactions on Graphics
  21~(3) (2002) 355--361.

\bibitem{HoufacialGV}
J.~Hou, L.-P. Chau, M.~Zhang, N.~Magnenat-Thalmann, Y.~He, A highly efficient
  compression framework for time-varying 3-d facial expressions, IEEE
  Transactions on Circuits and Systems for Video Technology 24~(9) (2014)
  1541--1553.
\newblock \href {http://dx.doi.org/10.1109/TCSVT.2014.2313890}
  {\path{doi:10.1109/TCSVT.2014.2313890}}.

\bibitem{KGV2014}
J.~Hou, L.~Chau, N.~Magnenat-Thalmann, Y.~He, Compressing 3-d human motions via
  keyframe-based geometry videos, IEEE Transactions on Circuits and Systems for
  Video Technology 25~(1) (2015) 51--62.
\newblock \href {http://dx.doi.org/10.1109/TCSVT.2014.2329376}
  {\path{doi:10.1109/TCSVT.2014.2329376}}.

\bibitem{preda2007optimized}
M.~Preda, B.~Jovanova, I.~Arsov, F.~Pr{\^e}teux, Optimized mpeg-4 animation
  encoder for motion capture data, in: Proceedings of the International
  Conference on 3D Web Technology, 2007, pp. 181--190.

\bibitem{beaudoin2007}
P.~Beaudoin, P.~Poulin, M.~van~de Panne, Adapting wavelet compression to human
  motion capture clips, in: Proceedings of Graphics Interface, 2007, pp.
  313--318.

\bibitem{Firouzmanesh2011}
A.~Firouzmanesh, I.~Cheng, A.~Basu, Perceptually guided fast compression of 3-d
  motion capture data, IEEE Transactions on Multimedia 13~(4) (2011) 829--834.
\newblock \href {http://dx.doi.org/10.1109/TMM.2011.2129497}
  {\path{doi:10.1109/TMM.2011.2129497}}.

\bibitem{zhu2012quaternion}
M.~Zhu, H.~Sun, Z.~Deng, Quaternion space sparse decomposition for motion
  compression and retrieval, in: Proceedings of the ACM SIGGRAPH/Eurographics
  SCA, 2012, pp. 183--192.

\bibitem{skodras2001jpeg}
A.~Skodras, C.~Christopoulos, T.~Ebrahimi, The jpeg 2000 still image
  compression standard, IEEE Signal Processing Magazine 18~(5) (2001) 36--58.

\bibitem{wang2012introduction}
R.~Wang, Introduction to orthogonal transforms: with applications in data
  processing and analysis, Cambridge University Press, 2012.

\bibitem{lowdimension}
A.~Safonova, J.~K. Hodgins, N.~S. Pollard, Synthesizing physically realistic
  human motion in low-dimensional, behavior-specific spaces, ACM Transaction on
  Graphics~(3) (2004) 514--521.

\bibitem{tan2013human}
C.-H. Tan, J.~Hou, L.-P. Chau, Human motion capture data recovery using
  trajectory-based matrix completion, Electronics Letters 49~(12) (2013)
  752--754.

\bibitem{tan2014motion}
C.-H. Tan, J.~Hou, L.-P. Chau, Motion capture data recovery using skeleton
  constrained singular value thresholding, The Visual Computer 31~(11) (2015)
  1521--1532.

\bibitem{donoho1998data}
D.~L. Donoho, M.~Vetterli, R.~A. DeVore, I.~Daubechies, Data compression and
  harmonic analysis, IEEE Trans. Information Theory 44~(6) (1998) 2435--2476.

\bibitem{cohen2002importance}
A.~Cohen, I.~Daubechies, O.~G. Guleryuz, M.~T. Orchard, On the importance of
  combining wavelet-based nonlinear approximation with coding strategies, IEEE
  Trans. Information Theory 48~(7) (2002) 1895--1921.

\bibitem{vavsa2009cobra}
L.~V{\'a}{\v{s}}a, V.~Skala, Cobra: Compression of the basis for pca
  represented animations, Computer Graphics Forum 28~(6) (2009) 1529--1540.

\end{thebibliography}

\end{document}